\documentclass{article}

\usepackage{epsfig}
\def\ligne#1{\hbox to \hsize{#1}}
\def\PlacerEn#1 #2 #3 {\rlap{\kern#1\raise#2\hbox{#3}}}
\def\cqfd{\hbox{\kern 2pt\vrule height 6pt depth 2pt width 8pt\kern 1pt}}
\def\leurre{\noindent\leftskip0pt\small\baselineskip 10pt}

\newtheorem{thm}{Theorem}
\newtheorem{lem}{Lemma}

\newtheorem{fig}{Figure}
\newtheorem{tab}{Table}

\font\ttviii=cmtt8

\font\ttvi=cmtt6

\def\grostrait{\ligne{\vrule height 1pt depth 1pt width \hsize}}
\def\demitrait{\ligne{\vrule height 0.5pt depth 0.5pt width \hsize}}

\begin{document}
\begin{center}
{\bf \Large A weakly universal cellular automaton in the hyperbolic $3D$ space with
three states
}
\vskip 5pt
Maurice {\sc Margenstern}
\vskip 2pt
Universit\'e Paul Verlaine $-$ Metz, IUT de Metz,\\
LITA EA 3097, UFR MIM,\\
Campus du Saulcy,\\
57045 METZ C\'edex 1, FRANCE\\
{\it e-mail}: {\tt margens@univ-metz.fr}
{\it Web page}: {\tt http;//www.lita.sciences.univ-metz.fr/\~{}margens}
\end{center}
\vskip 5pt
{\parindent 0pt\leftskip 20pt\rightskip 20pt
{\bf Abstract} $-$ In this paper, we significantly improve a previous result by the
same author showing the existence of a weakly universal cellular automaton with
five states living in the hyperbolic $3D$-space. Here, we get such a cellular automaton
with three states only. 
\par}

\section{Introduction}
\label{intro}
   In this paper, we follow the track of previous papers by the same author, with
various collaborators or alone, 
see~\cite{fhmmTCS,mm3DJCA,mmsyENTCS,mmsyPPL,mmarXiv4st,mmbook2}, 
which make use of the same basic model, the {\it railway model}, 
see~\cite{stewart,mmCSJMtrain,mmbook2} which we again define  in Section~\ref{railway}
for the self-containdness of the paper. Also for self-containdness, in 
Section~\ref{geometry}, we remind the reader what is needed to know about hyperbolic
geometry, focusing on the tilings of the hyperbolic plane and on the dodecagrid,
the tiling of the hyperbolic $3D$~space which we use for our implementation in
this paper.

   In~\cite{mmarXiv4st}, I proved the existence of a weakly cellular 
automaton on the heptagrid, a tiling of the hyperbolic plane, with four states only.
This result improves a published former result of the same author with a co-author,
see~\cite{mmsyENTCS}, in the same tiling of the hyperbolic plane, using a cellular 
automaton with 6~states.

   The reduction for 6~states to 4~states, using the same model, was obtained by
replacing the implementation of the tracks of the railway model. In all previous papers,
the track is implemented as a kind of one-dimensional structure: each cell of the track 
has two other neighbours on the track exactly, considering that the cell also belongs
to its neighbourhood. The locomotive follows the track by successively replacing two cells
of the track: the cells occupied by the front and by the rear of the locomotive. The 
locomotive has its own colours and the track has another one which is also different from
the blank, the colour of the quiescent state. In the mentioned paper, this traditional 
implementation is replaced by a new one, where the track in the traditional meaning is 
not materialized but suggested only. It is suggested by {\it milestones} which delimit 
it. It is important to notice that the milestone may not define a continuous structure.

   At this point, my attention was drawn by a referee of a submission to a journal
explaining the 4-state result that it is easy to implement rule~110{} in the
heptagrid, using three states only. This is true, but this trick produces an automaton
which is not really a planar automaton and does not improve our knowledge neither on
rule~110 nor on cellular automata in the hyperbolic plane. This implementation with 
three states is can also be easily adapted to the dodecagrid of the hyperbolic $3D$~space 
and suffers the same defect of bringing in no new idea.

   In this paper, we follow the same idea as in the heptagrid, which consists in changing 
the definition of the track. The milestones are also implemented in
two versions, as in the planar case. However, due to a particularity of the implementation,
the same pattern can be used to change directions, either inside a plane of the
hyperbolic $3D$~space or to switch from one plane to another one. This configuration is
used to avoid crossings, replacing them by bridges, as this was already performed
in~\cite{mm3DJCA}. Sections~\ref{the_tracks} and~\ref{the_switches} thoroughly
describe the implementation of the model in the hyperbolic $3D$~space. 
Section~\ref{the_rules} explains how the rules implement the model,
focusing on their rotation invariance, and also giving a short account on the computer
program which we used to perform the simulation and to check the correctness of the
rules.

   This will conclude our proof of the following result:

\begin{thm}\label{mainthm} {\rm (Margenstern)} $-$ 
There is a cellular automaton in the dodecagrid of the hyperbolic~$3D$ space which
is weakly universal and which has $3$~states. Moreover, the cellular automaton is rotation
invariant and its motion actually makes use of the three dimensions.
\end{thm}

   By the latter expression, we mean that the automaton cannot reduced to a lower 
dimension by a simple projection.

\section{The railway model}
\label{railway}

The model which we use in this paper is a schematizing of a railway circuit
initially devised by Ian Stewart, see~\cite{stewart}. This model was later
intensively used by the present author and several co-authors in
in~\cite{mmCSJMtrain,fhmmTCS,mm3DJCA,mmsyPPL,mmsyENTCS,mmbook2}, following the simulation
of a register machine defined in~\cite{mmCSJMtrain} in various tilings of the
hyperbolic plane and in one tiling of the hyperbolic $3D$ space.

   In this section, we represent a Euclidean implementation of the circuit
in order to facilitate the assimilation of the model by the reader. We turn to
the implementation in a hyperbolic context later, after a short reminder
of hyperbolic geometry and what is needed to remember about tilings in these
spaces.

The circuit uses tracks represented by lines and switches. Both make use of straight 
segments and quarters of circles. There are three kinds of switches: the {\bf fixed}, 
the {\bf memory} and the {\bf flip-flop} switches. They are represented by the schemes 
given in Figure~\ref{aiguillages}.

   A switch is an oriented structure: on one side, it has a single
track~$u$ and, on the the other side, it has two tracks~$a$ and~$b$. This
defines two ways of crossing a switch. Call the way from~$u$ to~$a$ or~$b$
{\bf active}. Call the other way, from~$a$ or~$b$ to~$u$ {\bf passive}. The
names comes from the fact that in a passive way, the switch plays no role on
the trajectory of the locomotive. On the contrary, in an active
crossing, the switch indicates which track between~$a$ and~$b$ will be followed by
the locomotive after running on~$u$: the new track is called the {\bf selected}
track.

\vskip 20pt
\vtop{
\setbox110=\hbox{\epsfig{file=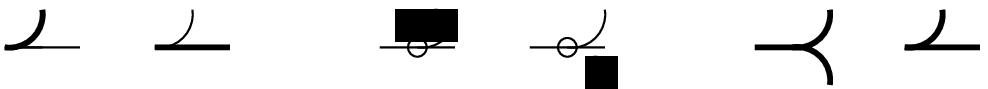,width=300pt}}
\ligne{\hfill
\PlacerEn {-325pt} {0pt} \box110
}
\vspace{-5pt}
\begin{fig}
\label{aiguillages}
\leurre
The three kinds of switches. From left to right: fixed, flip-flop and memory switches.
\end{fig}
}
\vskip 10pt

   With the help of these three kinds of switches, we define an
{\bf elementary circuit} as in~\cite{stewart}, which exactly contains one bit of
information. The circuit is illustrated by Figure~\ref{element}, below.
It can be remarked that the working of the circuit strongly depends on how
the locomotive enters it. 

\vtop{
\setbox110=\hbox{\epsfig{file=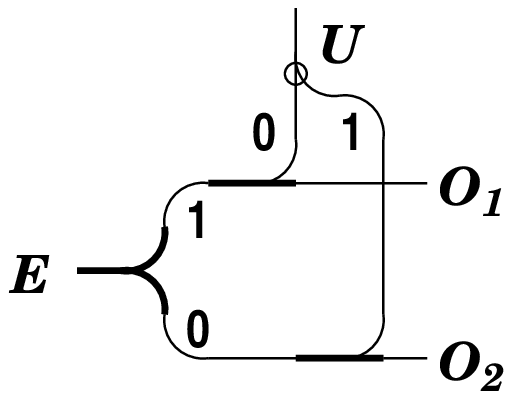,width=160pt}}
\ligne{\hfill
\PlacerEn {-285pt} {0pt} \box110
}
\vspace{-15pt}
\begin{fig}
\label{element}
\leurre
The elementary circuit.
\end{fig}
}
\vskip 10pt

If the locomotive enters the circuit through~$E$, it leaves it through~$O_1$ or~$O_2$, 
depending on the selected track of the memory switch which stands near~$E$. If the 
locomotive enters through~$U$, the application of the given definitions shows that 
the selected track at the switches near~$E$ and~$U$ are both changed: the switch 
at~$U$ is a flip-flop which is changed by the very active passage of the locomotive 
and the switch at~$E$ is a memory one which is changed because the locomotive is sent 
by the track imposed by~$U$ on a way which make it passively cross the switch through 
its non-selected track. The just described actions of
the locomotive correspond to a {\bf read} and a {\bf write} operation on the
bit contained by the circuit which consists of the configurations of the
switches at~$E$ and at~$U$. It is assumed that the write operation is triggered
when we know that we have to change this very bit.

   From this element, it is easy to devise circuits which represent different
parts of a register machine. As an example, Figure~\ref{unit} illustrates
an implementation of a unit of a register.

\vskip 10pt
   As indicated by its name, the {\bf fixed switch} is left unchanged by the
passage of the locomotive. It always remains in the same position: when
actively crossed by the locomotive, the switch always sends it onto the same
track. The flip-flop switch is assumed to be crossed actively only. Now,
after each crossing by the locomotive, it changes the selected track.
The memory switch can be crossed by the locomotive actively and passively.
In an active passage, the locomotive is sent onto the selected track. But the
selected track is defined by the track of the last passive crossing by the
locomotive. Of course, at initial time, the selected track is fixed.

Other parts of the needed circuitry are described
in~\cite{mmCSJMtrain,fhmmTCS}. The main idea in these different parts is
to organize the circuit in possibly visiting several elementary circuits
which represent the bits of a configuration which allow the whole system
to remember the last visit  of the locomotive. The use of this technique is
needed for the following two operations.

\vskip 10pt
\vtop{
\setbox110=\hbox{\epsfig{file=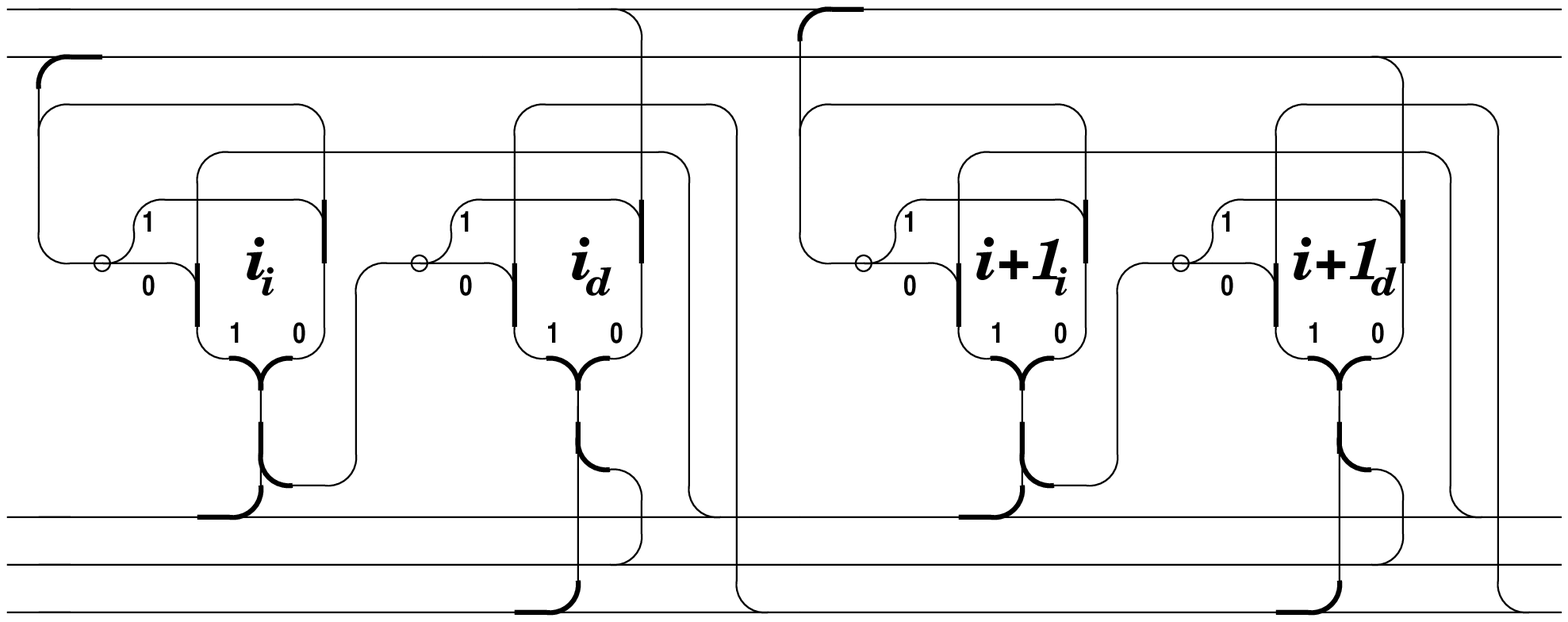,width=330pt}
\PlacerEn {-328pt} {24pt} {\small $i$}
\PlacerEn {-328pt} {14.5pt} {\small $d$}
\PlacerEn {-328pt} {5pt} {\small $r$}
\PlacerEn {-328pt} {131pt} {\small $j_1$}
\PlacerEn {-328pt} {121pt} {\small $j_2$}
}
\ligne{\hfill
\PlacerEn {-335pt} {0pt} \box110
\hskip 10pt}
\begin{fig}
\label{unit}
\leurre
Here, we have two consecutive units of a register. A register contains
infinitely many copies of units. Note the tracks $i$, $d$, $r$, $j_1$ and~$j_2$.
For incrementing, the locomotive arrives at a unit through~$i$ and it leaves the
unit through~$r$. For decrementing, it arrives though~$d$ and it leaves
also through~$r$ if decrementing the register was possible, otherwise, it leaves
through~$j_1$ or~$j_2$.
\end{fig}
}

   When the locomotive arrives to a register~$R$, it arrives either to
increment~$R$ or to decrement it. As can be seen on Figure~\ref{unit}, when the
instruction is performed, the locomotive goes back from the register by the
same track. Accordingly, we need somewhere to keep track of the fact whether
the locomotive incremented~$R$ or it decremented~$R$. This is one type of control.
The other control comes from the fact that several instructions usually apply
to the same register. Again, when the locomotive goes back from~$R$,
in general it goes back to perform a new instruction which depends on the one
it has just performed on~$R$. Again this can be controlled by what we called
the {\bf selector} in~\cite{mmCSJMtrain,fhmmTCS}.

At last, the dispatching of the locomotive on the right track for the next
instruction is performed by the {\bf sequencer}, a circuit whose main structure
looks like its implementation in the classical models of cellular automata such
as the game of life or the billiard ball model. The reader is referred to the
already quoted papers for full details on the circuit. Remember that this
implementation is performed in the Euclidean plane, as clear from
Figure~\ref{example} which illustrates the case of a few lines of a program of
a register machine.

\vskip 10pt
\vtop{
\setbox110=\hbox{\epsfig{file=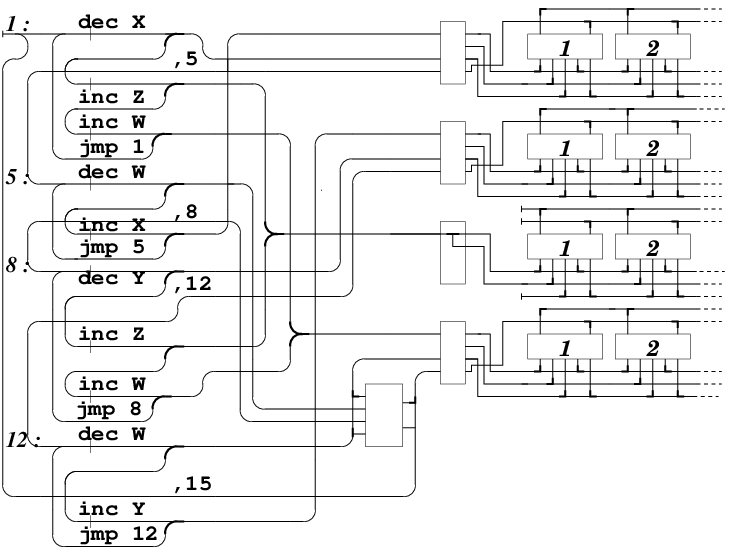,width=330pt}}
\ligne{\hfill
\PlacerEn {-335pt} {0pt} \box110
}
\vskip-15pt
\begin{fig}
\label{example}
\leurre
An example of the implementation of a small program of a register machine.
On the left-hand side of the figure, the part of the sequencer. It can be noticed
how the tracks are attached to each instruction of the program. Note that there
are four decrementing instructions for~$W$: this is why a selector gathers
the arriving tracks before sending the locomotive to the
control of the register. On the way back, the locomotive is sent on the right
track.
\end{fig}
}
\vskip 10pt
   Now, we turn to the implementation in the hyperbolic plane, which first
requires some features of hyperbolic geometry.

\section{Reminder of hyperbolic geometry}
\label{geometry}

   In this section, we give the essential features needed by the reader to 
understand the implementation. Of course, a reader who is well informed on hyperbolic
geometry may skip this section and go directly to Section~\ref{the_tracks}.

\subsection{Hyperbolic geometry}
\label{hypgeom}

Hyperbolic geometry appeared in the first half of the 19$^{\rm th}$ century,
in the last attempts to prove the famous parallel axiom of Euclid's {\it Elements}
from the remaining ones. Independently, Lobachevsky and Bolyai discovered a new geometry
by assuming that in the plane, from a point out of a given line, there are at
least two lines which are parallel to the given line. Later, models of the new
geometry were found, in particular Poincar\'e's model, which is the frame of
all this study.

\vskip 7pt
   In this model, the hyperbolic plane is the set of points
which lie in the open unit disc of the Euclidean plane whose border is the
unit circle. The lines of the hyperbolic plane in Poincar\'e's disc
model are either the trace of diametral lines or the trace of circles
which are orthogonal to the unit circle, see Figure~\ref{model}.
We say that the considered lines or circles {\bf support} the hyperbolic
line, simply {\bf line} for short, when there is no ambiguity, $h$-{\bf line}
when it is needed to avoid it. Figure~\ref{model}
illustrates the notion of {\bf parallel} and {\bf non-secant} lines in
this setting. In the model, parallels are characterized by the fact that they share
a common point on the border of the disc: this common point does not belong to
the hyperbolic plane and is called a {\bf point at infinity}.

\vskip 14pt
\vtop{
\setbox110=\hbox{\epsfig{file=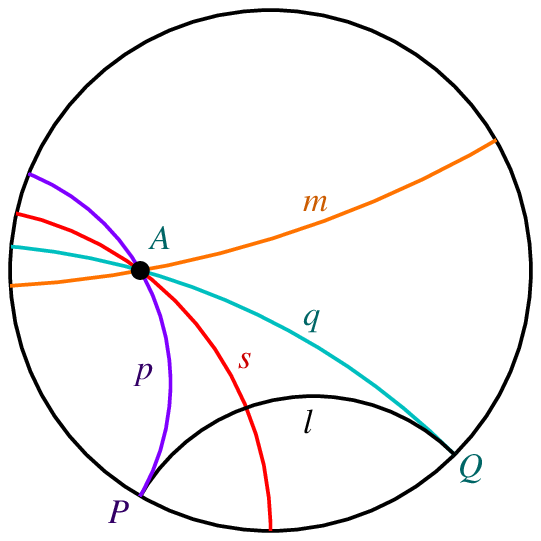,width=200pt}}
\ligne{\hfill
\PlacerEn {-130pt} {0pt} \box110
\hfill}
\vskip 0pt
\begin{fig}\label{model}
\leurre
The lines $p$ and $q$ are {\bf parallel} to the line~$\ell$, with points at
infinity~$P$ and~$Q$, on the border of the unit disc. The $h$-line $m$ is
{\bf non-secant} with $\ell$: it can be seen that there are infinitely
many such lines.
\end{fig}
}

   The angle between two $h$-lines are defined as the Euclidean angle between
the tangents to their support. The reason for choosing the Poincar\'e's model
is that hyperbolic angles between $h$-lines are, in a natural way, the
Euclidean angle between the corresponding supports. In particular, orthogonal circles
support perpendicular $h$-lines.

\subsection{Tilings in hyperbolic geometry}
\label{tilings}

   In this paper, we shall not give a general definition of a tiling. We shall
simply deal with a very special kind of tilings which is called a {\bf tessellation}.
Tessellations are themselves a particular member of the family of {\bf finitely
generated} tilings. This means that we have a finite set of closed bounded subsets
of the plane, called the {\bf prototiles} and that we can define a sequence of 
{\bf copies} of prototiles, the {\bf tiles}, in such a way that the union of 
the tiles cover the whole plane and that the interior of two copies neither intersect. 
By {\bf copy} of a prototile, we mean an isometric image, where a set of admissible
isometries has been fixed in advance. In a tessellation, there is a single prototile
which is a polygon and it is required that the copies are obtained from the prototile 
by reflection in its sides and, recursively, by reflection of the images in their sides.

From a famous theorem established by Poincar\'e in the late 19$^{\rm th}$ century,
it is known that there are infinitely many tilings in the hyperbolic plane, each one
being a tessellation based on a regular polygon~$P$. It is enough that the number~$p$ 
of sides of~$P$ and the number~$q$ of copies of~$P$ which can be put around a point~$A$ 
and exactly covering a neighbourhood of~$A$ without overlapping satisfy the relation:
\hbox{$\displaystyle{1\over p}+\displaystyle{1\over q}<\displaystyle{1\over2}$}.
The numbers $p$ and~$q$ characterize the tiling which is denoted $\{p,q\}$ and
the condition says that the considered polygons live in the hyperbolic plane. Note
that the three tilings of the Euclidean plane which can be defined up to similarities
can be characterized by the relation obtained by replacing~$<$ with~$=$ in the above
expression. We get, in this way, $\{4,4\}$ for the square, $\{3,6\}$ for the equilateral
triangle and $\{6,3\}$ for the regular hexagon.

\vskip 5pt
\vtop{
\setbox110=\hbox{\epsfig{file=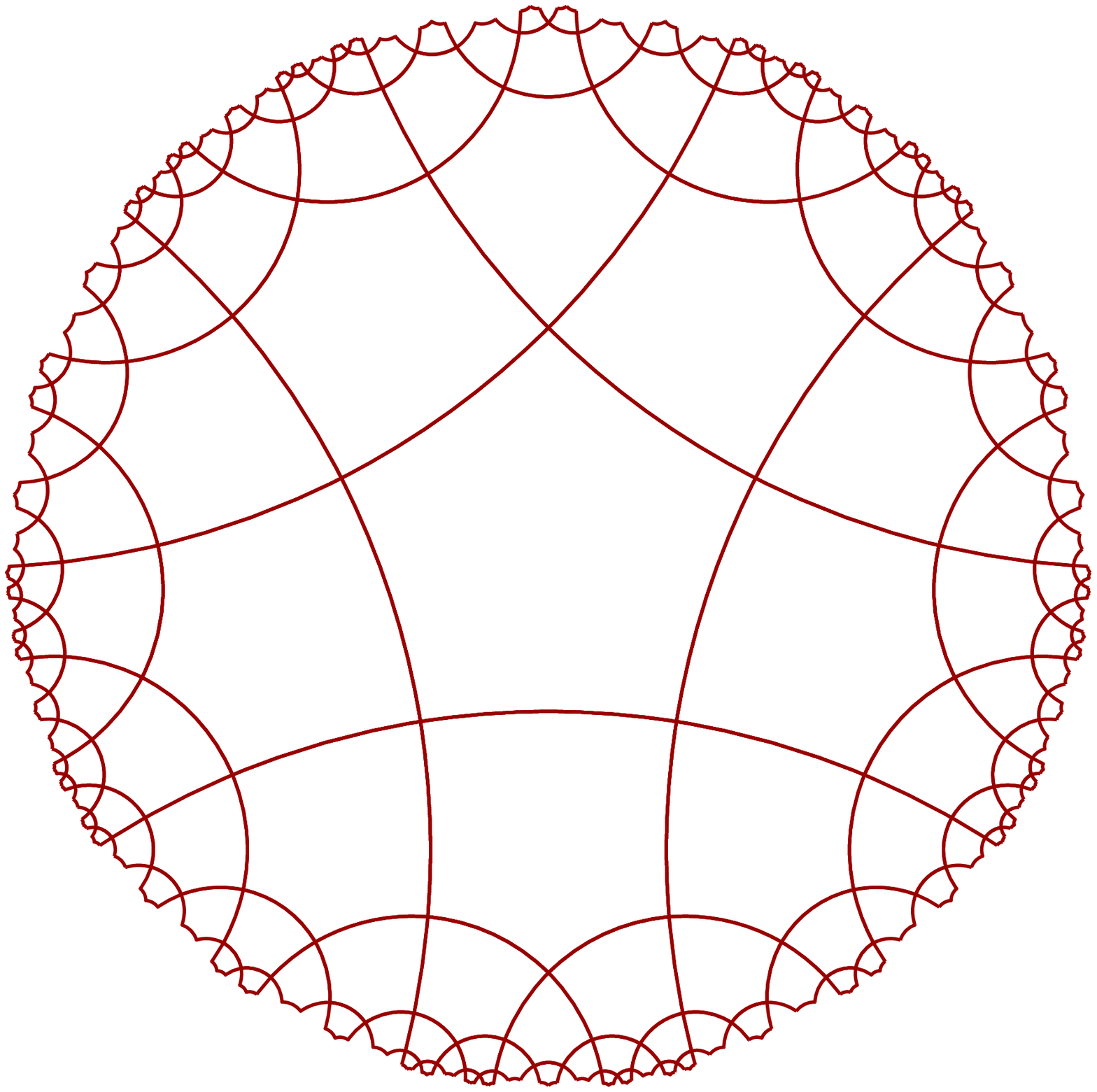,width=200pt}}
\ligne{\hfill
\PlacerEn {-130pt} {0pt} \box110
\hfill}
\vskip 0pt
\begin{fig}\label{tiling_54}
\leurre
An illustration of the pentagrid.
\end{fig}
}

   In the paper, we shall focus our attention on a single of these tilings, the simplest
one with a polygon with right angles: the tiling $\{5,4\}$ which is called 
the {\bf pentagrid}, see Figure~\ref{tiling_54}.

   Even in the small size of Figure~\ref{tiling_54} it is not that easy to distinguish each
tile of the tiling. This requires {\bf navigation tools} which we sketchily present in
Subsection~\ref{navigation}. 

\subsection{The dodecagrid}
\label{dodecagrid}

   Now, we turn to the case of the hyperbolic $3D$~space. Poincar\'e's theorem
tells us that there are infinitely many tessellations in the hyperbolic plane.
Geometers of the early 20$^{\rm th}$ century, see~\cite{sommerville} established
that there are only four tessellations in the hyperbolic $3D$~space. We shall
deal with one of them, which I call the {\bf dodecagrid} which is based on
Poincar\'e's dodecahedron: it is the dodecahedron which is obtained by assembling
twelve regular pentagons with right angles in a spatial closed figure, 
see~\ref{dodec}.

\vskip 14pt
\setbox110=\hbox{\epsfig{file=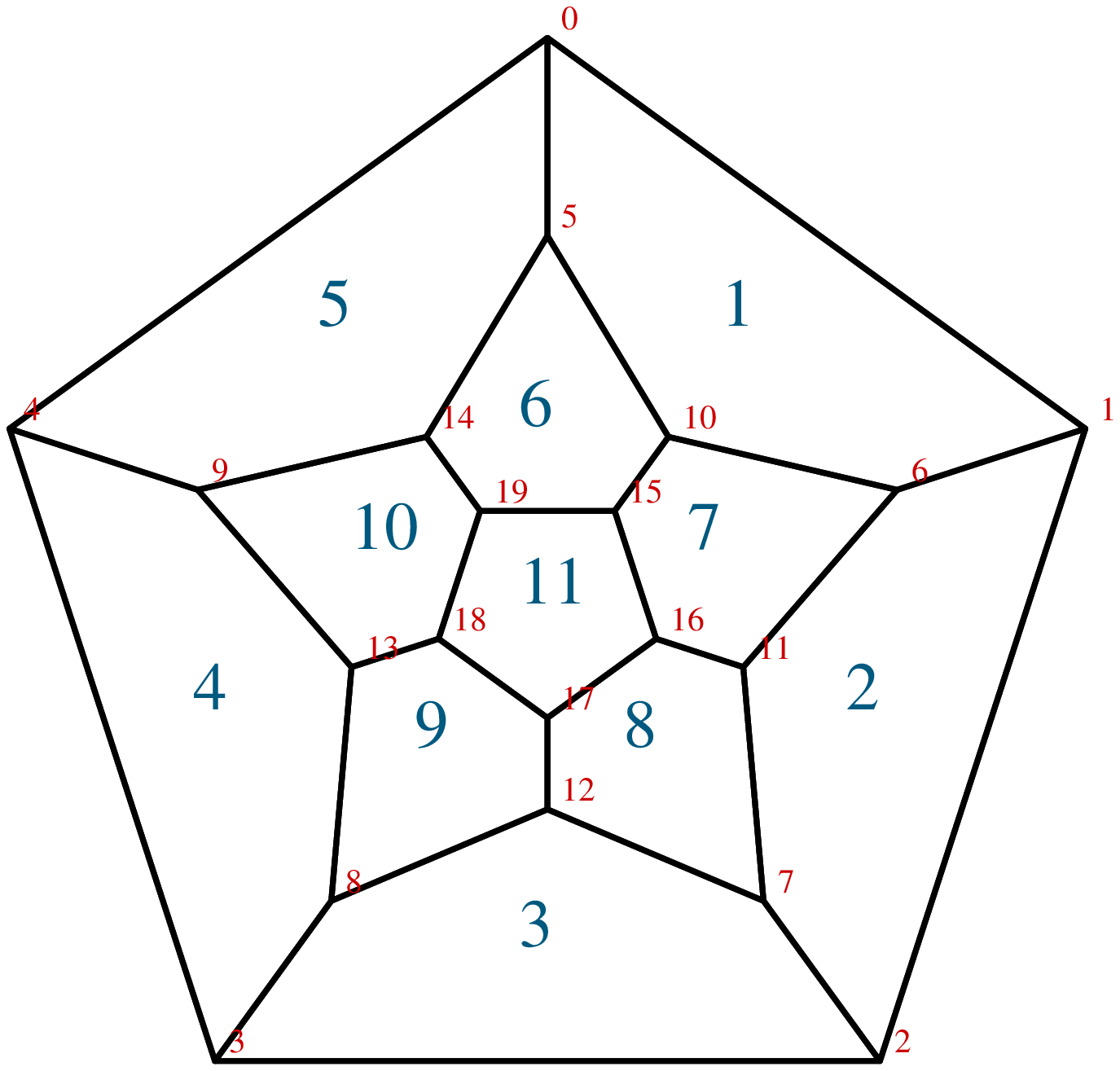,width=200pt}}
\vtop{
\vspace{-25pt}
\ligne{\hfill
\PlacerEn {-130pt} {0pt} \box110
\hfill}
\vspace{-30pt}
\begin{fig}\label{dodec}
\leurre
A dodecahedron. The faces are numbered from~$0$ to~$11$ and the vertices are numbered
from~$0$ to~$19$. Number~$0$ is not displayed for the corresponding face:
Face~$0$ is defined by vertices~$0$ up to~$5$.
\end{fig}
}
\vskip 7pt

   It can be proved that Poincar\'e's dodecahedron generates a tessellation of
the hyperbolic $3D$~space which is denoted by $\{5,3,4\}$ and which is called
the {\bf dodecagrid}. In $\{p,q,r\}$, $p$~indicates the number of sides of a face,
$q$~denotes how many faces share a common point and $r$~says how many polyhedra share
a common side. Note that there is another regular dodecahedron in the
hyperbolic $3D$~space which is called the {\bf big dodecahedron} as it is bigger
than Poincar\'e's one. It is based on another regular pentagon making a smaller
dihedral angle. The big dodecahedron also generates a tessellation of the hyperbolic
$3D$~space which is denoted by $\{5,3,5\}$. Accordingly, we can see that in
the tessellation generated by the big dodecahedron, there are five dodecahedra around
an edge instead of~4 of them in the dodecagrid.

   Figure~\ref{dodec} illustrates one dodecahedron. The numbering of the faces is a 
convention which we shall follow through out the paper. Figure~\ref{dodec}  makes use 
of a central projection. We can imagine that the centre of the projection is a point 
which is on the axis of
symmetry which is orthogonal to face~11 and which crosses both faces~11 and~0. We can 
imagine that this centre is outside the dodecahedron and above face~11, considering that
face~0 is below face~11. The projection is done onto the plane of face~0. This representation
is called a {\bf Schlegel diagram} after the name of the geometer of the 19$^{\rm th}$
who devised it in his study of various tessellations, including spatial ones.

   The representation of the dodecagrid is not an easy problem. In~\cite{mmgsFI} I have
shown how to use Schlegel diagrams in order to represent the dodecagrid itself inside
a single dodecahedron. As we shall not use this representation in this paper and we
refer the reader to~\cite{mmgsFI,mmbook1} for more details on this topic.

   We conclude this subsection by mentioning that in higher hyperbolic dimensions,
the problem almost vanishes. There are five tessellations in the hyperbolic
$4D$~space. One of them is thoroughly studied in~\cite{mmJUCSh4D,mmbook1}. However,
starting from dimension~5, there are no more tessellations in hyperbolic spaces,
again see~\cite{sommerville}.

\subsection{Navigation in the pentagrid and in the dodecagrid}
\label{navigation}

   Now, we turn to the question of locating tiles in the tessellations we shall
use in this paper: the pentagrid and the dodecagrid. This is the key point in order
to navigate in these spaces.

   The navigation in a tiling of the hyperbolic space can be compared to the flight
of a plane with instruments only. Indeed, we are in the same situation as a pilot
in this image as long as the representation of hyperbolic spaces in the Euclidean ones
entail such a distortion that only a very limited part of the hyperbolic space 
is actually visible.

\vskip 5pt
\vtop{
\setbox110=\hbox{\epsfig{file=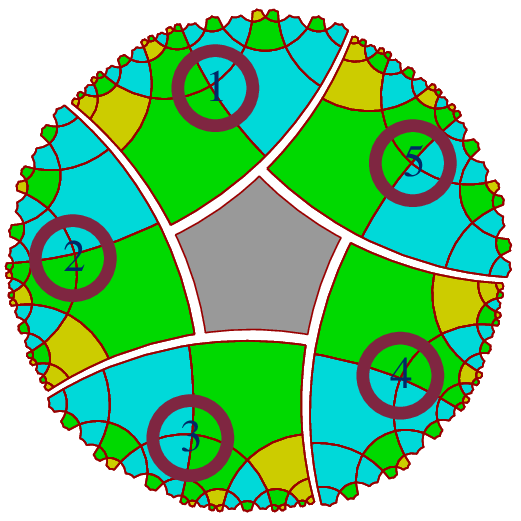,width=180pt}}
\ligne{\hfill
\PlacerEn {-100pt} {0pt} \box110
\hfill}
\vskip 0pt
\begin{fig}\label{eclate_54}
\leurre
The central tile and the five quarters around it, splitting exactly the hyperbolic plane.
\end{fig}
}
\vskip 5pt

   However, tools can be used which make this very limited visibility still useful,
constituting a true $GPS$ allowing to navigate in these spaces.

   The principle of the navigation algorithms rely on three ideas which can be applied in
many tilings, see~\cite{mmbook1} which we illustrate on the pentagrid by 
Figures~\ref{eclate_54}, \ref{split_54} and~\ref{fibo}.
 
   For what is the pentagrid, we first split the hyperbolic plane into six regions:
a central tile and five {\bf quarters} which are dispatched around the central tile
as illustrated by Figure~\ref{eclate_54}. The figure also indicates a numbering of 
the quarters which we shall use in the remaining sections of the paper.

   Then, we split each quarter as indicated in Figure~\ref{split_54}. The recursive 
structure of this splitting defines a tree which spans the tiling. The third idea 
consists in numbering the nodes of the tree level after level and, remarking that the 
number of nodes on the level~$n$ is $f_{2n+1}$ defined by the Fibonacci sequence 
where $f_0=f_1=1$, to represent the numbers in the numbering basis defined by this 
sequence. Also, as the representation is not unique, we fix it by choosing the longest one.

   On Figure~\ref{split_54}, we can notice that the Fibonacci tree has two kinds of nodes:
the {\bf white} ones, which have three sons, and the black ones, which have two sons.
In both cases, the leftmost son is black, the others are white. Figure~\ref{fibo}
represents the tree in a more traditional way together with the numbering of the nodes
and their representation in the Fibonacci basis. Later on, we shall call {\bf coordinate}
this representation of the number of a node.

\vskip 7pt
\setbox110=\hbox{\epsfig{file=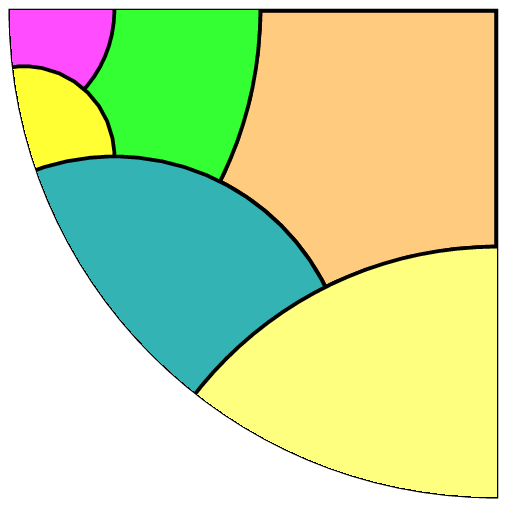,width=150pt}}
\setbox112=\hbox{\epsfig{file=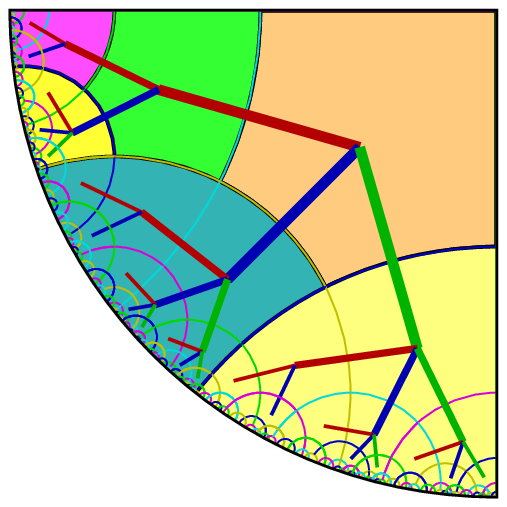,width=150pt}}
\vtop{
\ligne{\hfill
\PlacerEn {-330pt} {0pt} \box110
\PlacerEn {-170pt} {0pt} \box112
}
\begin{fig}
\label{split_54}
\leurre
Second part of the splitting: splitting a sector, here a quarter of the hyperbolic plane.
On the left-hand side: the first two steps of the splitting. On the right-hand side:
expliciting the tree which spans a sector.
\end{fig}
}

   There are important properties which are illustrated by Figure~\ref{fibo}
and which I called the {\bf preferred son properties}. It consists in the fact that 
for each node of a Fibonacci tree, among the coordinates of its sons, there is exactly 
one of them which is obtained
from the coordinate of the node by appending two 0's and which is called the preferred
son. Moreover, the place of the preferred son is always the same among the sons of a node:
the leftmost son for a black node, the second one for a white node.

\vskip 7pt
\setbox110=\hbox{\epsfig{file=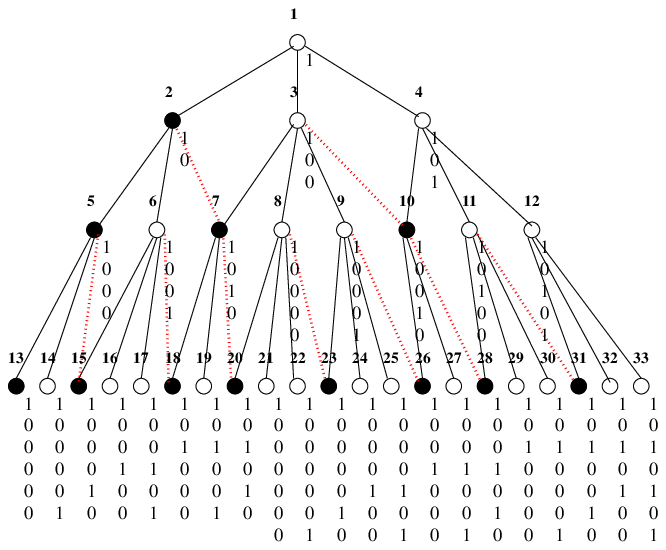,width=280pt}}
\vtop{
\ligne{\hfill
\PlacerEn {-140pt} {0pt} \box110
\hfill}
\vspace{-20pt}
\begin{fig}
\label{fibo}
\leurre
The Fibonacci tree: the representation of the numbers of the nodes in the Fibonacci
basis.
\end{fig}
}
\vskip 7pt

   From the preferred son properties, it was possible to devise an algorithm which
computes the path from the root to a node in a linear time in the length of the
coordinate of the node, see \cite{mmASTC}. From this, we also get that the coordinates
of the neighbours of a given tile can be computed from the coordinate of the tile in
linear time too.

\vskip 7pt
\setbox110=\hbox{\epsfig{file=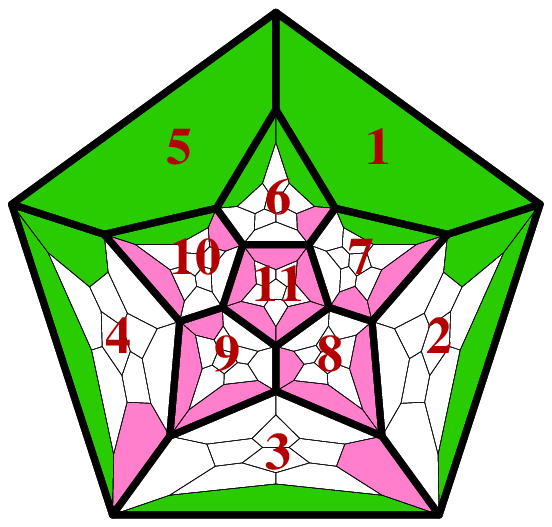,width=120pt}}
\setbox112=\hbox{\epsfig{file=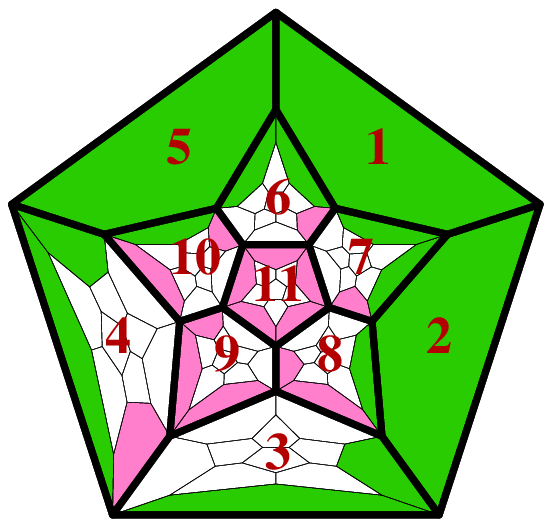,width=120pt}}
\setbox114=\hbox{\epsfig{file=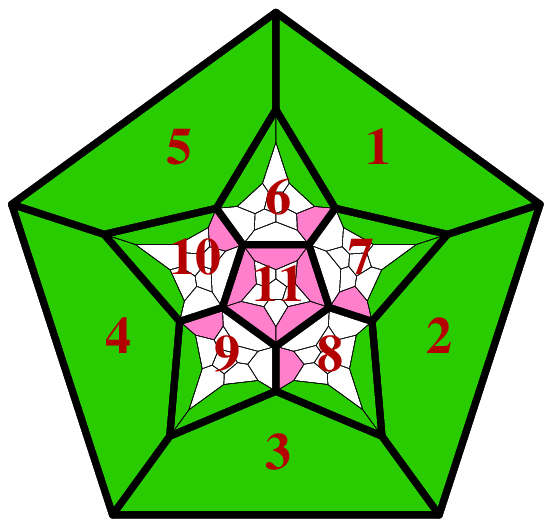,width=120pt}}
\vtop{
\ligne{\hfill
\PlacerEn {-330pt} {0pt} \box110
\PlacerEn {-220pt} {0pt} \box112
\PlacerEn {-110pt} {0pt} \box114
}
\vspace{-20pt}
\begin{fig}
\label{split_dod}
\leurre
Splitting a corner of the dodecagrid. On the left-hand side, the splitting of a corner.
On the middle, the splitting of a half-octant. On the right-hand side, the splitting
of a tunnel. Note that the faces are numbered according to the convention introduced
in Figure~{\rm\ref{dodec}}.
\end{fig}
}

    The same ideas can be used to define navigation tools for the dodecagrid.
Using Schlegel diagrams, we can also define a splitting of the hyperbolic $3D$~space,
this time into 8~corners around a point which is a common vertex. Next, we split the corner
as indicated in Figure~\ref{split_dod}. 

   The idea of the representation is, as in the pentagrid, to fix rules which allow to get
a bijection of a tree with the tiling restricted to the corner. If we allow the reflection
of any dodecahedron of the tiling in its faces, we shall get many doubled replications
as explained in Figure~\ref{explic3D}.

\setbox110=\hbox{\epsfig{file=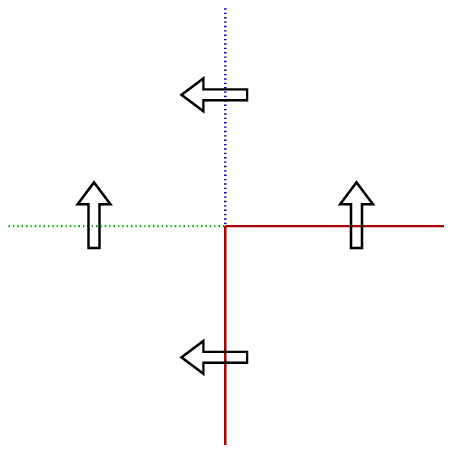,width=150pt}
                 \PlacerEn {-30pt} {77pt} 6
                 \PlacerEn {-82pt} {50pt} {10}
                 \PlacerEn {-60pt} {95pt} {reflected}
                 \PlacerEn {-140pt} {35pt} {reflected}
                 \PlacerEn {-140pt} {110pt} {collision}
                 \PlacerEn {-82pt} {125pt} {$A$}
                 \PlacerEn {-137pt} {72pt} {$B$}
                 \PlacerEn {-47pt} {35pt} {$D$}
                 \PlacerEn {-47pt} {110pt} {$D_6$}
                 \PlacerEn {-120pt} {50pt} {$D_{10}$}
}
\ligne{\hfill
\PlacerEn {-250pt} {-145pt} \box110
}
\begin{fig}\label{explic3D}
Why we can see the necessity of inhibition rules prohibiting
each second reflection producing an already produced dodecahedron.
\end{fig}
\vskip 7pt

Look at faces~10 and~6. Consider the bisector plane of the edge~$e$ shared by both those 
faces. In this plane, the trace of the faces defines two lines which intersect at 
right angles on the projection of~$e$. The reflection of the dodecahedron in faces~6 
and~10 produce two dodecahedra $D_6$ and~$D_{10}$. Let~$A$ be the face~$D_6$ which is 
in contact with~$e$ and let~$B$ be that of~$D_{10}$ which is also in contact with~$e$,
see Figure~\ref{explic3D}. 
We know that around~$e$ there are exactly four dodecahedra in the tiling. We have already
three of them with our initial dodecahedron, $D_6$ and~$D_{10}$. The last one can be
obtained either by the reflection of~$D_6$ in~$A$ or by the reflection of~$D_{10}$ 
in~$B$. In order to get a bijection, one reflection must be ruled out and the other 
permitted. This is the reason of the coloured faces in Figure~\ref{split_dod}. The 
reflection of a dodecahedron is forbidden in a coloured faces. It is mandatory in a white
face.

   The tree which we obtain has not the nice properties of his cousin of the pentagrid.
But still, it allows to obtain a system of coordinates which works in polynomial time, in
fact in cubic time at most.

   We refer the reader to~\cite{mmbook1}. However, the splitting suggested by 
Figure~\ref{split_dod} is a bit different from that indicated in~\cite{mmbook1}. The
difference is that the splitting of Figure~\ref{split_dod} is more symmetric and it
involves three basic regions instead of four ones in the splitting of~\cite{mmbook1}.

   In the paper, we shall not directly use the tree. Taking into account that
most of the circuitry will occur in a plane, we shall use projections onto that plane.
Now, we can chose as plane of projection, the plane of a face of a fixed dodecahedron.
We obtain that the restriction of the dodecagrid to this plane is a copy of the
pentagrid. And so, for the projections we have in mind, we can use the pentagrid.
\vskip 24pt
\setbox110=\hbox{\epsfig{file=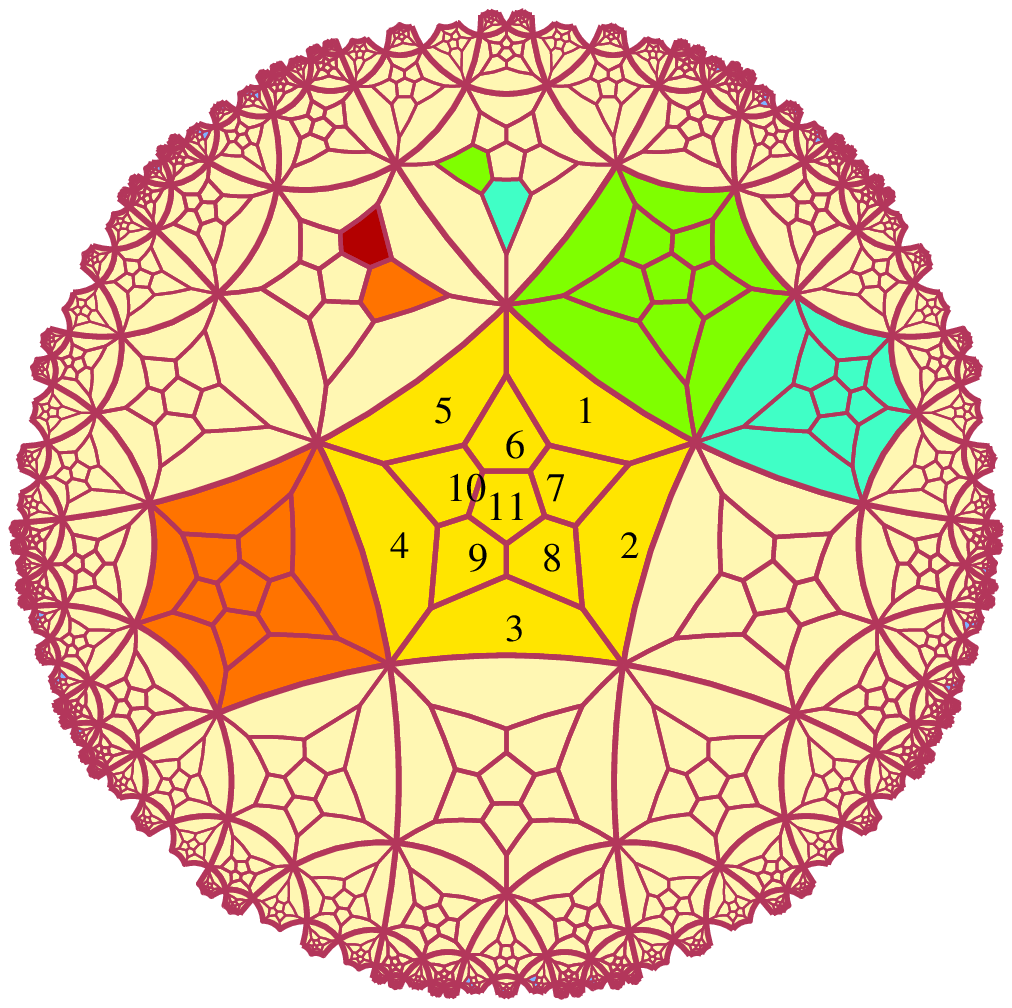,width=190pt}}
\setbox112=\hbox{\epsfig{file=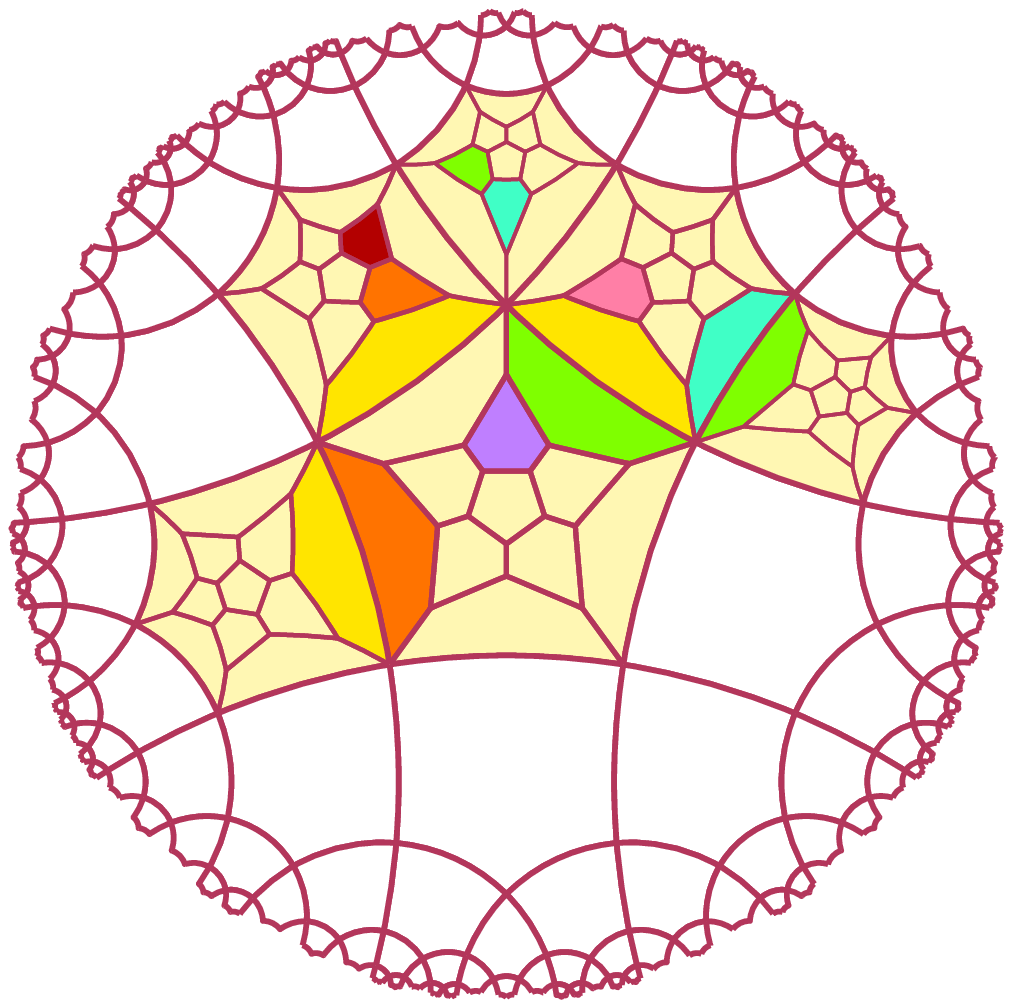,width=190pt}}
\vtop{
\vspace{-30pt}
\ligne{\hfill
\PlacerEn {-360pt} {0pt} \box110
\PlacerEn {-190pt} {0pt} \box112
}
\vspace{-15pt}
\begin{fig}
\label{projections}
\leurre
Two different ways for representing a pseudo-projection on the pentagrid.
On the left-hand side: the tiles have their colour. On the right-hand side: the colour of 
a tile is reflected by its neighbours only.
\end{fig}
}

   Fix a plane~$\Pi_0$ of the dodecagrid which supports one face of a dodecahedron. We
know that the trace of the dodecagrid on this plane is a copy of the pentagrid. We each
tile of the pentagrid is a face of exactly one tile of the dodecagrid over~$\Pi_0$.
We draw a Schlegel diagram of the corresponding dodecahedron within its face which lies
on~$\Pi_0$. We shall call this a {\bf pseudo-projection onto~$\Pi_0$}. Imagine that we 
have four tiles~$O$, $Y$, $G$ and~$B$ defined by their 
respective colour, orange, yellow, green and blue. Imagine that another tile~$W$, a 
white one, sees~$Y$ through its face~1, the same face being numbered~5{} in~$Y$, 
the face~1 of~$Y$ being that which is shared with~$G$. Then, we can see two other 
tiles on~$W$, a red one and an orange one, on faces~10 and~6 respectively.  

  On the left-hand side of Figure~\ref{projections}, the tiles keep their colour.

We can see that this raises a problem with the tiles which are put on the faces~6 and~10
of the tile~$W$, in case $W$~would be blue, for instance. For: what colour should be that
of face~6? The colour of~$W$ or the colour of the other dodecahedron which shares it 
with~$W$? Another problem is given, for instance by the faces~1 of~$Y$ and~$G$, assuming
that the face~1 of~$G$ is that which sees~$Y$. In fact, as can easily be seen by the fact
that these faces are both perpendicular to~$\Pi_0$ and that they share a common edge
lying in~$\Pi_0$, these faces coincide. The fact that they have different colour is a bit
misleading.  

   In order to fix a choice allowing us to keep as much information as possible
in the pseudo-projection, we shall consider that a face of tile does not show the colour
of the tile but the colour of its neighbour sharing the same face. The right-hand side
of Figure~\ref{projections} shows the same configuration as the one of the left-hand side,
but under the new convention. Also, to make the figure more readable, we do not
draw the pseudo-projection of a tile who would be white with only white neighbours among 
those of its neighbours which do not touch~$\Pi_0$. Now we cans ee that we can use the 
fact that two different faces coincide in the $3D$~space by indicating the colour of
the other tile.  

   Later on, we shall adopt this second solution to represent our 
cellular automaton. Indeed, the cells of the cellular automaton are the dodecahedra of
the dodecagrid and the colour of a tile is given by the state of the cellular automaton
at the considered dodecahedron.

   Before turning to the next section, we have an important remark.

   We have already indicated that, in the pseudo-projection, faces which share an edge 
but which belong to different dodecahedra do coincide in the hyperbolic $3D$~space.
The consequences are important with respect to the neighbours of a tile~$\tau$ where, 
by neighbour, we mean a polyhedron which shares a face with~$\tau$. On the left-hand 
side part of Figure~\ref{projections}, we can see four small faces coloured with~$r$,
$o$, $g$ and~$b$ on two white dodecahedra which we call~$W_1$ and~$W_2$, with $W_1$
being a neighbour of the central cell and $W_2$ a neighbour of the green neighbour of
the central cell. We can view these coloured faces as dodecahedra obtained 
from the dodecahedron to which the face belongs by reflection in the very face. Call 
these dodecahedra by the colour of their defining faces. Considering the planes of the 
faces and their relations with~$\Pi_0$, it is not difficult to see that the 
dodecahedra~$r$ and~$g$ are neighbours as well as the dodecahedra~$o$ and~$b$. However, 
despite the fact that the corresponding faces share an edge, the dodecahedra~$r$ and~$o$ 
are not neighbours. However, $r$, $o$ and~$W_1$ share a common edge: this means that there
is a fourth dodecahedron~$\delta$ sharing this edge which is not represented in the figure.
Now, $\delta_1$ plays an important role for both~$r$ and~$o$ as it is a neighbour for both
of them. The same remark holds for the dodecahedra~$g$ and~$b$ for which there is
a dodecahedron~$\delta_2$, a neighbour of both dodecahedra, sharing a common edge also
with~$W_2$. Moreover, it can be seen that $\delta_1$ and~$\delta_2$ are also neighbours:
their common face is in the plane of the common face of~$W_1$ and~$W_2$, which also
contains the common face of~$r$ and~$g$ as well as the common face of~$o$ and~$b$.

   Both couples $r$ with~$g$ and $o$ with~$b$ can also be seen on the right-hand side 
part of Figure~\ref{projections}. There are also two other coloured small faces: a 
purple one on the central tile, call it~$p$ as well as the dodecahedron which it defines. 
There is also a pink one on the green dodecahedron which is a neighbour of the central 
one. Call the pink dodecahedron $\pi$. It is not difficult to see that the following 
pairs of dodecahedra are neighbours in the dodecagrid: $b$ and~$\pi$, $\pi$ and~$p$ 
as well as $p$ and~$o$. Moreover, the four dodecahedra $o$, $b$, $\pi$ and~$p$ share 
a common edge which belongs to the same line as the one which supports the edge 
shared by~$W_1$, $W_2$, the central tile and~$G$.

   At last, remark that the plane of a face of a dodecahedron~$D$ defines two half-spaces: 
the half-space which does not contain~$D$ contains one neighbour of~$D$ exactly. The
half-space which contains~$D$ contains all the other neighbours of~$D$ also. This can
be seen as a consequence of the convexity of the dodecahedron.

   Now, we can turn to the implementation of the railway model in the dodecagrid.

\section{Implementation of the tracks}
\label{the_tracks}

   The implementation of the model is much more difficult in the hyperbolic $3D$-space than
in the hyperbolic plane. Speaking about implementations in the hyperbolic plane,
I often use the metaphor of a pilot flying with instruments only. This can be reinforced
in the case of the hyperbolic $3D$-space by saying that this time we are in the situation
of an astronaut who can do no other thing than fly with instruments only: sometimes,
the astronaut may look at the earth. It is a fantastic image, however of no help for
the navigation in cosmos. For the dodecagrid, we hope that the method explained
in Subsection~\ref{navigation} shows that the situation is after all a bit better than
in cosmos. The figures which we can obtain from the projections defined in 
Subsection~\ref{navigation} may help the reader to have a satisfactory view of the 
situation. We have to never forget that the views we can obtain are dramatically 
simplified images of what actually happens. However, always bearing in mind that the 
images are always a local view, a good training based on rigorous principles may 
transform then into an efficient tool. 

   Remember that in most its parts, the track followed by the locomotive
runs on a fixed plane~$\Pi_0$ of the hyperbolic $3D$~space. Only occasionally, it switches 
to other planes, perpendicular to~$\Pi_0$. In particular, this is the case for the 
implementation of crossings: as in~\cite{mm3DJCA}, we take advantage of the third 
dimension in order to replace them by bridges. For the sensors too, we shall take 
benefit of the third dimension to differentiate the configurations of the various switches.

   In this section, we deal with the tracks only, postponing the implementation of the
switches to Section~\ref{the_switches}.

\subsection{The pieces}
\label{elements}

   Below, Figure~\ref{track_straight} illustrates a copy of the most common element of the 
tracks. Later on, we call this element the {\bf straight} one. This element consists 
of a single dodecahedron, the track itself, marked by four blue dodecahedra, the 
{\bf milestones}, which are neighbours of this dodecahedron.

   Note the numbering of the faces on the Figure: it follows the convention 
introduced in Subsection~\ref{navigation}. In Figure~\ref{track_straight}, 
pictures~$(a)$ and~$(b)$, face~0 is not visible but it is visible in the other pictures. 
Similarly, face~5 and face~2 respectively, are not visible in pictures~$(c)$ with~$(d)$ 
and $(e)$ with~$(f)$ respectively. Usually, this not visible face will be called the 
face of the plane of the element or its {\bf back}, as we did in~\cite{mmJUCSh4D,mmbook1}, 
or also its {\bf bottom}. Now, due to the role of the elements in the circuit, we shall 
say that face~1 is the {\bf entry} of the element and that faces~3 and~4 are its 
{\bf exits} in the case of pictures~$(a)$ and~$(b)$. In the case of pictures~$(c)$ 
with~$(d)$ and $(e)$ with~$(f)$ respectively, the entries are face~4 with face~10 and 
face~3 with face~8 respectively. We shall say {\bf exit~3}, {\bf exit~4}, {\bf exit~8} or 
{\bf exit~10} if we need to make it more accurate. It is important to notice that
exits and entries can be exchanged: we can have exit~1 and entry~3 but not exit~3
and entry~4. Such a change of direction is necessary, but it will be realized by
another element. As the role of entry and exits can be exchanged, we shall use
the word {\bf exit} in general descriptions with the possible meaning of both an
entry or an exit through the possibly indicated face.

\setbox110=\hbox{\epsfig{file=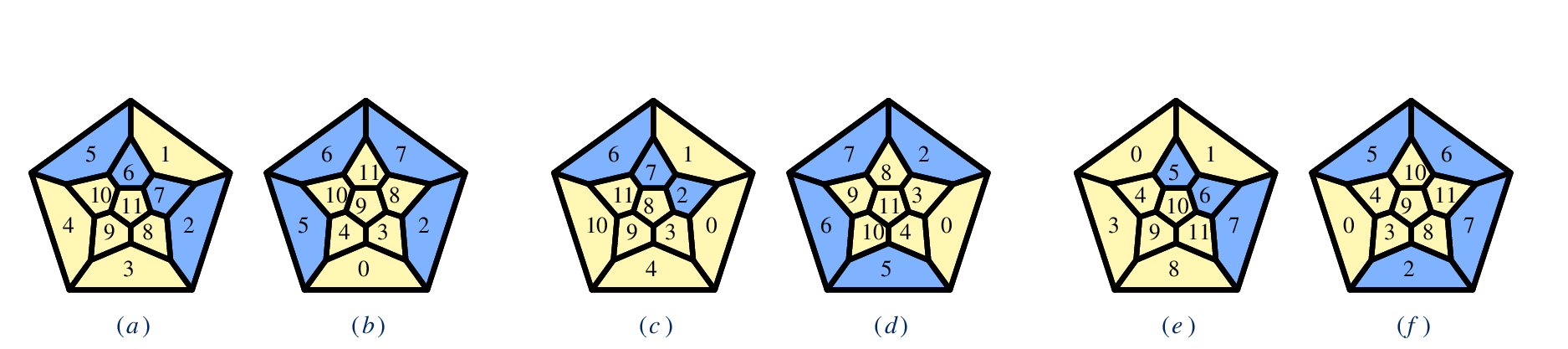,width=340pt}}
\vtop{
\vspace{-15pt}
\ligne{\hfill
\PlacerEn {-340pt} {0pt} \box110
}
\vspace{-10pt}
\begin{fig}\label{track_straight}
\leurre
An ordinary element of the track. Figure~$(a)$ is a view from above. Figure~$(b)$ 
is a view from the back of face~$1$. The locomotive enters the element via face~$1$
and exits via face~$3$ or face~$4$. It also may enter via face~$3$ or face~$4$ and then
it exits through face~$1$.\vskip 2pt
In Figures~$(c)$ and~$(d)$, the element is a bit turned around face~$1$ and the exits
are now face~$(4)$ and~$(10)$. In Figures~$(e)$ and~$(f)$, the element is turned
around face~$(1)$ too, but in the opposite direction, and the exits are now faces~$3$
and~$8$.\vskip 2pt The motion in the opposite direction is always possible.
\end{fig}
}

Remember the convention we introduced in Subsection~\ref{navigation}. 
Ff not otherwise mentioned, the face~$F$ of a cell has the colour 
of the state of the neighbour of the cell sharing~$F$ with it. Accordingly, in most
figures of the paper, the colour of a cell can be deduced from the colours of the face of
its neighbours. As an example, in the pictures of Figure~\ref{track_straight}, the
milestones are blue and they are neighbours of the element.

   As the name suggests, the milestones are usually fixed elements: they are not changed 
by the passage of the locomotive. This means that the milestones always remain blue, 
while the track is white as most cells of the space itself: the white state plays the role
of the quiescent state: if a cell is white as well as all its neighbours, then it 
remains white. 

\setbox110=\hbox{\epsfig{file=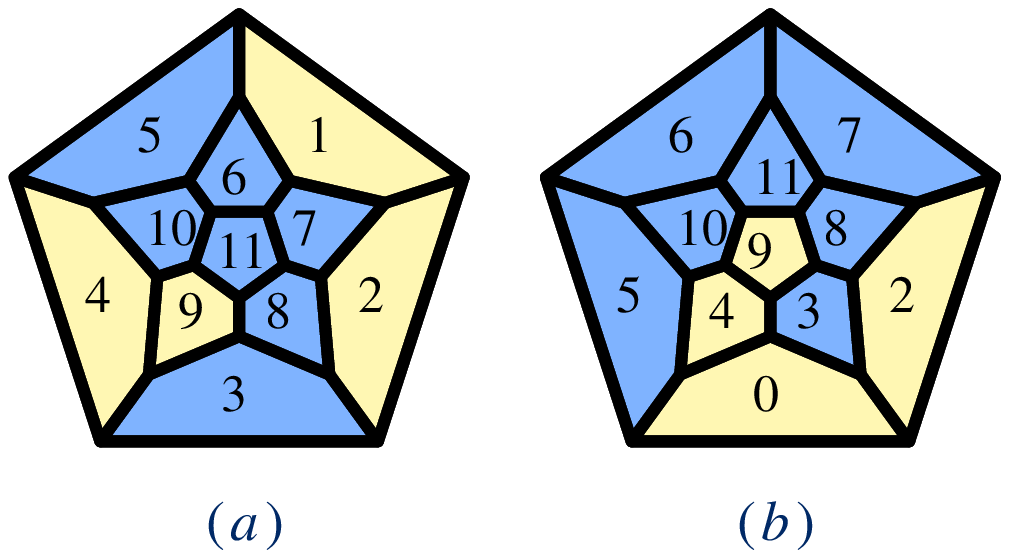,width=200pt}}
\vtop{
\vspace{-15pt}
\ligne{\hfill
\PlacerEn {-280pt} {0pt} \box110
}
\vspace{-10pt}
\begin{fig}\label{track_corner}
\leurre
The corner element of the track. Figure~$(a)$ is a view from above. Figure~$(b)$ 
is a view from the back of face~$1$. The locomotive enters the element via face~$1$
and exits via face~$2$. It also may enter via face~$2$ and then
it exits through face~$1$.
\end{fig}
}

In Figure~\ref{track_straight}, the pictures represent various positions of the same 
elements which can be obtained from each other by a rotation a face of the dodecahedron or 
by a product of such rotations. We refer the reader to Subsection~\ref{rotation_invariance}
where this problem is examined. In the figure, pictures~$(a)$ and~$(b)$ show
a situation where the plane of the railways is that of face~0. In the pictures~$(c)$
and~$(d)$, the plane is that of face~5. In the pictures~$(e)$ and~$(f)$, it is that of
face~2. The milestones can be viewed as the materialization of a catenary over the
track itself, assumed to be put on the plane of the element.

 Figure~\ref{track_corner} illustrates another element of the track which we
call a {\bf corner}. This element allows the locomotive to perform a turn at a right angle.
This possibility is very important and absolutely needed, as we shall see later.

   As we can see from Figure~\ref{track_corner}, the corner has more milestones around it
then a straight element: 7 milestones instead of 4~ones. However, the back of a corner 
is white, while that of a straight element is usually blue. 

\subsection{Vertical segments}
\label{vert}
   
   When finitely many straight elements are put one after each other, with the entry of 
one of them shared by the exit of the previous one, we say that these elements are set 
into a {\bf vertical segment}, {\bf vertical} for short, provided that the plane
of these elements is the same and that there is a line of this plane which
supports one side of each element which we call the {\bf guideline}. 
Figure~\ref{idle_vert} illustrates the basic example of a vertical. The guideline
supports a side of the faces~0 of the elements and the common plane is that of
the faces~5.

In the representation of Figure~\ref{idle_vert}, the dodecahedra are projected
on the plane of their back, face~5.

\vskip 14pt
\setbox110=\hbox{\epsfig{file=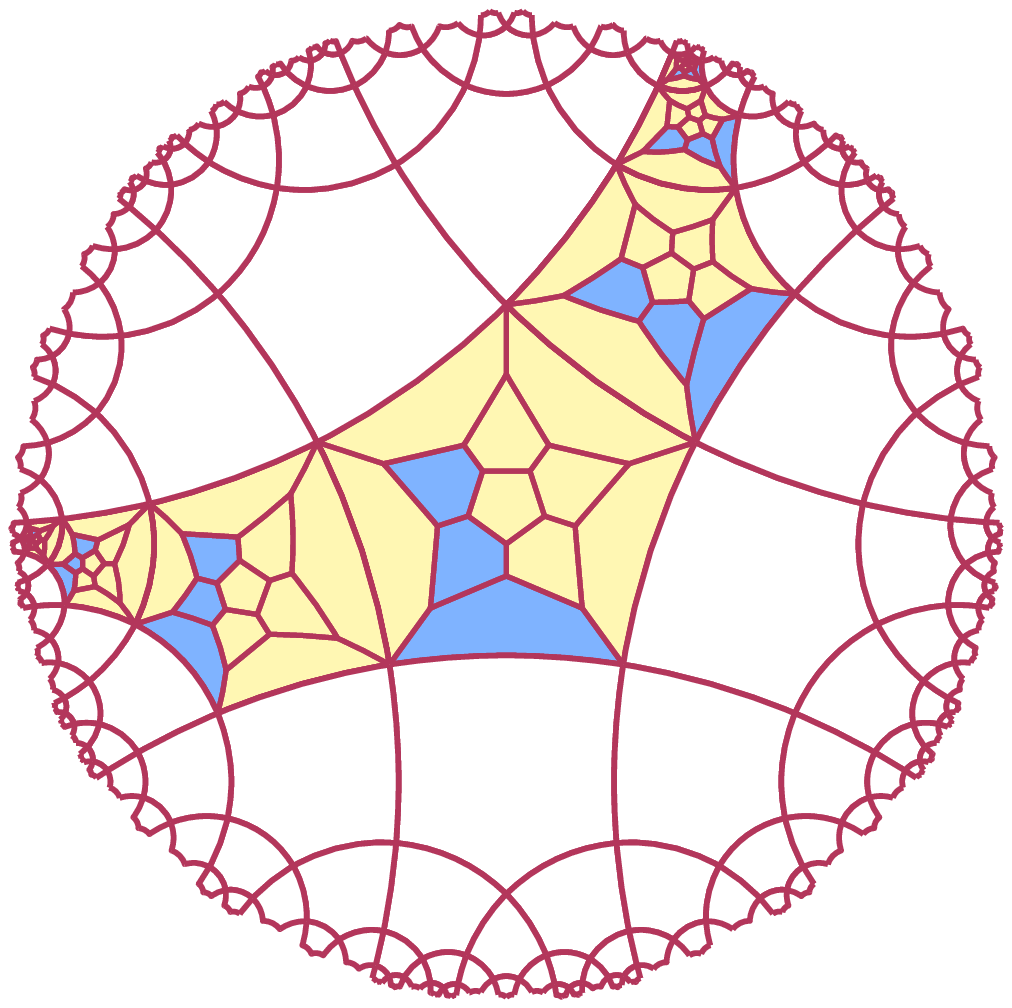,width=240pt}}
\vtop{
\ligne{\hfill
\PlacerEn {-130pt} {0pt} \box110
\hfill}
\vspace{-20pt}
\begin{fig}\label{idle_vert}
\leurre
Pseudo-projection on the plane of the track of its elements in the case of a vertical
segment.
\end{fig}
}

Number the elements of the figure from~1 to~7. We can see that the element~$i$ is in 
contact with the element~$i$+1 with $i\in\{1..6\}$. Consider elements~3 and~4, the 
latter one occupying the central pentagon of the figure. The exit~4 of element~4
and the entry~1 of element~3 appear as different faces of dodecahedra: each one is 
projected inside the face~5 of the dodecahedron. Now, by definition, the entry~1 of
element~3 and the exit~4 of the element~4 coincide. Indeed: elements~3 and~4 have 
their faces~5 on a common plane. They also have their sides~0 on the guideline. 
The entry~1 of element~3 and the exit~4 of element~4 are perpendicular to the 
guideline and they share a common side: they are the same face.
   As we stressed in Subsection~\ref{navigation}, this situation is important and 
we shall not repeat this point systematically. It is a property of the 
hyperbolic $3D$~space which we have to bear in mind while looking at the figures.

   Note that in the figure, the entry~1 of an element is connected with the exit~4 
of the previous one. Of course, the segment can be run in the opposite direction: then 
an exit~4 becomes an entry~4 and an entry~1 becomes an exit~1.

\subsection{Horizontal segments}
\label{horiz}

   In Figure~\ref{idle_horiz}, we represent another kind of track which we shall
call {\bf horizontal segments}. Such tracks consists of finitely many elements which can
be written as a word of the form $(SeC)^k$, where $Se$~denotes a straight element
and $C$~denotes a corner. The entry of the corner abuts an exit of the straight
element. It is not always the same exit. In fact, there is an alternation of the
exits which makes a Fibonacci word: if we associate to $SeC$ the number of the exit
of the straight element which abuts the entry of the corner, then this defines a 
homomorphism 
\ligne{\hfill}
\setbox110=\hbox{\epsfig{file=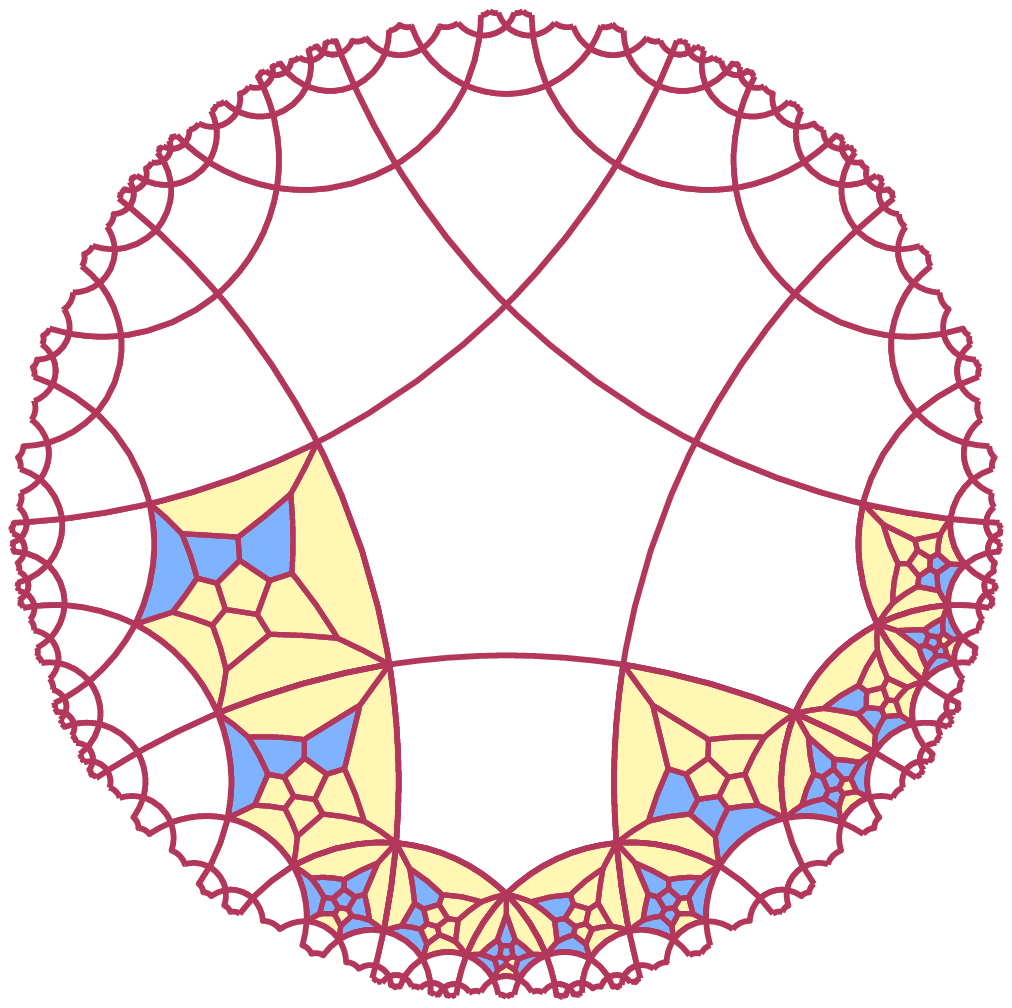,width=240pt}}
\vtop{
\ligne{\hfill
\PlacerEn {-130pt} {0pt} \box110
\hfill}
\vspace{-20pt}
\begin{fig}\label{idle_horiz}
\leurre
Pseudo-projection on the plane of the track of its elements in the case of a vertical
segment.
\end{fig}
}

\noindent
of $(SeC)^k$ on a factor of length~$k$ of the infinite Fibonacci word.
Indeed, all corners are put on a black node of the Fibonacci tree. Straight elements
are put on either white or black nodes. This can be made more accurate as follows.
The straight elements of the segment are in contact of cells of
the level~$n$ of the tree while the straight elements themselves are in the level~$n$+1.
The corners of the segment are all in the level~$n$+2. Now, when the straight element
is put on a white node, the exits are through faces~1 and~4. When it is put
on a black node, the exits are through faces~1 and~10. This explains the connection
of a horizontal segment with the infinite Fibonacci word, also see~\cite{mmfiboTUCS}. 

   In Figure~\ref{idle_horiz}, all the cells, the leftmost one excepted, constitute
an illustration of a horizontal segment. Note that the leftmost element 
does not belong to the horizontal segment but it realizes the connection 
with a vertical segment.

\subsection{Bridges}
\label{bridge}

   As already mentioned in this section, crossings of the planar railway circuit are 
replaced by bridges. We can arrange the crossing in such a way that two vertical
segments~$V_0$ and~$V_1$ cross each other. Assume that~$V_0$ will remain in the 
plane~$\Pi_0$ of its faces~5 while $V_1$ will make a detour in the plane~$\Pi_1$,
perpendicular to~$\Pi_0$,  which contains the guideline of its projection onto~$\Pi_0$. 
In~$\Pi_1$ the track will follow a horizontal segment which will take the cells
of two circles of cells in~$\Pi_1$: at a distance~2 or~3 from the cell~$c_0$ of~$V_0$ 
which has a contact with both~$\Pi_0$ and~$\Pi_1$. Figure~\ref{bridge_proj} represent 
such a bridge using two pseudo-projections: one onto the plane~$\Pi_0$,
on the left-hand side of the figure, and onto the plane~$\Pi_1$ on its right-hand side.
We shall say that the projection onto the plane~$\Pi_0$ is the view from above
and that the projection onto the pane~$\Pi_1$ is the frontal view, both ways of
views referring to the bridge itself.

   Let us have a closer look at the figures. 

   In the view from above, we can see two vertical segments: one goes from the right-up 
part of the figure to the left-bottom one. It can be easily recognized as a copy
of the vertical segment illustrated by Figure~\ref{idle_vert}. Here, it contains two 
tiles coloured with light brown. We shall call this track the top-down track. The other 
track goes from the left-upper part of the figure and goes to the right-bottom one. We 
shall call it the left-right track. We can see the guideline of the top-down track. 
It is the intersection of the planes~$\Pi_0$ and~$\Pi_1$. 

   Still in the view from above, we can see golden yellow marks on the light brown tiles
and two green marks on the central tile. The golden marks indicate that the top-down track 
goes on these tiles. The green marks indicate the two piles of the bridge.
Number the cells of the projection of the top-down track in the view from above
from~1 to~7, 1 being the number of the topmost cell. Cell~4 is the central cell and
it belongs to the left-right track: the top-down track follows a horizontal 
segment on the plane~$\Pi_1$ which can be seen in the frontal view, see the
right-hand side part of Figure~\ref{bridge_proj}. The departure/arrival of the horizontal
segment is defined by cells~6 and~2 which can be seen on both views. In the frontal
view, the trace of the plane~$\Pi_0$ can easily be seen: it is the border between the 
coloured tiles and the others which remain blank, on the bottom part of the figure.
There are seven coloured cells along this line in the frontal view: they are exactly
the cells number from~1 to~7{} in the view from above. In the frontal view, cell~6
is on the left-hand side. 

\vskip 15pt
\setbox110=\hbox{\epsfig{file=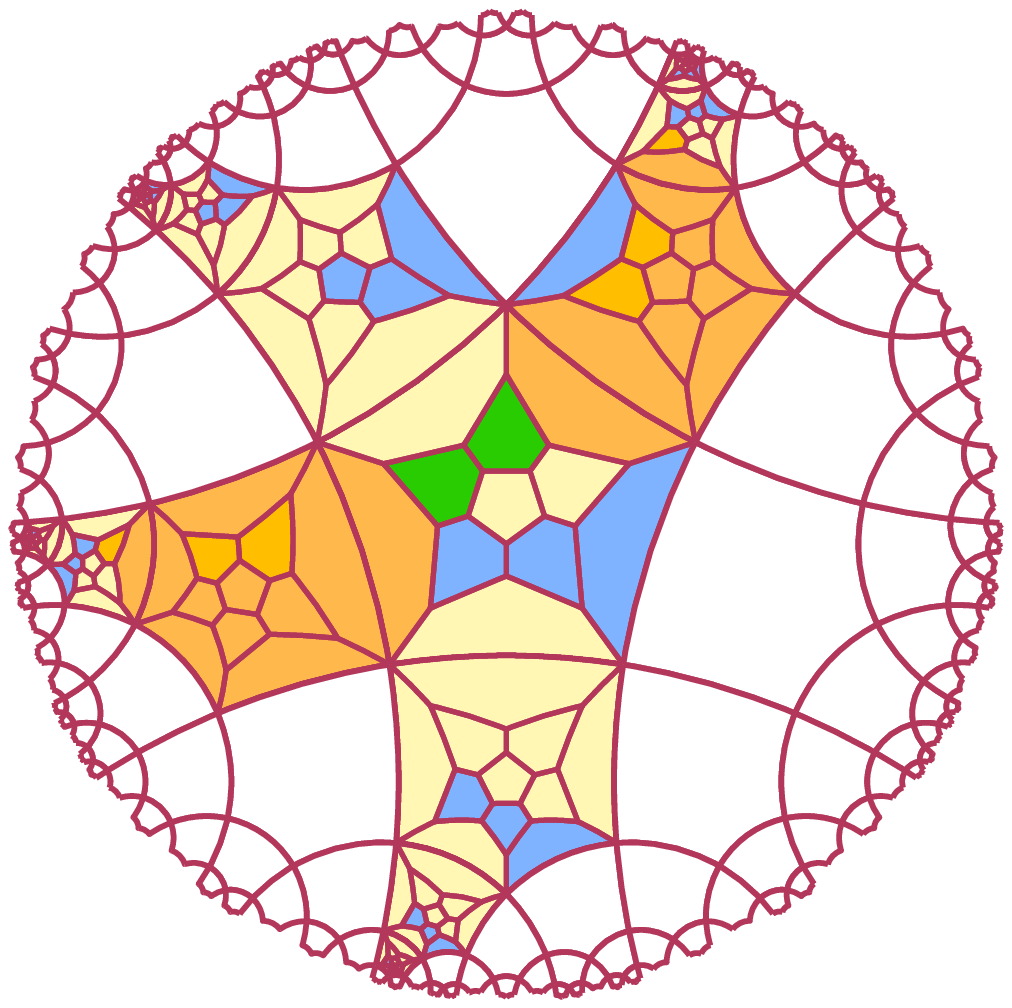,width=180pt}}
\setbox112=\hbox{\epsfig{file=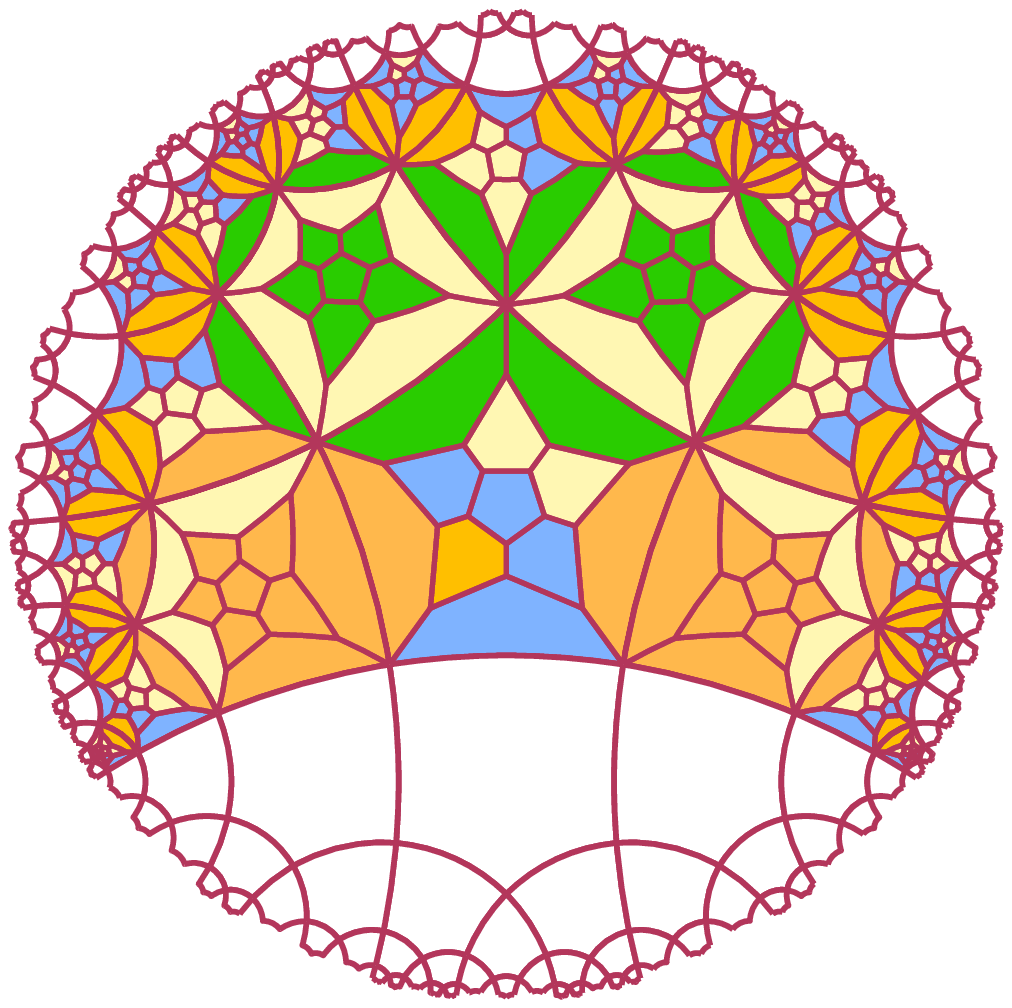,width=180pt}}
\vtop{
\vspace{-10pt}
\ligne{\hfill
\PlacerEn {-360pt} {0pt} \box110
\PlacerEn {-180pt} {0pt} \box112
}
\vspace{-10pt}
\begin{fig}\label{bridge_proj}
\leurre
Pseudo-projections of two tracks crossing through a bridge. On the left-hand side:
pseudo-projection on the plane~$\Pi_0$; on the right-hand side: pseudo-projection on
the plane~$\Pi_1$. The exit faces are marked by a golden yellow colour on
the right-hand side figure. The piles of the bridge are marked with green on the 
central cell. In the frontal view, we have exceptionally given the colour of the
cell to its faces, in fact its faces which are not in contact with a cell of a track.
This is to underline the elements of the bridge on which the track relies.
\end{fig}
}

   Cells~6 and~2 are an application of what we have mentioned in Subsection~\ref{elements}, 
about the different ways to rotate a straight element in order to access to another
plane. Cell~6 is alike the picture~$(c)$ of Figure~\ref{track_straight}. Its faces~0
and~5 are in contact with the guideline. Now, exit~3 may be used to go into the plane 
of face~0 which is perpendicular to the plane of face~5. This will be the starting point
of our bridge. Note that face~0 of cell~6 is on the plane~$\Pi_1$.
From this tile, the bridge follows a horizontal segment until it arrives at
cell~2 which is in contact with both~$\Pi_0$ and~$\Pi_1$. Notice that cell~2
is alike the picture~$(e)$ of Figure~\ref{track_straight}, and that its face~4 is the face
where the track of the bridges again joins the top-down track. Note that in cell~2, face~0
is on the plane~$\Pi_1$ as this is the case for cell~6.
Looking at the cells of the horizontal segment in the frontal view, we can notice that
the straight elements have a milestone which is below the plane~$\Pi_1$: this means that
there are milestones on both half-spaces defined by the plane~$\Pi_1$. Now,
for the corners, all the milestones are in a same half-space defined by the plane~$\Pi_1$. 
For the straight elements, their exits are most often the faces~1 and~4. However, from
time to time, the exits are the faces~1 and~10. In Subsection~\ref{horiz} we
have seen the reason of these variants. As can be seen on the frontal view, 
all corners are put on a black node of the Fibonacci tree and straight elements 
can be either on a black node or on a white one. From Subsection~\ref{horiz}, we know
that when a straight element is on a white tile, the exits are the faces~1 and~4.
When it is on a black tile, it is the faces~1 and~10. 

\subsection{The motion of the locomotive on the tracks}
 
   Presently, we describe the motion of the locomotive on its tracks.
  
   This motion is very different from the simulation of~\cite{mm3DJCA}. There, the track
were materialized by a specific colour and the locomotive simply occupied two contiguous
cells of the track. Here, as this is the case in the planar simulation described 
in~\cite{mmarXiv4st}, the set of cells occupied at some time by the locomotive consists
of blank cells. They belong to the track because they are marked in a specific way by
the milestones. However, outside this particular point, the motion of the locomotive
is the same as in the previous simulations explained 
in~\cite{mmbook2,mmsyPPL,mmsyENTCS,mmarXiv4st}. If we restrict our attention to
the cells of the track, we have the following one-dimensional rules for the motion of the
locomotive:
\vskip 7pt
\vtop{

\ligne{\hfill\tt B\ W\ W\ $\rightarrow$ B\hskip 40pt W\ W\ B $\rightarrow$ B\hfill} 
\ligne{\hfill\tt R\ B\ W\ $\rightarrow$ R\hskip 40pt W\ B\ R $\rightarrow$ R\hfill} 
\ligne{\hfill\tt W\ R\ B\ $\rightarrow$ W\hskip 40pt B\ R\ W $\rightarrow$ W\hfill} 
\ligne{\hfill\tt W\ W\ R\ $\rightarrow$ W\hskip 40pt R\ W\ W $\rightarrow$ W\hfill} 
}
\vskip 7pt

   As can easily be deduced from the rules, the locomotive consists of two
contiguous cells: one is blue, the front, the other is red, the rear. 

   With the just mentioned principles in mind, we can easily device the rules for 
the motion of the locomotive. We postpone the systematic writing of the rules to 
Section~\ref{the_rules}.

   Figures~\ref{loco_vert} and~\ref{loco_horiz} illustrate the application of the rules
deduced from the one-dimensional rules which we have mentioned above. In these 
representations and the one we shall use for the switches in Section~\ref{the_switches},
we do not represent the colour of the state of the cell, as already mentioned in
Section~\ref{the_tracks}. We shall use the convention which we introduced there according
to which, a face~$F$ of a cell~$C$ represents the state of the neighbour sharing~$F$
with~$C$. As in our pseudo-projection the same face is represented twice as it is shared
by two cells, each face represents the colour of the other cell. This way allows us
to also materialize the motion of the locomotive as can be seen in Figures~\ref{loco_vert}
and~\ref{loco_horiz}, as well as in a part of the figures of Section~\ref{the_switches}.

\vskip 15pt
\setbox110=\hbox{\epsfig{file=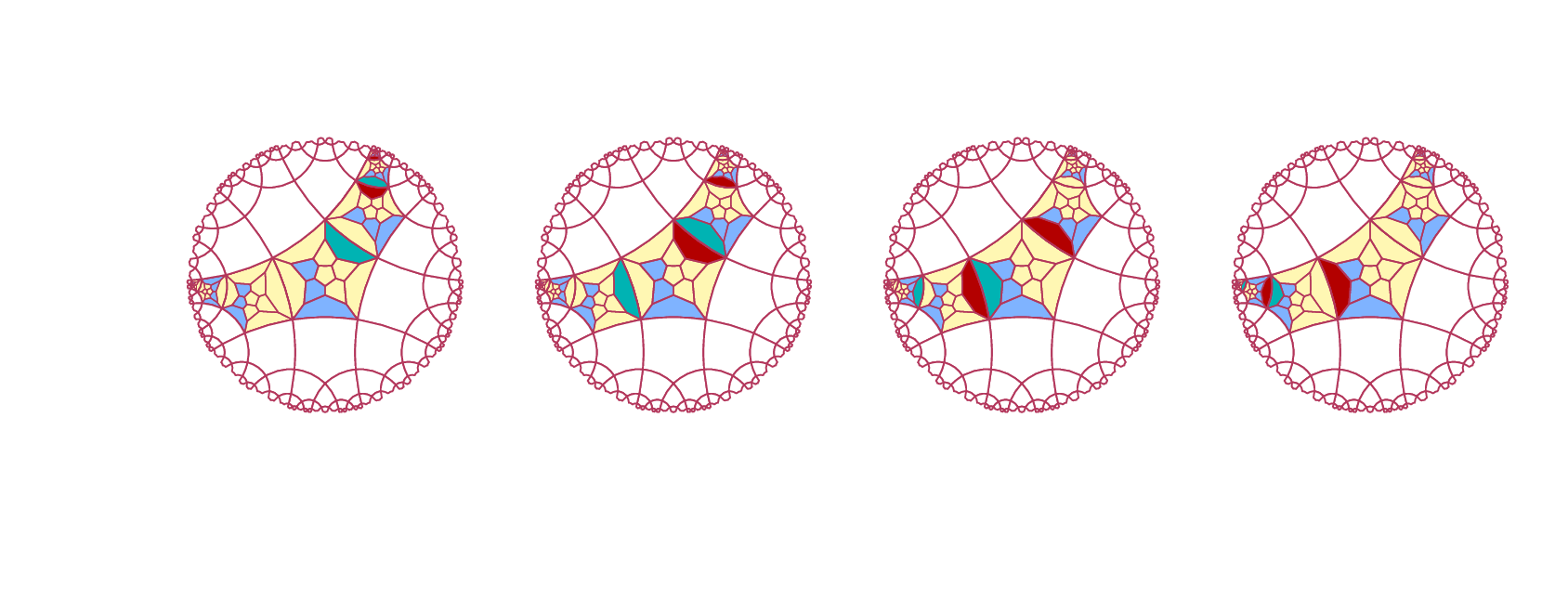,width=380pt}}
\vtop{
\vspace{-40pt}
\ligne{\hfill
\PlacerEn {-380pt} {0pt} \box110
}
\vspace{-40pt}
\begin{fig}\label{loco_vert}
\leurre
Sequence of pseudo-projections on the plane of the track in the case of a vertical
segment when the locomotive runs over it.
\end{fig}
}

It can be noticed that the elements can be freely assembled, provided that they observe 
the principle which we have fixed: exits of an element are 1~and~2 for corners, they 
are 1~and~3, 1~and~4, 1~and~8 or 1~and~10 for a straight element. Other combinations 
are ruled out by the rules, we shall come back to this point in Subsection~\ref{the_rules}. 

\vskip 5pt 
\vskip 15pt
\setbox110=\hbox{\epsfig{file=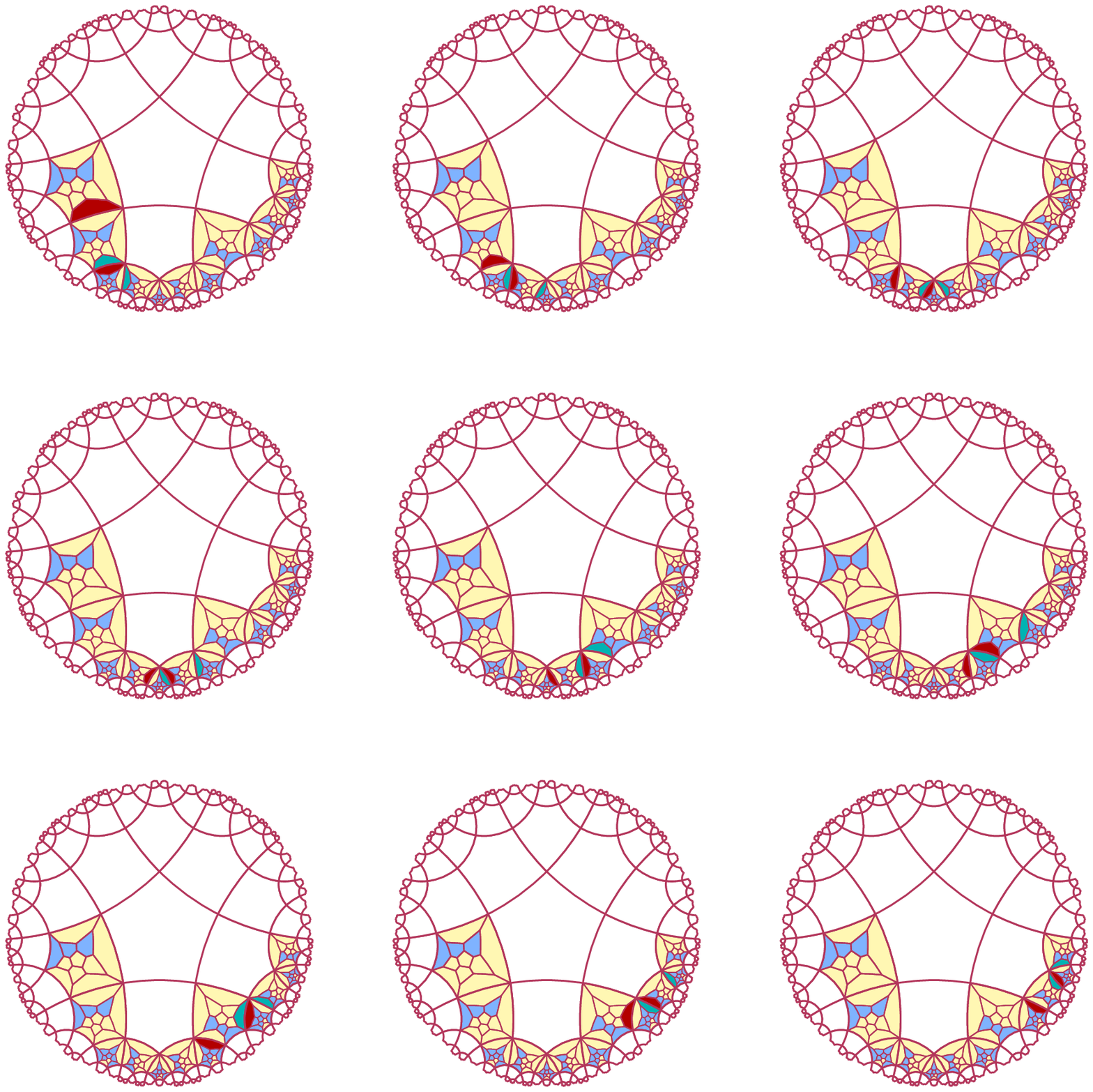,width=380pt}}
\vtop{
\vspace{-40pt}
\ligne{\hfill
\PlacerEn {-340pt} {0pt} \box110
}
\vspace{-40pt}
\begin{fig}\label{loco_horiz}
\leurre
Sequence of pseudo-projections on the plane of the track in the case of a horizontal
segment when the locomotive runs over it.
\end{fig}
}

\section{Implementation of the switches}
\label{the_switches}

   In order to describe the switches, we shall focus on the memory switch which has 
the most complex mechanism among the switches. In fact, this mechanism consists of two
connected parts~$A$ and~$B$. Notice from now on that the fixed switch will use the
mechanism~$A$ alone and that the flip-flop switch will use the mechanism~$B$ alone.

   All switches will share the following common features. They are assumed to be on
the same plane $\Pi_0$. However, certain parts of the above mechanisms are on both 
half-spaces defined by~$\Pi_0$. This is why we shall present two figures for each switch:
one is a pseudo-projection from above onto~$\Pi_0$, the other is a pseudo-projection
onto the same plane, from below. We have to remember that in such a case, the left-hand side
and the right-hand side are exchanged. 

   Next, for two of them, the switches have a left-hand side version and a right-hand side
one. In the left-hand side, right-hand side version respectively, the active passage sends 
the locomotive to the left-hand side, right-hand side track respectively. However, for the 
fixed switch, a left-hand side version is enough. A right-hand side fixed switch is
obtained from a left-hand side one as follows: after the switch, the left-hand side 
track crosses the right-hand side one in order to exchange the directions. Thanks to 
the bridge which we have implemented, this is easily performed.

   The switches will be presented according to a similar scheme.

   First, we describe what we call the {\bf idle configuration}: it is the situation
of the switch when it is not visited by the locomotive. All switches are the meeting
of three tracks. The meeting tile is a straight element and, in the figures, it is
placed at the central tile. The track which arrives to the entry~1 of this element
represents the arrival for an active crossing of the switch. Exit~3 gives access to 
the track which goes to the left and exit~4 gives access to the track going to the right.
The cells of the tracks are numbered from~2 to~10 and from~12 to~15. Cells~2 to~5
constitute the arriving track. They follow a vertical segment which arrives to the
leading tile of a quarter constructed around the central cell. We shall number this 
sector by~1, as the exit to which the track leads. Cell~2 is the farthest from the central
cell, cell~5 is the leading tile of sector~1. The central cell is cell~6. Cells~7 to~10
constitute the track which leaves the switch through exit~3. They are displayed in a 
vertical segment included in a sector lead by cell~7 and which is called sector~3, after 
exit~3. Cells~12 to~15 constitute the vertical segment which leaves the switch
through exit~4. These cells belong to sector~4 headed by cell~12.

   The numbers come from the numbers of the cells given in the computer program which
was used to check the correctness of the rules we shall describe in Section~\ref{the_rules}.

   A closer look shows that the tracks are not exactly along a vertical: the cell which is
in contact with an exit of the central cell, is a straight element whose face~0 is on the
plane~$\Pi_0$. The next cell, cell~4, 8 and~13 respectively is a corner, again with its
face~0 on the plane~$\Pi_0$. The remaining two cells constitute a vertical segment in 
the way we have defined them with a milestone below the plane~$\Pi_0$ with respect to
the other milestone which consider as upon this plane.

   With these conventions, we can start the study of each switch. We shall see the memory 
switches, the fixed switch and the flip-flop switches in this order.

\subsection{Memory switches}
\label{memory}

    As mentioned in the beginning of this section, the memory switches are the
most complex construction in our implementation.

    Figures~\ref{idle_memo_gauche} and~\ref{idle_memo_droit} respectively represent the 
left-hand side, right-hand side memory switches respectively. In each figure, there is
a big disc and a smaller one. The big disc is a pseudo-projection onto the 
plane~$\Pi_0$ from above, while the smaller one is a pseudo-projection onto the
same plane from below. In the big disc, we apply the convention about the colour
of the cells. We also do this for the small disc, except for the locomotive. Indeed,
in the projection from below, the locomotive is hidden by the cell which is below its track.
Accordingly, in order to locate the position of the locomotive, we imagine that the cells
below its track are a bit transparent so that a pale hue of the colour is visible.

\setbox110=\hbox{\epsfig{file=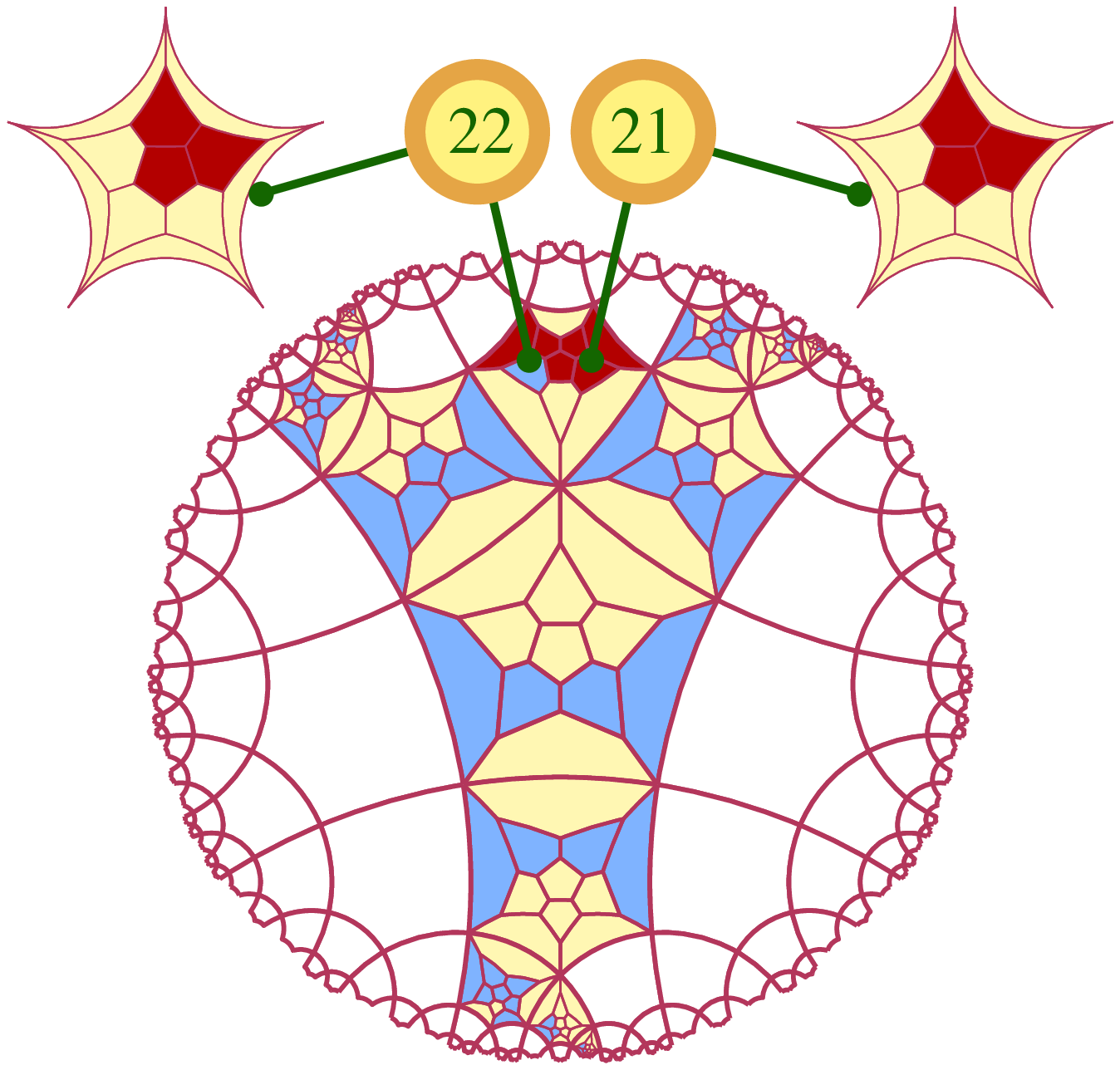,width=210pt}}
\setbox112=\hbox{\epsfig{file=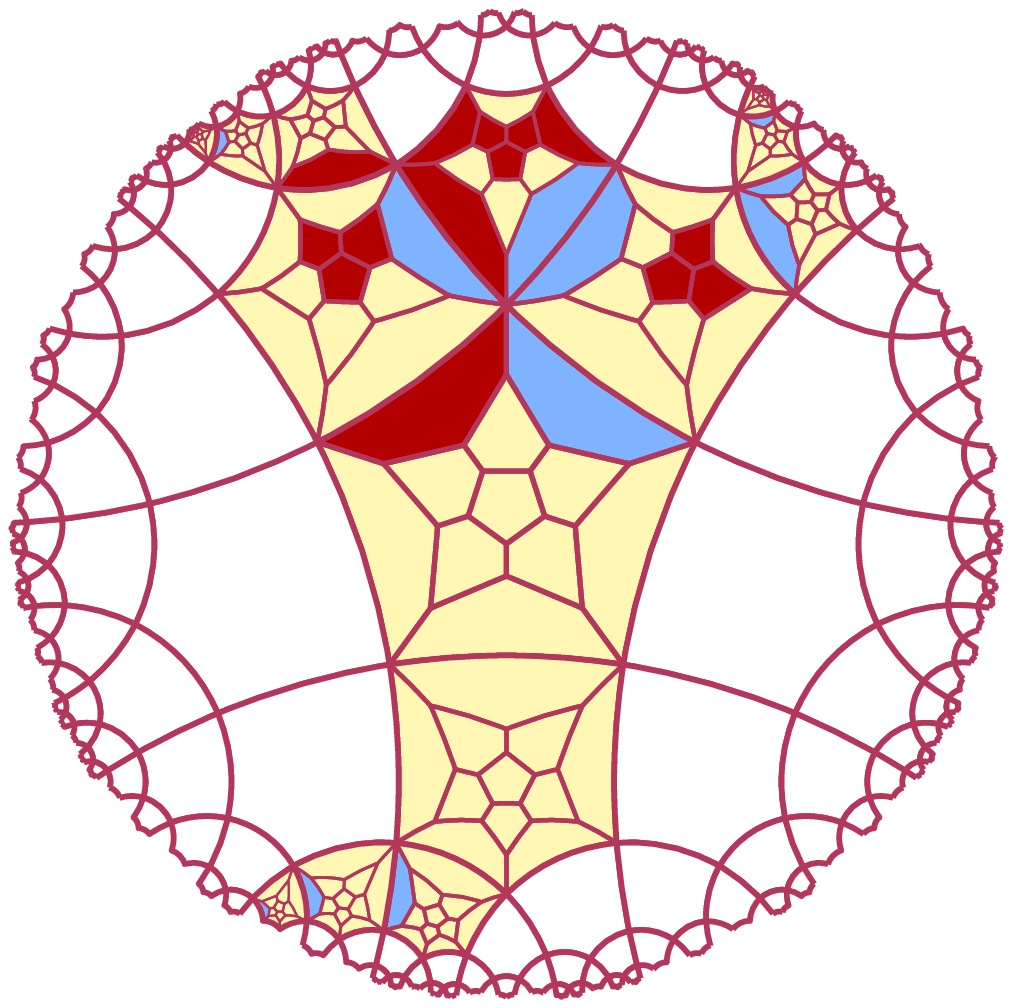,width=160pt}}
\vtop{
\ligne{\hfill
\PlacerEn {-220pt} {0pt} \box110
\PlacerEn {-350pt} {0pt} \box112
}
\vspace{-10pt}
\begin{fig}\label{idle_memo_gauche}
\leurre
The idle configuration of a left-hand side memory switch, represented by the two
pseudo-projections, one from above: the big disc; the other from below: the small disc. 
\end{fig}
}

   In the memory switch, there are two {\bf sensors}, two {\bf markers} and 
two {\bf controllers}. The sensors are cells~17 and~18 which are neighbours
of the cells~7 and~12 respectively through their faces~0. Cells~7 and~12 are
called the {\bf scanned cells}, inspected by their sensors. The {\bf upper} controller
is cell~20 which is the second common neighbour of cells~7 and~12 above the 
plane~$\Pi_0$: the first common neighbour is the central cell. We consider that
cell~20 has its face~0 on the plane~$\Pi_0$. The {\bf lower} controller is cell~19 
which is the neighbour of cell~20 through its face~0: cell~19 is thus below the
plane~$\Pi_0$ and we also consider that its face~0 is on the plane~$\Pi_0$.
The two markers are cells~21 and~22: they are neighbours of cell~20 through its
faces~8 and~10 respectively. Now, the sensor of cell~7 is blue and that of cell~12
is red. Similarly, cell~21 is red and cell~22 is blue. The colours of the sensors
and of the markers allow to identify the left-hand side memory switch. As easily
seen in Figure~\ref{idle_memo_droit}, in a right-hand side memory switch, the colours of
the sensors and the markers are exchanged: cells~21 and~18 are blue, cells~22 and~17
are red.

   The working of the memory switch is the following.

   A blue sensor is indifferent to the direction of the locomotive: it may cross
the cell it scans in both ways. A red sensor does not behave the same. First, it 
prevents the locomotive to enter the cell it scans in an active passage. In a 
passive passage, it detects the passage of the locomotive, it allows it
to pass through the cell it scans, but it reacts to the passage by changing its
colour: as the blue sensor does not see the red one, the red sensor cannot change
its colour to blue. It changes it to white. This is detected by the lower controller,
usually blue, which becomes red. When the lower controller is red, both sensors 
change their colour: the blue one to red the now white one to blue. And the lower
controller goes back to blue. Now, the upper controller, usually blue, also detects 
the passive passage through the non-selected track: its markers allow it to 
differentiate cell~7 from cell~12. And so, when the front of the locomotive leaves 
cell~7 or cell~12 when this cell is on the non-selected track, the upper controller 
becomes white and then red. It becomes white to prevent the locomotive from being 
duplicated on the selected track: the locomotive must go through entry~1. Then it 
becomes red, at the same times as the lower controller becomes red. When the 
upper controller is red, both markers exchange their colour and at the next time, 
the upper controller returns to blue.
\vskip 7pt
   In the rest of this subsection, we show figures which we illustrate the crossing
of a memory switch by the locomotive. Also, with each figure, we reproduce, as a
table, a trace of the execution of the computer program which simulated the various 
motions. In a heading line, the trace indicates the visited cells by their number as 
well as their immediate neighbours, also numbered, when they may be changed during 
the visit. Below this leading line, for each time, the state of a cell is given in 
the column corresponding to its number. Looking at the numbers, we can see that the 
simulation program considered two more cells on the track arriving to the switch
than what is shown by the figures. But the figures cannot show more.

\setbox110=\hbox{\epsfig{file=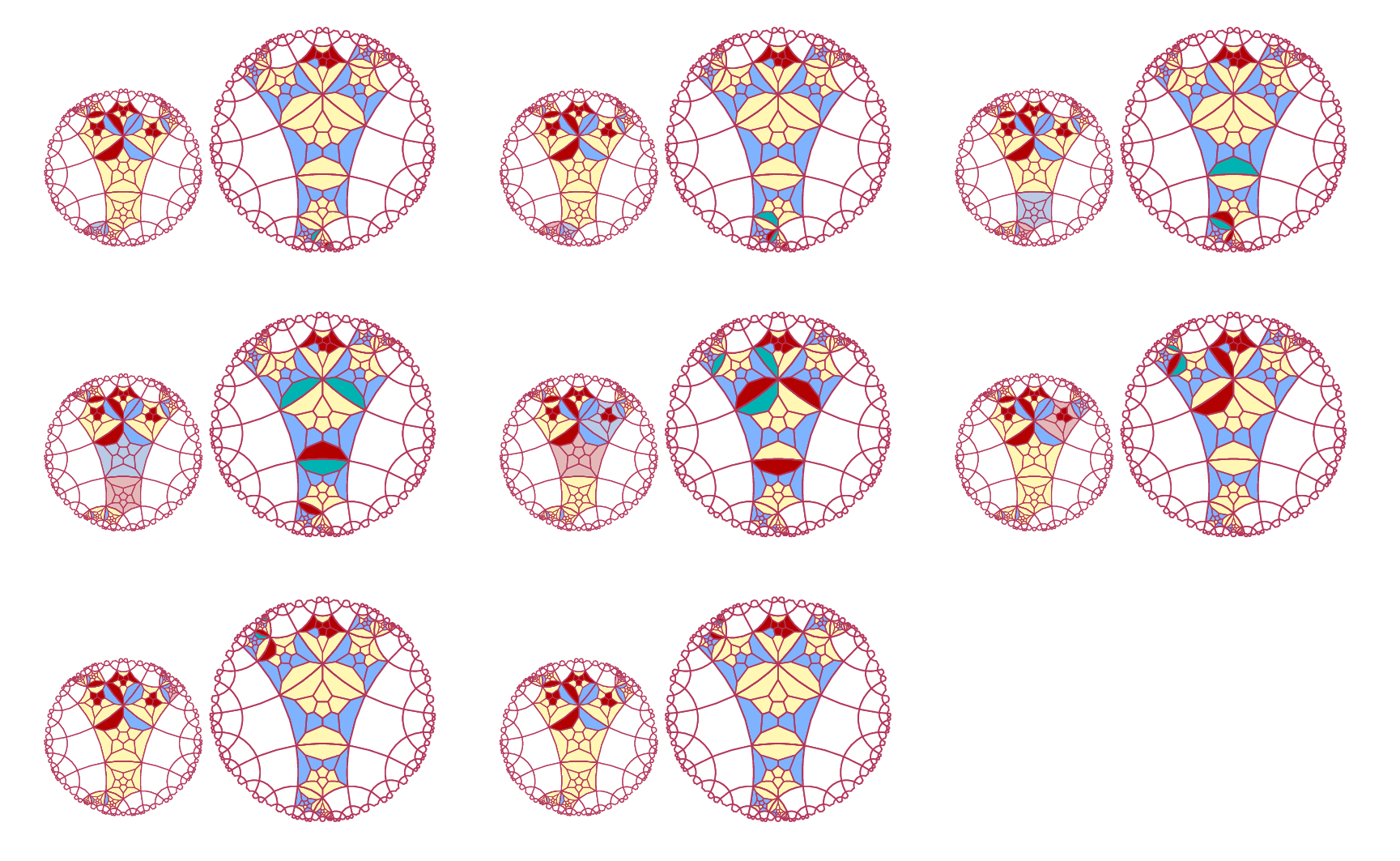,width=360pt}}
\vtop{
\ligne{\hfill
\PlacerEn {-350pt} {0pt} \box110
}
\vspace{-10pt}
\begin{fig}\label{loco_memo_gauche_1}
\leurre
The active passage of the locomotive through a left-hand side memory switch.
\end{fig}
}
\vskip 7pt

   Figure~\ref{loco_memo_gauche_1} illustrates the motion of the locomotive during
an active passage through a left-hand side memory switch and Table~\ref{exec_memog1}
gives the trace of the corresponding execution. Due to the symmetry of the projections 
onto~$\Pi_0$ used to establish the big and the small discs, it can be noticed that the 
locomotives seem to go closer and closer from each other. Also, we can see that the blue 
and the red sensors remain unchanged during the passage of the locomotive and that the 
presence of the red sensor below cell~12 prevents the locomotive from going in this 
direction. 

\vtop{
\begin{tab}\label{exec_memog1}
\leurre
Run of the simulation programme corresponding to Figure~{\rm\ref{loco_memo_gauche_1}}.
The active passage correspond to cells~$1$ up to~$11$, in this order.
\end{tab}
\vspace{-12pt}
\grostrait
{\ttviii
\obeylines
\leftskip 0pt
\obeyspaces\global\let =\ \parskip=-2pt
active crossing of a memory switch, left-hand side :

          1  2  3  4  5  6  7  8  9 10 11 12 13 14 15 16 17 18 19 20 21 22

time 0 :  W  R  B  W  W  W  W  W  W  W  W  W  W  W  W  W  B  R  B  B  R  B
time 1 :  W  W  R  B  W  W  W  W  W  W  W  W  W  W  W  W  B  R  B  B  R  B
time 2 :  W  W  W  R  B  W  W  W  W  W  W  W  W  W  W  W  B  R  B  B  R  B
time 3 :  W  W  W  W  R  B  W  W  W  W  W  W  W  W  W  W  B  R  B  B  R  B
time 4 :  W  W  W  W  W  R  B  W  W  W  W  W  W  W  W  W  B  R  B  B  R  B
time 5 :  W  W  W  W  W  W  R  B  W  W  W  W  W  W  W  W  B  R  B  B  R  B
time 6 :  W  W  W  W  W  W  W  R  B  W  W  W  W  W  W  W  B  R  B  B  R  B
time 7 :  W  W  W  W  W  W  W  W  R  B  W  W  W  W  W  W  B  R  B  B  R  B
\par}
\demitrait
\vskip 7pt
}
\vskip 10pt
\setbox110=\hbox{\epsfig{file=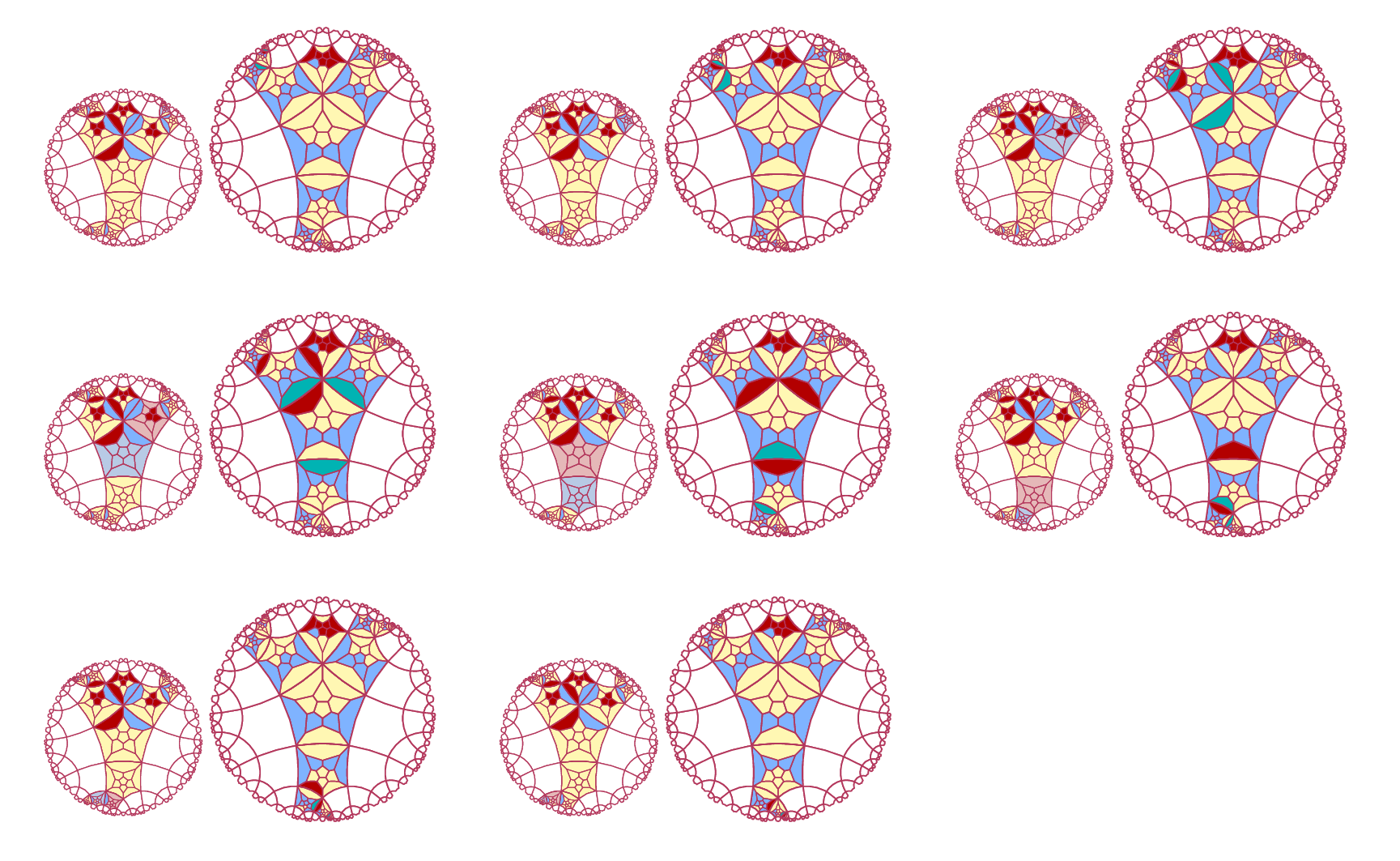,width=360pt}}
\vtop{
\vspace{10pt}
\ligne{\hfill
\PlacerEn {-350pt} {0pt} \box110
}
\vspace{-10pt}
\begin{fig}\label{loco_memo_gauche_3}
\leurre
The passive crossing of a left-hand side memory switch by the locomotive through 
the selected track.
\end{fig}
}
\vskip 7pt
   Figure~\ref{loco_memo_gauche_3} and Table~\ref{exec_memog3} do the same
for a passive crossing of a left-hand side memory switch by the locomotive through 
the selected track. No particularities happen here. Again, the red sensor prevents 
the locomotive from going into the non-selected track.

   Figure~\ref{loco_memo_gauche_4} and Table~\ref{exec_memog4} represent
a passive crossing of a left-hand side memory switch by the locomotive through 
the non-selected track. With this situation, we have a big difference with the previous
figures and tables where the sensors and markers had a passive role. Now, the sensors
and the markers play an active role and, at the end of the role, they are exchanged.
Accordingly, the locomotive entered a left-hand side memory switch and it leaves
a right-hand side memory switch.

\vtop{
\begin{tab}\label{exec_memog3}
\leurre
Run of the simulation programme corresponding to Figure~{\rm\ref{loco_memo_gauche_3}}.
The passive crossing through the selected track correspond to cells~$1$ up to~$11$, 
in the reverse order.
\end{tab}
\vspace{-12pt}
\grostrait
{\ttviii
\obeylines
\leftskip 0pt
\obeyspaces\global\let =\ \parskip=-2pt
passive crossing of a memory switch, left-hand side,
  through the selected track :

          1  2  3  4  5  6  7  8  9 10 11 12 13 14 15 16 17 18 19 20 21 22

time 0 :  W  W  W  W  W  W  W  W  B  R  W  W  W  W  W  W  B  R  B  B  R  B
time 1 :  W  W  W  W  W  W  W  B  R  W  W  W  W  W  W  W  B  R  B  B  R  B
time 2 :  W  W  W  W  W  W  B  R  W  W  W  W  W  W  W  W  B  R  B  B  R  B
time 3 :  W  W  W  W  W  B  R  W  W  W  W  W  W  W  W  W  B  R  B  B  R  B
time 4 :  W  W  W  W  B  R  W  W  W  W  W  W  W  W  W  W  B  R  B  B  R  B
time 5 :  W  W  W  B  R  W  W  W  W  W  W  W  W  W  W  W  B  R  B  B  R  B
time 6 :  W  W  B  R  W  W  W  W  W  W  W  W  W  W  W  W  B  R  B  B  R  B
time 7 :  W  B  R  W  W  W  W  W  W  W  W  W  W  W  W  W  B  R  B  B  R  B
\par}
\demitrait
\vskip 7pt
}

\setbox110=\hbox{\epsfig{file=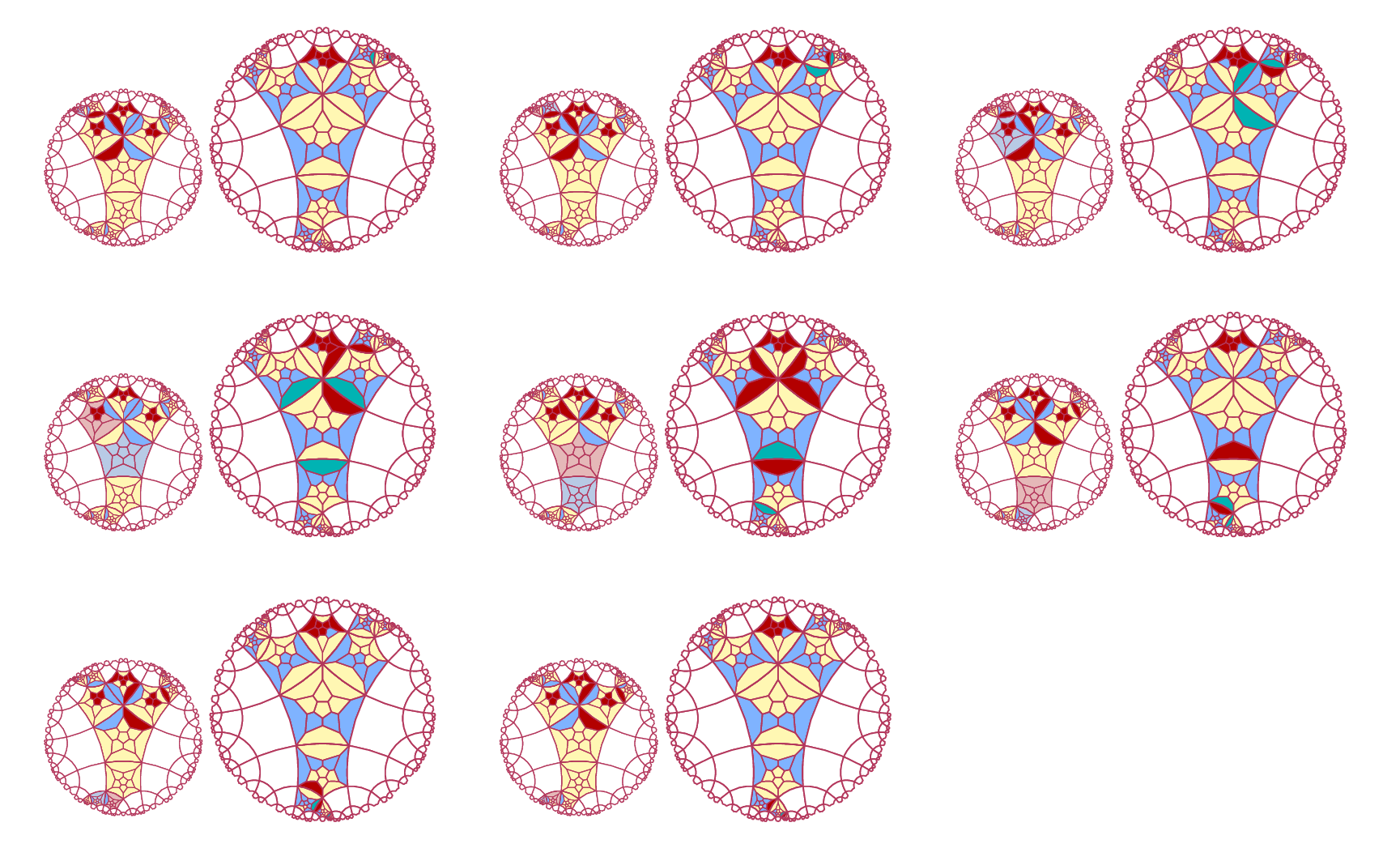,width=360pt}}
\vtop{
\vspace{10pt}
\ligne{\hfill
\PlacerEn {-350pt} {0pt} \box110
}
\vspace{-10pt}
\begin{fig}\label{loco_memo_gauche_4}
\leurre
The passive crossing of a left-hand side memory switch by the locomotive through 
the non-selected track. Note that after the fifth step, the sensors and markers 
have exchanged their colours: the memory switch is now a right-hand side one.
\end{fig}
}
\vskip 7pt
   Note that Table~\ref{exec_memog4} shows exactly when each sensor and controller is
triggered. The front of the locomotive is in cell~12 at time~2. This makes 
cell~20 and~18 becoming white. As already noticed, the red sensor cannot change
to blue as the blue sensor, which cannot see neither cell~12 nor cell~18, did not yet 
realized that a change must occur. At time~3, the front of the locomotive
is now in the central cell, cell~6, and cells~20 and~18 are now white. This is the signal
for both controllers to flash the red signal which will trigger the exchange of colours
in the sensors and in the markers. The signal is sent at time~4 and the exchange of colours
happens at time~5: starting from that time, the memory switch is now a right-hand side
one.

\vtop{
\vspace{-10pt}
\begin{tab}\label{exec_memog4}
\leurre
Run of the simulation programme corresponding to Figure~{\rm\ref{loco_memo_gauche_4}}.
The passive crossing through the non-selected track correspond to cells~$12$ up to~$16$, 
in the reverse order and then to cells~$1$ up to~$6$ in the reverse order too.
\end{tab}
\vspace{-12pt}
\grostrait
{\ttviii
\obeylines
\leftskip 0pt
\obeyspaces\global\let =\ \parskip=-2pt
passive crossing of a memory switch, left-hand side,
  through the NON selected track :

          1  2  3  4  5  6  7  8  9 10 11 12 13 14 15 16 17 18 19 20 21 22

time 0 :  W  W  W  W  W  W  W  W  W  W  W  W  W  B  R  W  B  R  B  B  R  B
time 1 :  W  W  W  W  W  W  W  W  W  W  W  W  B  R  W  W  B  R  B  B  R  B
time 2 :  W  W  W  W  W  W  W  W  W  W  W  B  R  W  W  W  B  R  B  B  R  B
time 3 :  W  W  W  W  W  B  W  W  W  W  W  R  W  W  W  W  B  W  B  W  R  B
time 4 :  W  W  W  W  B  R  W  W  W  W  W  W  W  W  W  W  B  W  R  R  R  B
time 5 :  W  W  W  B  R  W  W  W  W  W  W  W  W  W  W  W  R  B  B  B  B  R
time 6 :  W  W  B  R  W  W  W  W  W  W  W  W  W  W  W  W  R  B  B  B  B  R
time 7 :  W  B  R  W  W  W  W  W  W  W  W  W  W  W  W  W  R  B  B  B  B  R
\par}
\demitrait
\vskip 7pt
}

\setbox110=\hbox{\epsfig{file=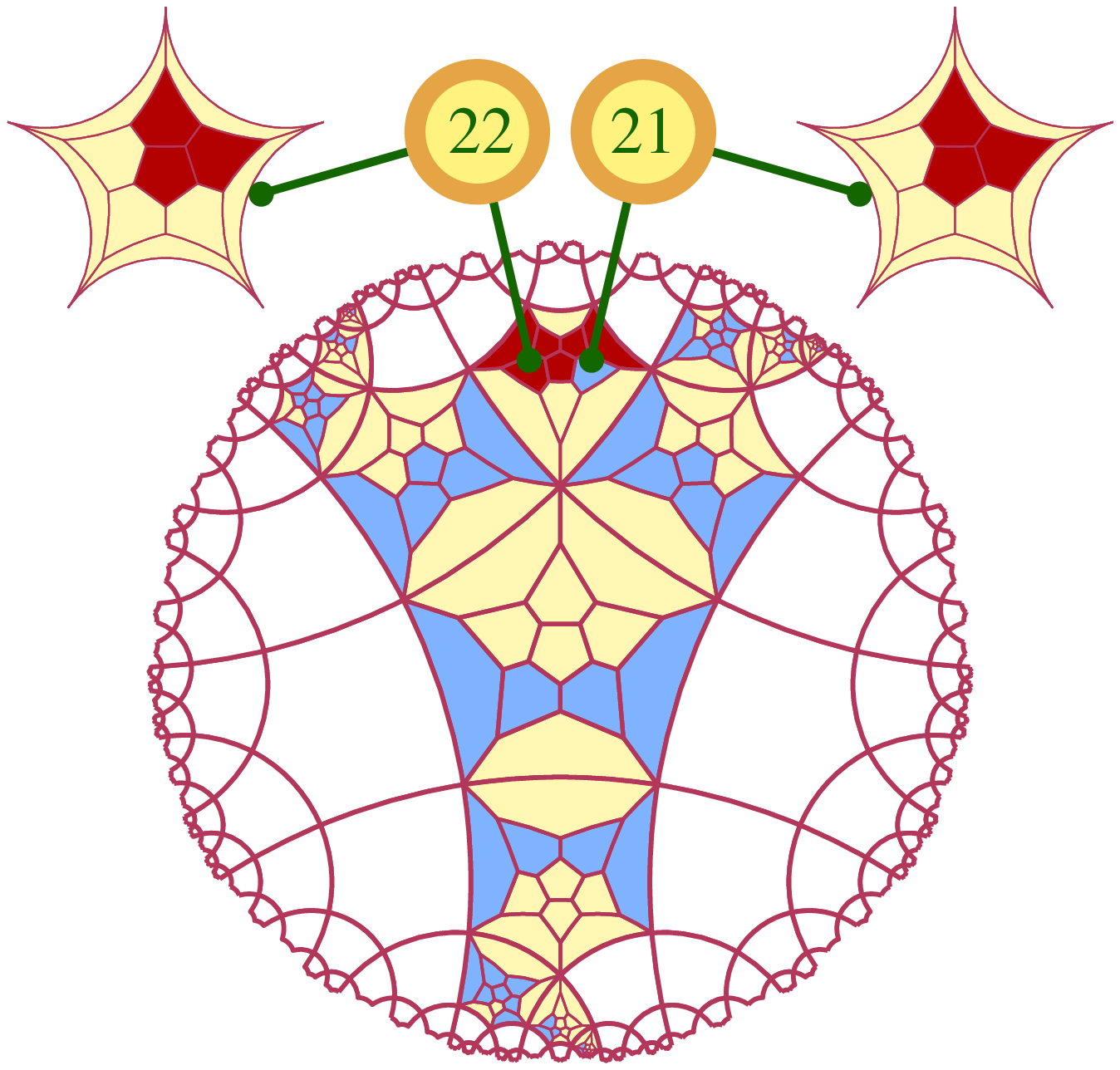,width=210pt}}
\setbox112=\hbox{\epsfig{file=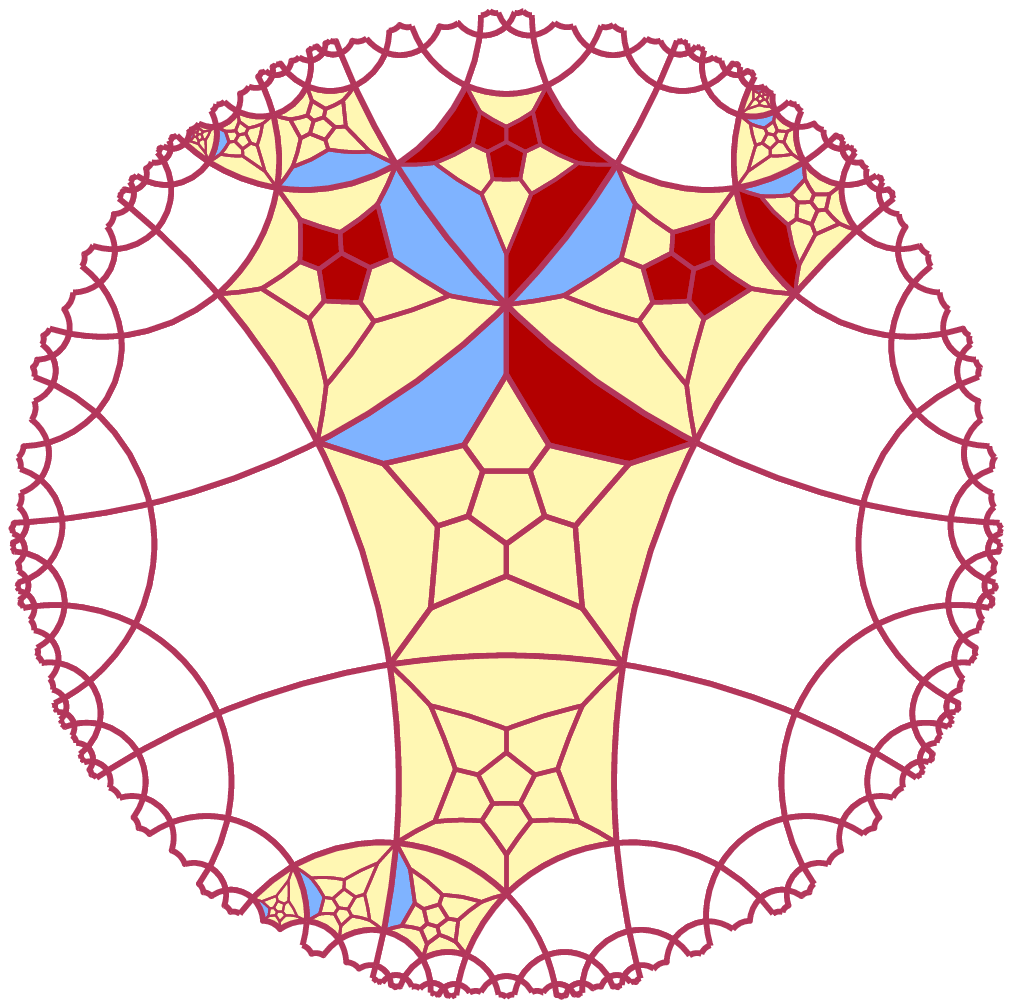,width=160pt}}
\vtop{
\ligne{\hfill
\PlacerEn {-220pt} {0pt} \box110
\PlacerEn {-350pt} {0pt} \box112
}
\vspace{-10pt}
\begin{fig}\label{idle_memo_droit}
\leurre
Idle configuration of a right-hand side memory switch. Note a situation almost symmetric
to that of Figure~{\rm\ref{idle_memo_gauche}}.
\end{fig}
}

   Figure~\ref{idle_memo_droit} illustrates the idle configuration of a right-hand side
memory switch. Note that the configuration is not exactly symmetric with
respect to that of Figure~\ref{idle_memo_gauche}. We can notice that the tracks leaving
the switch are not symmetric of each other in the common axis of symmetry of the central
cell and of cell~20. In fact, the right-hand side track leaving the switch is a rotated
image of the left-hand side one: the axis of the $3D$~rotation is the perpendicular
to~$\Pi_0$ raised from the centre of the central tile and the angle is
$\displaystyle{{2\pi}\over5}$. The configurations differ by the sensors and the markers
which simply exchanged their colours.

\setbox110=\hbox{\epsfig{file=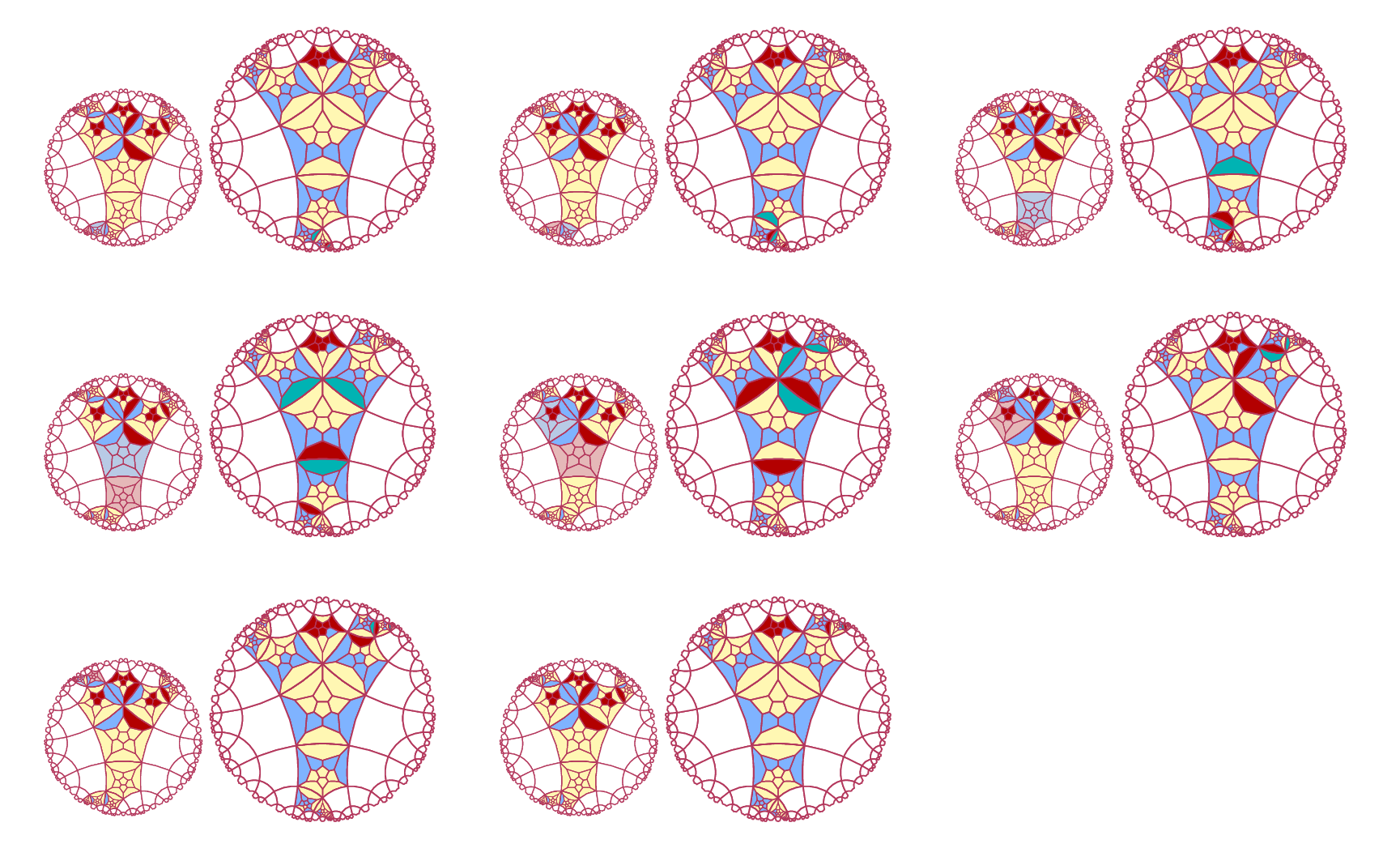,width=360pt}}
\vtop{
\ligne{\hfill
\PlacerEn {-350pt} {0pt} \box110
}
\vspace{-10pt}
\begin{fig}\label{loco_memo_droit_1}
\leurre
The active passage of a right-hand side memory switch by the locomotive.
Note that the sensors and markers are just the opposite with respect to
Figure~{\rm\ref{loco_memo_gauche_1}}.
\end{fig}
}
\vskip 7pt

\vtop{
\begin{tab}\label{exec_memod1}
\leurre
Run of the simulation programme corresponding to Figure~{\rm\ref{loco_memo_droit_1}}.
The active passage correspond to cells~$1$ up to~$6$ and then to cells~$12$ up to~$16$,
always in this order. 
\end{tab}
\vspace{-12pt}
\grostrait
{\ttviii
\obeylines
\leftskip 0pt
\obeyspaces\global\let =\ \parskip=-2pt
active crossing of a memory switch, right-hand side :

          1  2  3  4  5  6  7  8  9 10 11 12 13 14 15 16 17 18 19 20 21 22

time 0 :  W  R  B  W  W  W  W  W  W  W  W  W  W  W  W  W  R  B  B  B  B  R
time 1 :  W  W  R  B  W  W  W  W  W  W  W  W  W  W  W  W  R  B  B  B  B  R
time 2 :  W  W  W  R  B  W  W  W  W  W  W  W  W  W  W  W  R  B  B  B  B  R
time 3 :  W  W  W  W  R  B  W  W  W  W  W  W  W  W  W  W  R  B  B  B  B  R
time 4 :  W  W  W  W  W  R  W  W  W  W  W  B  W  W  W  W  R  B  B  B  B  R
time 5 :  W  W  W  W  W  W  W  W  W  W  W  R  B  W  W  W  R  B  B  B  B  R
time 6 :  W  W  W  W  W  W  W  W  W  W  W  W  R  B  W  W  R  B  B  B  B  R
time 7 :  W  W  W  W  W  W  W  W  W  W  W  W  W  R  B  W  R  B  B  B  B  R
\par}
\demitrait
\vskip 7pt
}
\vskip 10pt

   Figures~\ref{idle_memo_droit}, \ref{loco_memo_droit_1}, \ref{loco_memo_droit_4} 
and~\ref{loco_memo_droit_3} illustrate the motions of the locomotive in the various 
crossings of a right-hand side memory switch, while Tables~\ref{exec_memod1}, 
\ref{exec_memod4} and~\ref{exec_memod3} give traces of executions by the simulation program.
There is no essential difference with respect to the previous figures and tables.
The change of sides give rise to the differences which we have already noticed.

\setbox110=\hbox{\epsfig{file=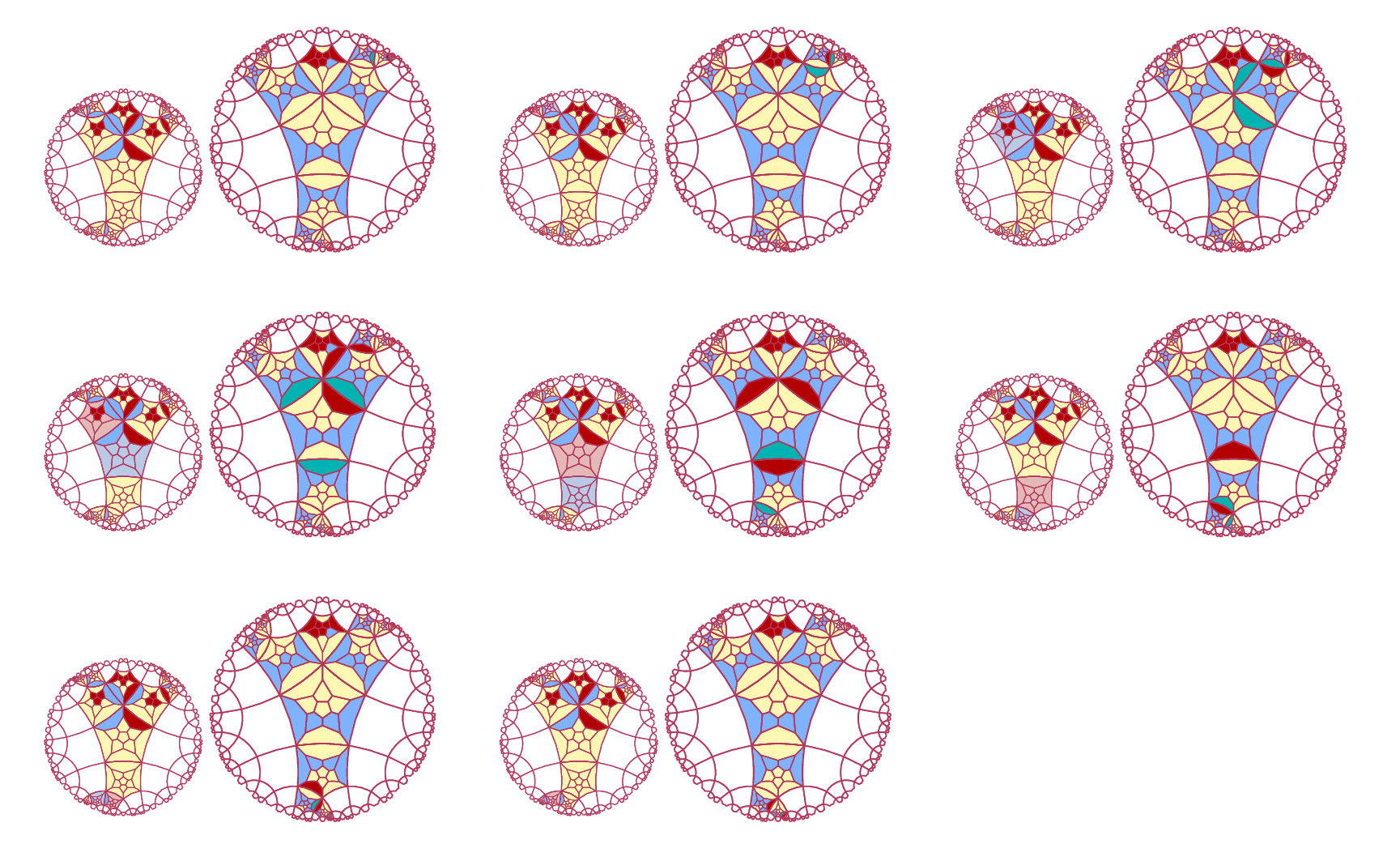,width=360pt}}
\vtop{
\ligne{\hfill
\PlacerEn {-350pt} {0pt} \box110
}
\vspace{-10pt}
\begin{fig}\label{loco_memo_droit_4}
\leurre
The passive crossing of a right-hand side memory switch by the locomotive through 
the selected track. Note that the sensors and markers are just the opposite with respect to
Figure~{\rm\ref{loco_memo_gauche_3}} and that the above figure is somehow symmetric
to the mentioned one.
\end{fig}
}
\vskip 7pt

   As there is not a true symmetry in the figures and also not in the rules, it is needed
to check the correctness of the above described scenario in order to ensure the
validity of the construction. In particular, it is important to check that when
the locomotive enters a right-hand side memory switch through the non-selected track,
it leaves the switch as a left-hand side one.

\vtop{
\vspace{-10pt}
\begin{tab}\label{exec_memod4}
\leurre
Run of the simulation programme corresponding to Figure~{\rm\ref{loco_memo_droit_4}}.
The passive crossing through the selected track correspond to cells~$12$ up to~$16$, 
in the reverse order and then to cells~$1$ up to~$6$ in the reverse order too.
\end{tab}
\vspace{-12pt}
\grostrait
{\ttviii
\obeylines
\leftskip 0pt
\obeyspaces\global\let =\ \parskip=-2pt
passive crossing of a memory switch, right-hand side,
  through the selected track :

          1  2  3  4  5  6  7  8  9 10 11 12 13 14 15 16 17 18 19 20 21 22

time 0 :  W  W  W  W  W  W  W  W  W  W  W  W  W  B  R  W  R  B  B  B  B  R
time 1 :  W  W  W  W  W  W  W  W  W  W  W  W  B  R  W  W  R  B  B  B  B  R
time 2 :  W  W  W  W  W  W  W  W  W  W  W  B  R  W  W  W  R  B  B  B  B  R
time 3 :  W  W  W  W  W  B  W  W  W  W  W  R  W  W  W  W  R  B  B  B  B  R
time 4 :  W  W  W  W  B  R  W  W  W  W  W  W  W  W  W  W  R  B  B  B  B  R
time 5 :  W  W  W  B  R  W  W  W  W  W  W  W  W  W  W  W  R  B  B  B  B  R
time 6 :  W  W  B  R  W  W  W  W  W  W  W  W  W  W  W  W  R  B  B  B  B  R
time 7 :  W  B  R  W  W  W  W  W  W  W  W  W  W  W  W  W  R  B  B  B  B  R
\par}
\demitrait
\vskip 7pt
}

   Figure~\ref{loco_memo_droit_3} and Table~\ref{exec_memod3} show that this is the case.
In particular, the table shows that the change of colour in the red sensor
also occurs at time~3, that the controller flashes at time~4 and that the sensors
exchange their colours at time~5. 

\setbox110=\hbox{\epsfig{file=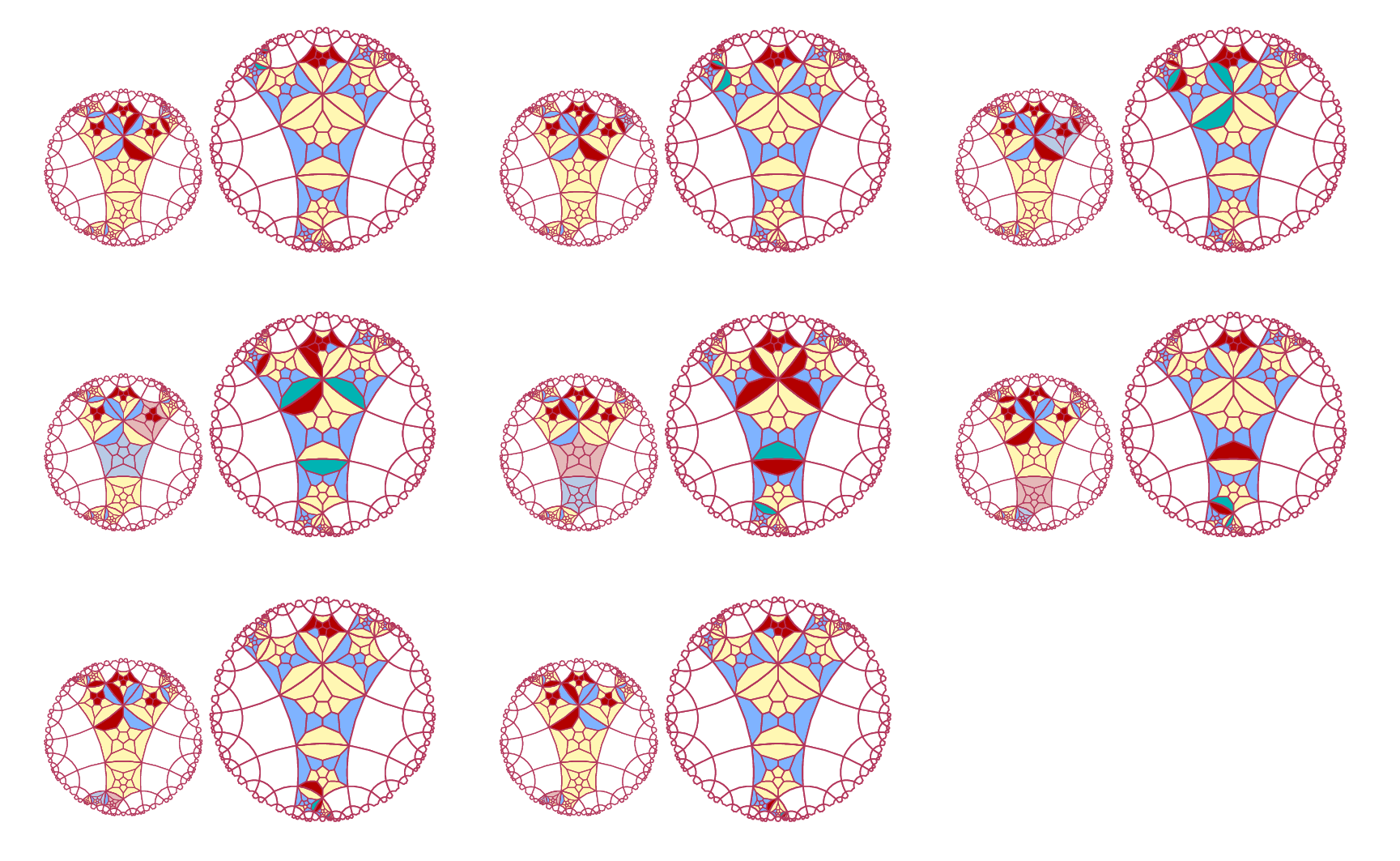,width=360pt}}
\vtop{
\vspace{5pt}
\ligne{\hfill
\PlacerEn {-350pt} {0pt} \box110
}
\vspace{-10pt}
\begin{fig}\label{loco_memo_droit_3}
\leurre
The passive crossing of a right-hand side memory switch by the locomotive through 
the non-selected track. Note that the sensors and markers are just the opposite with 
respect to Figure~{\rm\ref{loco_memo_gauche_4}} and that the above figure is 
somehow symmetric to the mentioned one. \vskip 0pt
\noindent
Note that after the fifth step, the sensors and markers 
have exchanged their colours: the memory switch is now a left-hand side one.
\end{fig}
}
\vskip 7pt

\vtop{
\begin{tab}\label{exec_memod3}
\leurre
Run of the simulation programme corresponding to Figure~{\rm\ref{loco_memo_droit_3}}.
The passive crossing through the non-selected track correspond to cells~$1$ up to~$11$, 
in the reverse order.
\end{tab}
\vspace{-12pt}
\grostrait
{\ttviii
\obeylines
\leftskip 0pt
\obeyspaces\global\let =\ \parskip=-2pt
passive crossing of a memory switch, right-hand side,
  through the NON selected track :

          1  2  3  4  5  6  7  8  9 10 11 12 13 14 15 16 17 18 19 20 21 22

time 0 :  W  W  W  W  W  W  W  W  B  R  W  W  W  W  W  W  R  B  B  B  B  R
time 1 :  W  W  W  W  W  W  W  B  R  W  W  W  W  W  W  W  R  B  B  B  B  R
time 2 :  W  W  W  W  W  W  B  R  W  W  W  W  W  W  W  W  R  B  B  B  B  R
time 3 :  W  W  W  W  W  B  R  W  W  W  W  W  W  W  W  W  W  B  B  W  B  R
time 4 :  W  W  W  W  B  R  W  W  W  W  W  W  W  W  W  W  W  B  R  R  B  R
time 5 :  W  W  W  B  R  W  W  W  W  W  W  W  W  W  W  W  B  R  B  B  R  B
time 6 :  W  W  B  R  W  W  W  W  W  W  W  W  W  W  W  W  B  R  B  B  R  B
time 7 :  W  B  R  W  W  W  W  W  W  W  W  W  W  W  W  W  B  R  B  B  R  B
\par}
\demitrait
\vskip 7pt
}

\subsection{Fixed switches}
\label{fixed}

   Figure~\ref{idle_fix} illustrates the idle configuration of a fixed switch.
As announced at the beginning of Section~\ref{the_switches} we can see on the
figures that the idle configuration of a fixed switch is, in some sense the half
of the configuration of a left-hand side memory switch. The point is that there is
no lower controller and that the sensors and markers are now fixed milestones.
Two of them, cell~18 and~21, are always red and the others, cell~17 and~22 are
always blue. Now, the fixed switch keeps the upper controller. As already noticed,
the red milestone prevents the locomotive to go through the non-selected track in
an active passage. However, as also noticed in the study of memory switches, this does not 
prevent the passage of the locomotive when it passively comes from the non-selected
track and it should work so. Now, the change of colour in the sensor of cell~12 is not 
enough to prevent the locomotive to go from the central cell both to cell~5, as required, 
and to cell~7 which should be avoided. Cell~7 cannot itself prevent such a passage 
because it sees cell~6 but it does not see at the same time cell~12. Now, we have seen in 
Subsection~\ref{memory} that the upper controller is able to perform this tasks:
as soon as it sees that the front of the locomotive is in cell~12, it becomes
white. As cell~7 sees this new colour at the same time when the front of the
locomotive is in cell~6, it allows it to reject the access of the locomotive
to the selected track.

   It is not difficult to check that Figure~\ref{idle_fix} implements this changes.
It was just enough to neutralize the lower controller by changing its colour from
blue to blank. Moreover, the cell itself has no other non-blank neighbour than the 
sensors. Similarly, as the sensors and markers are fixed milestones, this means
that their neighbours are all blank, except the cell of the switch with which they are
in contact: cell~7 or~12 for the sensors, cell~20 for the markers. Consequently, the
configuration of the fixed switch is a bit simpler than that of the left-hand side
memory switch. It also requires less non-blank cells.

   Figures~\ref{loco_fix_1}, \ref{loco_fix_3} and~\ref{loco_fix_4} illustrate the
three possible crossings of the switch by the locomotive. Each figure is accompanied
by a table which shows a trace of the execution of the simulating program corresponding
to that crossing: Table~\ref{exec_fix1} for an active passage, Table~\ref{exec_fix3}
for a passive crossing through the selected track and Table~\ref{exec_fix4}
for a passive crossing through the non-selected track.

\setbox110=\hbox{\epsfig{file=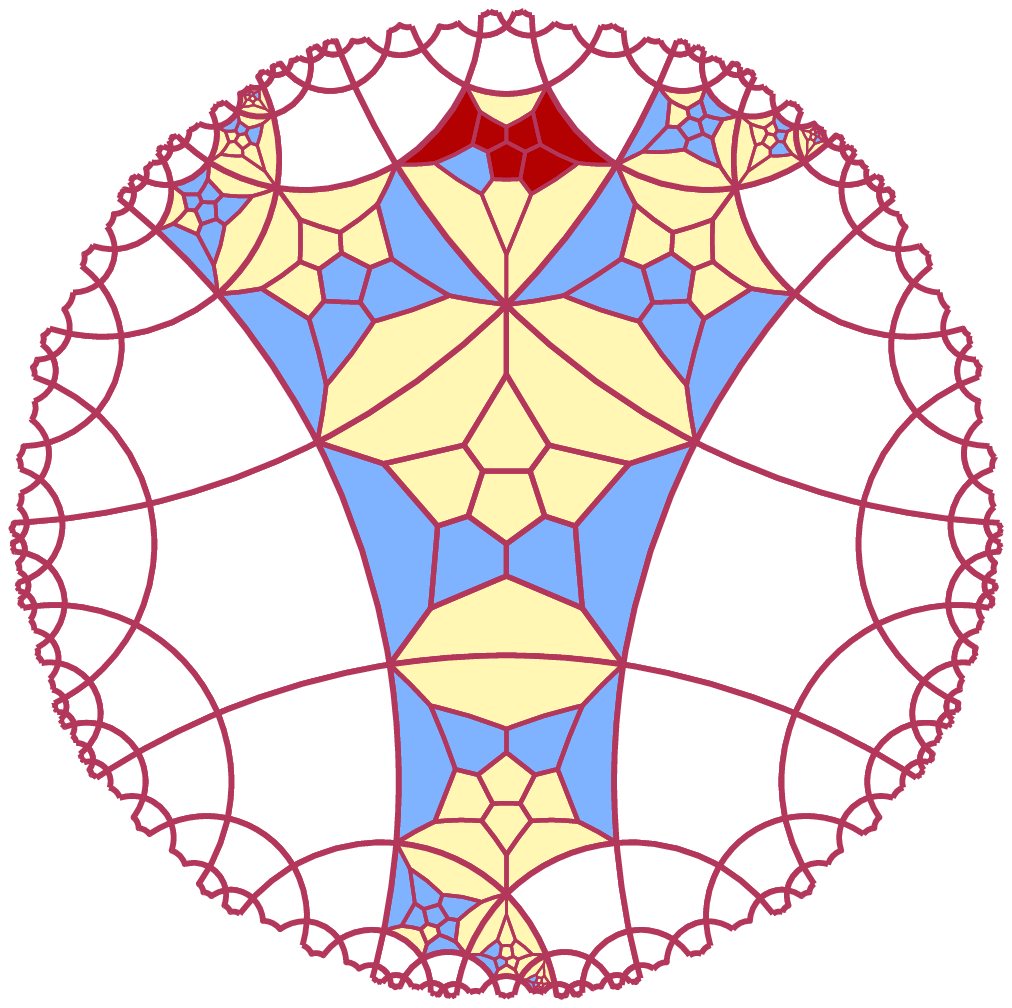,width=210pt}}
\setbox112=\hbox{\epsfig{file=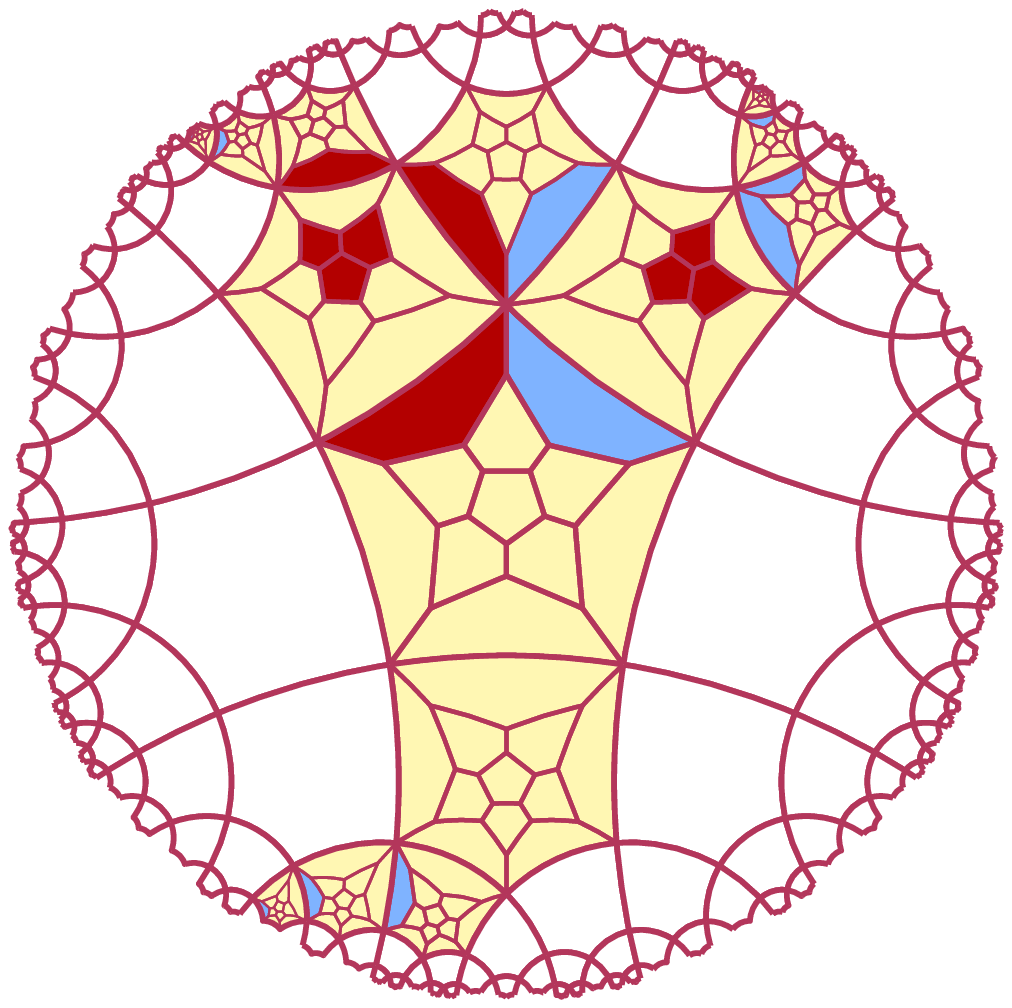,width=160pt}}
\vtop{
\vspace{-45pt}
\ligne{\hfill
\PlacerEn {-220pt} {0pt} \box110
\PlacerEn {-350pt} {0pt} \box112
}
\vspace{-10pt}
\begin{fig}\label{idle_fix}
\leurre
The idle configuration of a fixed switch. The big disc is a view from above, the small one
a view from below.
\end{fig}
}
\vskip 7pt

\setbox110=\hbox{\epsfig{file=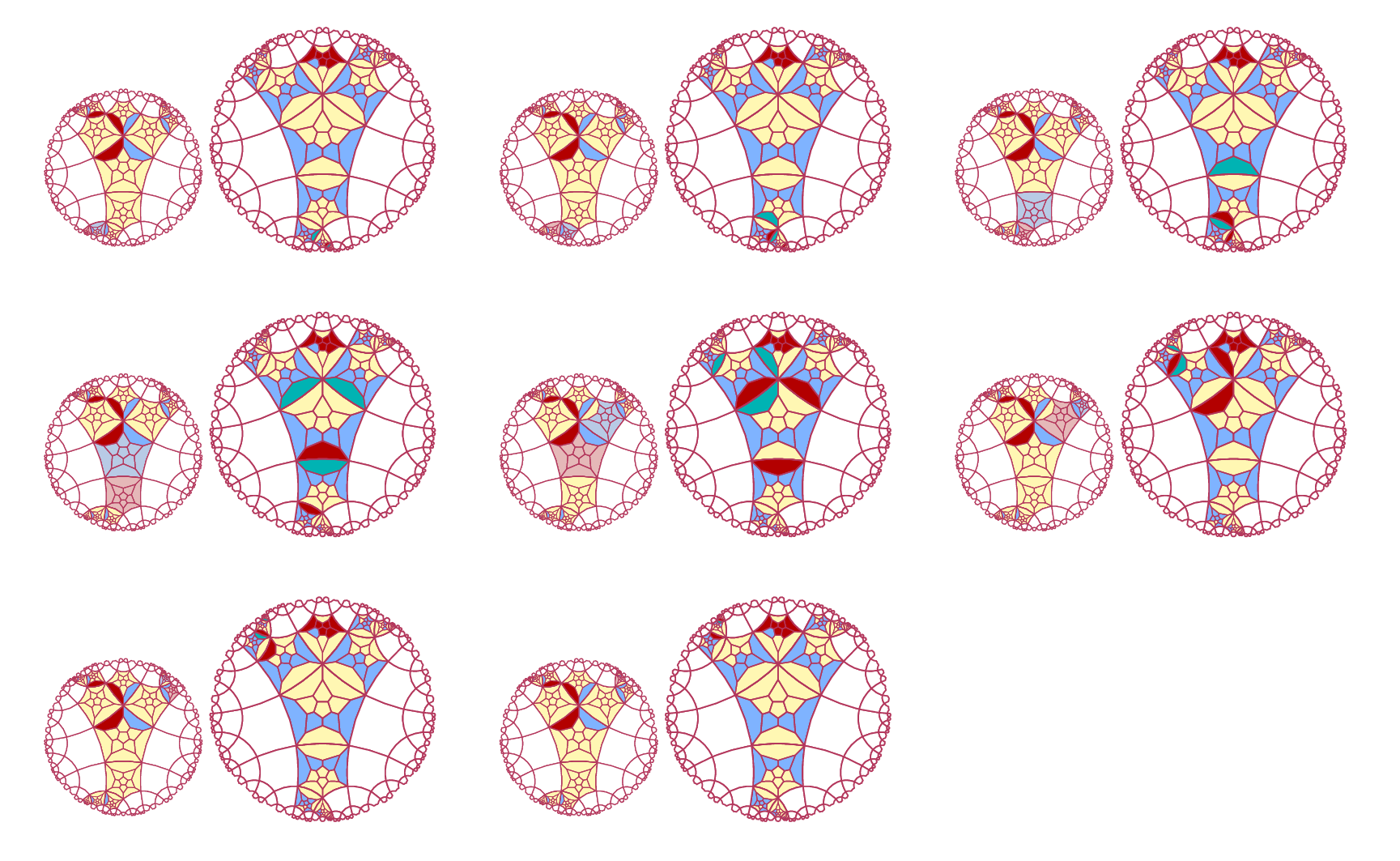,width=360pt}}
\vtop{
\vspace{10pt}
\ligne{\hfill
\PlacerEn {-350pt} {0pt} \box110
}
\vspace{-20pt}
\begin{fig}\label{loco_fix_1}
\leurre
The active passage of the locomotive through a fixed switch. Note that
the sensors and markers are unchanged.
\end{fig}
}
\vskip 7pt

\vtop{
\vspace{-10pt}
\begin{tab}\label{exec_fix1}
\leurre
Run of the simulation programme corresponding to Figure~{\rm\ref{loco_fix_1}}.
The active passage corresponds to the visit of cells~$1$ to~$11$ in this order.
\end{tab}
\vspace{-12pt}
\grostrait
{\ttviii
\obeylines
\leftskip 0pt
\obeyspaces\global\let =\ \parskip=-2pt
active crossing of a fixed switch :

          1  2  3  4  5  6  7  8  9 10 11 12 13 14 15 16 17 18 19 20 21 22

time 0 :  W  R  B  W  W  W  W  W  W  W  W  W  W  W  W  W  B  R  W  B  R  B
time 1 :  W  W  R  B  W  W  W  W  W  W  W  W  W  W  W  W  B  R  W  B  R  B
time 2 :  W  W  W  R  B  W  W  W  W  W  W  W  W  W  W  W  B  R  W  B  R  B
time 3 :  W  W  W  W  R  B  W  W  W  W  W  W  W  W  W  W  B  R  W  B  R  B
time 4 :  W  W  W  W  W  R  B  W  W  W  W  W  W  W  W  W  B  R  W  B  R  B
time 5 :  W  W  W  W  W  W  R  B  W  W  W  W  W  W  W  W  B  R  W  B  R  B
time 6 :  W  W  W  W  W  W  W  R  B  W  W  W  W  W  W  W  B  R  W  B  R  B
time 7 :  W  W  W  W  W  W  W  W  R  B  W  W  W  W  W  W  B  R  W  B  R  B
\par}
\demitrait
\vskip 7pt
}
\vskip 10pt
   The tables allow us to check that the sensors and markers are unchanged as
they should do and that for this purpose, it was enough to block their 
changing by transforming them into milestones. This is a point to which we
shall go back in the study of the rules.

   Table~\ref{exec_fix4} also allows us to check that a half-control, namely that of
cell~20 was enough to guarantee the correct working of the switch. This is an interesting
point which shows us another advantage which we can take from the third dimension.

\setbox110=\hbox{\epsfig{file=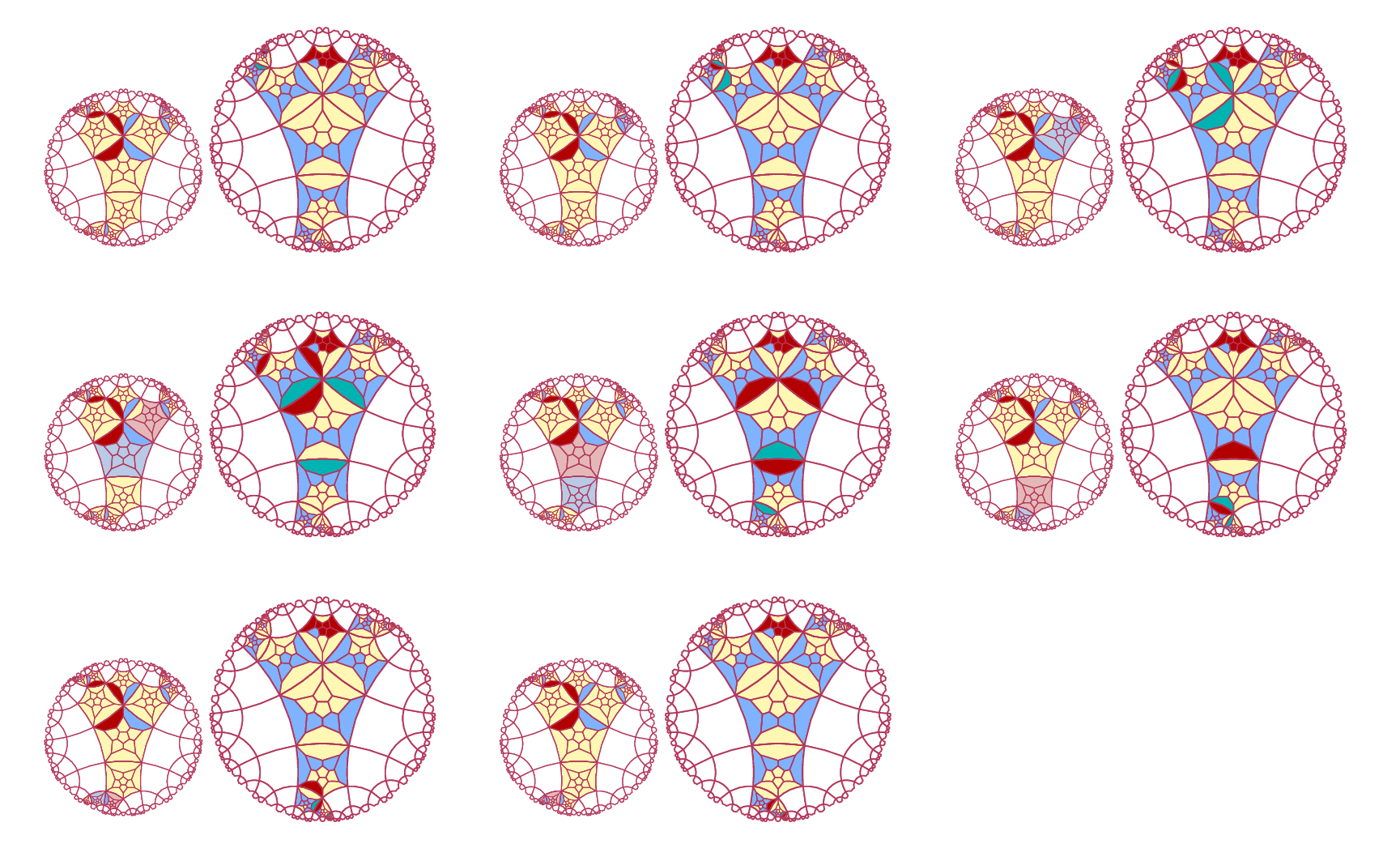,width=360pt}}
\vtop{
\vspace{10pt}
\ligne{\hfill
\PlacerEn {-350pt} {0pt} \box110
}
\vspace{-10pt}
\begin{fig}\label{loco_fix_3}
\leurre
A passive crossing of a fixed switch by the locomotive from the selected track.
Note that the sensors and markers are unchanged.
\end{fig}
}
\vskip 7pt

\vtop{
\vspace{-10pt}
\begin{tab}\label{exec_fix3}
\leurre
Run of the simulation programme corresponding to Figure~{\rm\ref{loco_fix_3}}.
The passive crossing through the selected track correspond to cells~$1$ up to~$11$, 
in the reverse order.
\end{tab}
\vspace{-12pt}
\grostrait
{\ttviii
\obeylines
\leftskip 0pt
\obeyspaces\global\let =\ \parskip=-2pt
passive crossing of a fixed switch, selected track :

          1  2  3  4  5  6  7  8  9 10 11 12 13 14 15 16 17 18 19 20 21 22

time 0 :  W  W  W  W  W  W  W  W  B  R  W  W  W  W  W  W  B  R  W  B  R  B
time 1 :  W  W  W  W  W  W  W  B  R  W  W  W  W  W  W  W  B  R  W  B  R  B
time 2 :  W  W  W  W  W  W  B  R  W  W  W  W  W  W  W  W  B  R  W  B  R  B
time 3 :  W  W  W  W  W  B  R  W  W  W  W  W  W  W  W  W  B  R  W  B  R  B
time 4 :  W  W  W  W  B  R  W  W  W  W  W  W  W  W  W  W  B  R  W  B  R  B
time 5 :  W  W  W  B  R  W  W  W  W  W  W  W  W  W  W  W  B  R  W  B  R  B
time 6 :  W  W  B  R  W  W  W  W  W  W  W  W  W  W  W  W  B  R  W  B  R  B
time 7 :  W  B  R  W  W  W  W  W  W  W  W  W  W  W  W  W  B  R  W  B  R  B
\par}
\demitrait
\vskip 7pt
}
\vskip 10pt
   Indeed, in previous simulations in the hyperbolic plane on the heptagrid, with six
or four states, we had a curious phenomenon during the active passage of the locomotive
and also during a passive crossing for the fixed switch too. In these simulations, 
the passage of the locomotive created a duplicate of its front towards the wrong 
direction. However, as this new front was not followed by a red rear, it was possible to
erase it, simply by appending a few rules.

\setbox110=\hbox{\epsfig{file=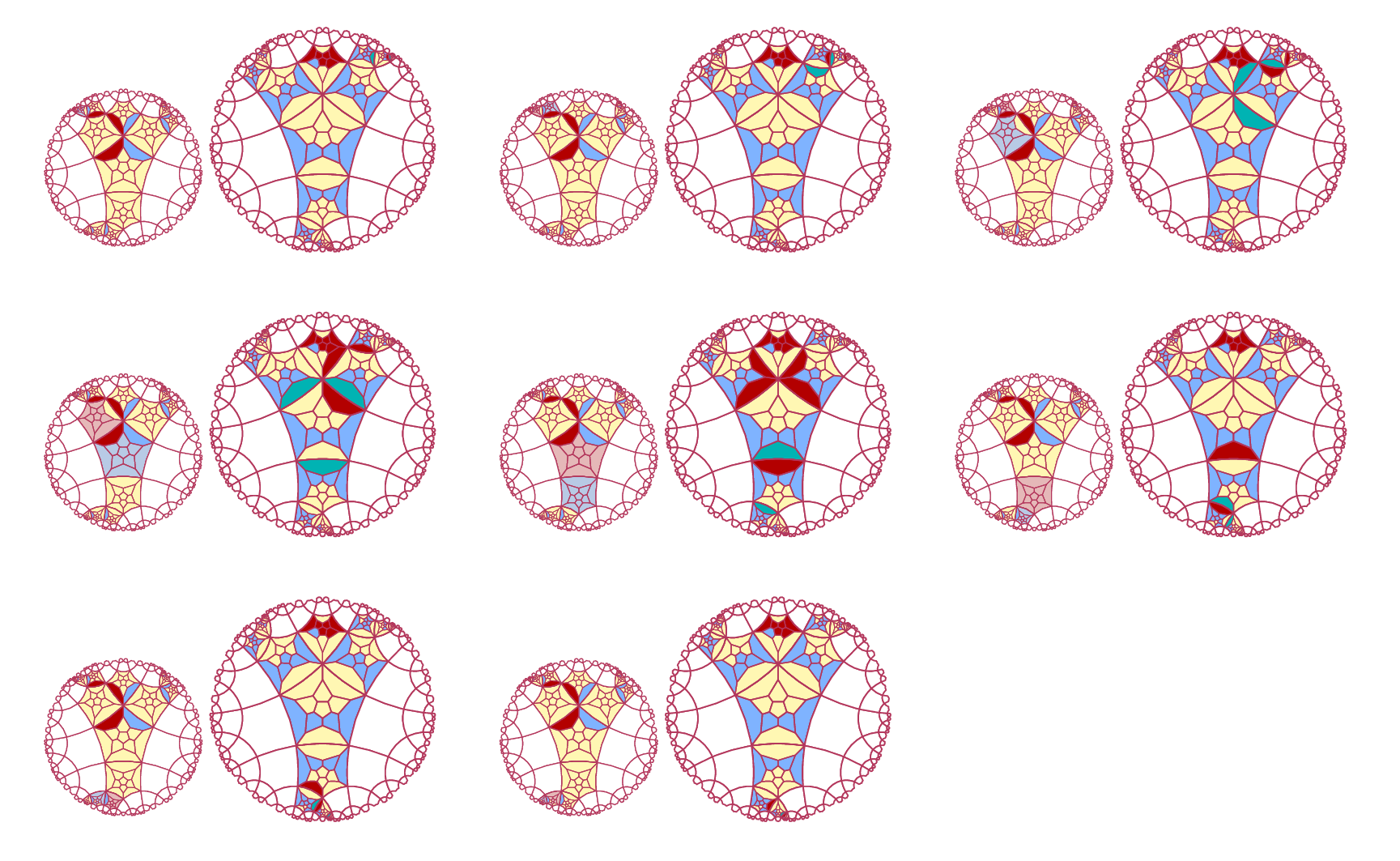,width=360pt}}
\vtop{
\vspace{10pt}
\ligne{\hfill
\PlacerEn {-350pt} {0pt} \box110
}
\vspace{-10pt}
\begin{fig}\label{loco_fix_4}
\leurre
A passive crossing of a fixed switch by the locomotive from the non-selected track.
Note that the sensors and markers are unchanged, although cell~$20$ reacts as in the
case of a memory switch. 
\end{fig}
}
\vskip 7pt

\vtop{
\vspace{-10pt}
\begin{tab}\label{exec_fix4}
\leurre
Run of the simulation programme corresponding to Figure~{\rm\ref{loco_fix_4}}.
The passive crossing through the selected track correspond to cells~$12$ up to~$16$, 
in the reverse order and then to cells~$1$ up to~$6$ in the reverse order too.
\end{tab}
\vspace{-12pt}
\grostrait
{\ttviii
\obeylines
\leftskip 0pt
\obeyspaces\global\let =\ \parskip=-2pt
passive crossing of a fixed switch, NON selected track :

          1  2  3  4  5  6  7  8  9 10 11 12 13 14 15 16 17 18 19 20 21 22

time 0 :  W  W  W  W  W  W  W  W  W  W  W  W  W  B  R  W  B  R  W  B  R  B
time 1 :  W  W  W  W  W  W  W  W  W  W  W  W  B  R  W  W  B  R  W  B  R  B
time 2 :  W  W  W  W  W  W  W  W  W  W  W  B  R  W  W  W  B  R  W  B  R  B
time 3 :  W  W  W  W  W  B  W  W  W  W  W  R  W  W  W  W  B  R  W  W  R  B
time 4 :  W  W  W  W  B  R  W  W  W  W  W  W  W  W  W  W  B  R  W  R  R  B
time 5 :  W  W  W  B  R  W  W  W  W  W  W  W  W  W  W  W  B  R  W  B  R  B
time 6 :  W  W  B  R  W  W  W  W  W  W  W  W  W  W  W  W  B  R  W  B  R  B
time 7 :  W  B  R  W  W  W  W  W  W  W  W  W  W  W  W  W  B  R  W  B  R  B
\par}
\demitrait
\vskip 7pt
}
\vskip 15pt

\subsection{Flip-flop switches}
\label{flip_flop}

   Figures~\ref{idle_flip_flop_G} and~\ref{idle_flip_flop_D} show the idle configuration
of the flip-flop switches which exist in two versions: the left- and the right-hand side
ones. As announced at the beginning of Section~\ref{the_switches} we can see on the
figures that the idle configuration of a flip-flop switch switch is, in some sense the half
of the configuration of a memory switch of the same laterality. The point is that there is
no upper controller and, consequently, no markers. However, the sensors and the lower
are still present. The lower controller is exactly the same as in the memory switches 
and it works in the same way. Contrarily to the fixed switch, the sensors are not 
milestones. They are true sensors like in the memory switch. 
\ligne{\hfill}
\vskip -12pt
\setbox110=\hbox{\epsfig{file=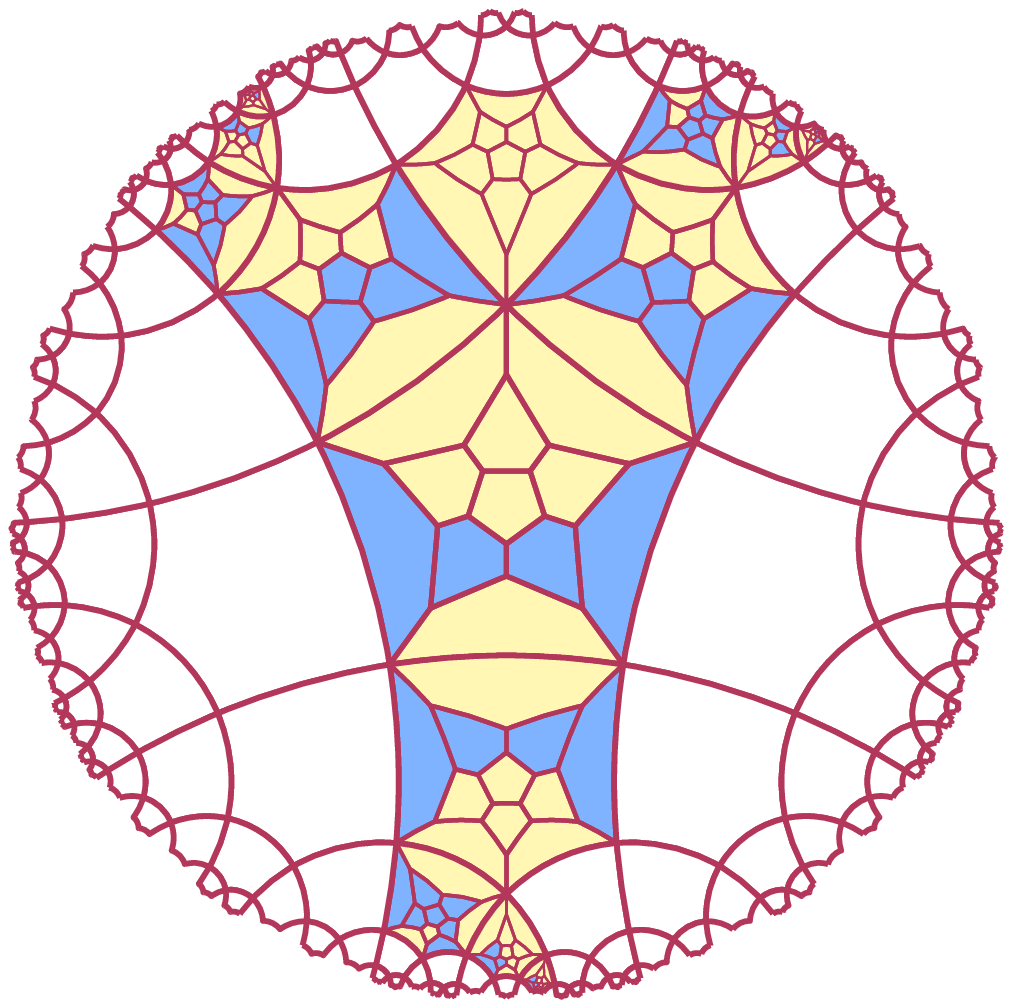,width=210pt}}
\setbox112=\hbox{\epsfig{file=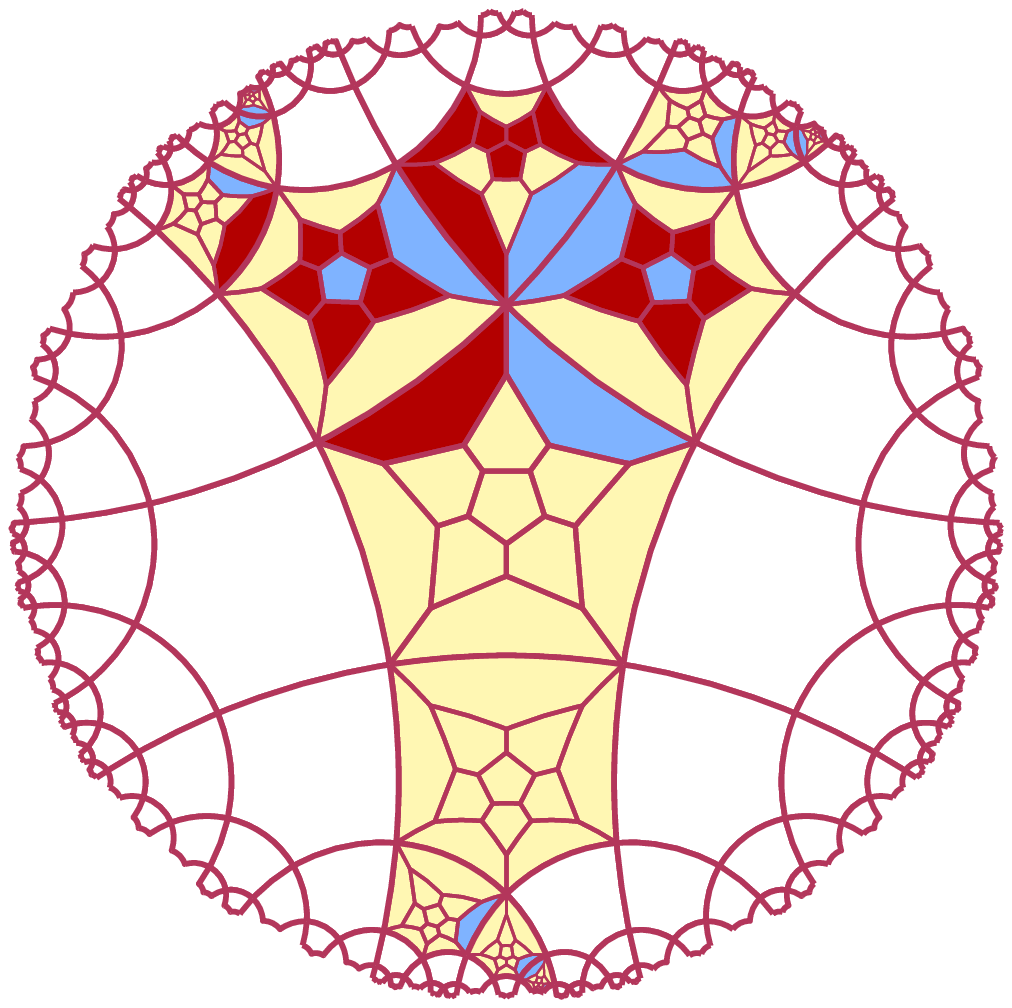,width=160pt}}
\vtop{
\vspace{-15pt}
\ligne{\hfill
\PlacerEn {-220pt} {0pt} \box110
\PlacerEn {-350pt} {0pt} \box112
}
\vspace{-10pt}
\begin{fig}\label{idle_flip_flop_G}
\leurre
The idle configuration of the left-hand side flip-flop switch. The big disc is
a view from above, the small one, a view from below.
\end{fig}
}

\setbox110=\hbox{\epsfig{file=idle_flip_flop_up.ps,width=210pt}}
\setbox112=\hbox{\epsfig{file=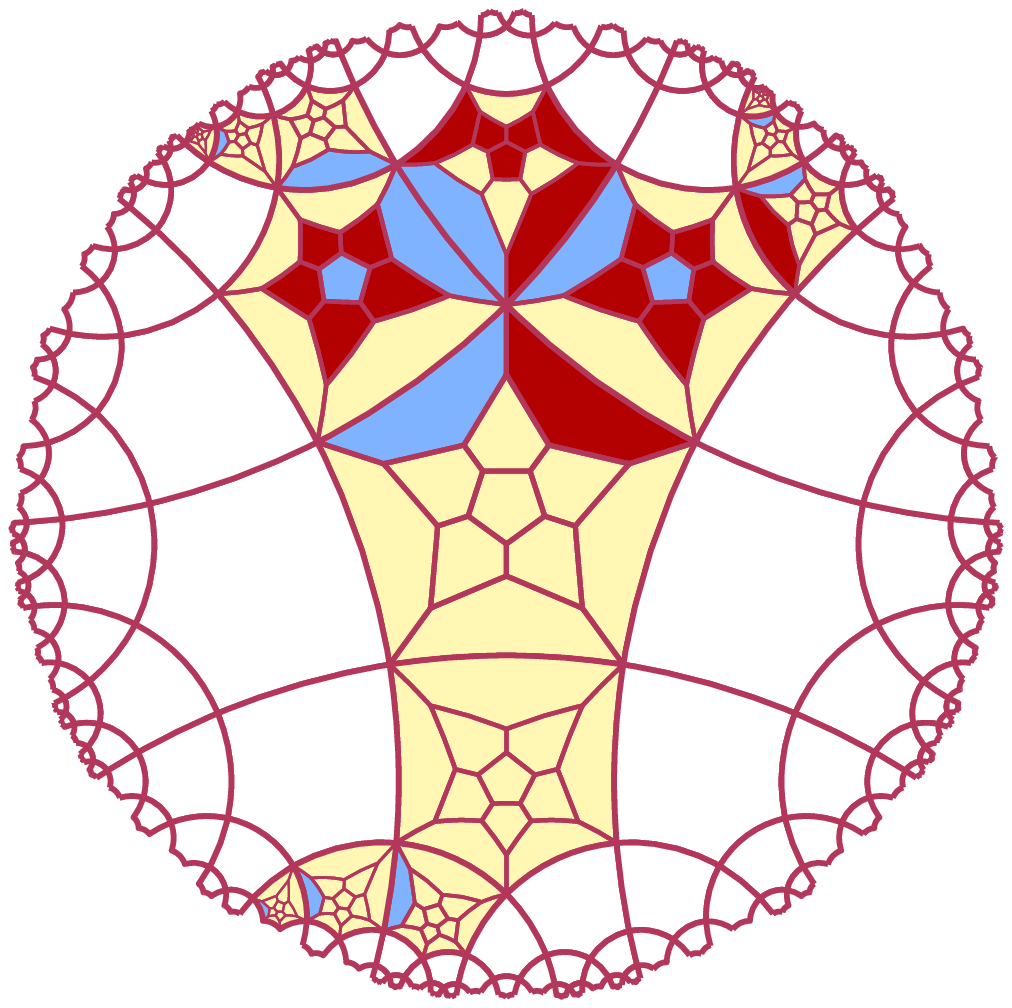,width=160pt}}
\vtop{
\vspace{-15pt}
\ligne{\hfill
\PlacerEn {-220pt} {0pt} \box110
\PlacerEn {-350pt} {0pt} \box112
}
\vspace{-10pt}
\begin{fig}\label{idle_flip_flop_D}
\leurre
The idle configuration of the right-hand side flip-flop switch. The big disc is
a view from above, the small one, a view from below.
\end{fig}
}
\vskip 7pt
\noindent
However, they are a bit different from the sensors of the memory switch as they 
work in a different way. The difference can be noticed in the small discs of 
Figures~\ref{idle_memo_gauche} and~\ref{idle_memo_droit} and those of 
Figures~\ref{idle_flip_flop_G} and~\ref{idle_flip_flop_D}. In 
Figures~\ref{idle_memo_gauche} and~\ref{idle_memo_droit}, the sensors are marked by
a group of three red milestones, on pairwise contiguous faces, one of them being the
face which is opposite to that in contact with the scanned cell. In 
Figures~\ref{idle_flip_flop_G} and~\ref{idle_flip_flop_D}, the sensors are marked
by a ring of five milestones whose contact faces with the sensor are around the face
which is opposite to the face in contact with the scanned cell. Also, the
face opposite to that which is shared with the scanned cell is blue as it is in contact
with a blue milestone.

   This difference is explained by the fact that the working of the sensors is
very different from that of the memory switch, quite the opposite: in the memory
switch, the blue sensor is passive and the red sensor blocks the access to
the non-selected track in the active passage and turns to white when the front of 
the locomotive appears in the scanned cell in a passive crossing. In the flip-flop 
switch, the red sensor only blocks the access to the non-selected track and the
blue sensor is active: when the front of the locomotive leaves the scanned cell,
it becomes white, triggering the flash of the lower controller at the next time
which, to its turn, makes the sensors exchange their colour.

\setbox110=\hbox{\epsfig{file=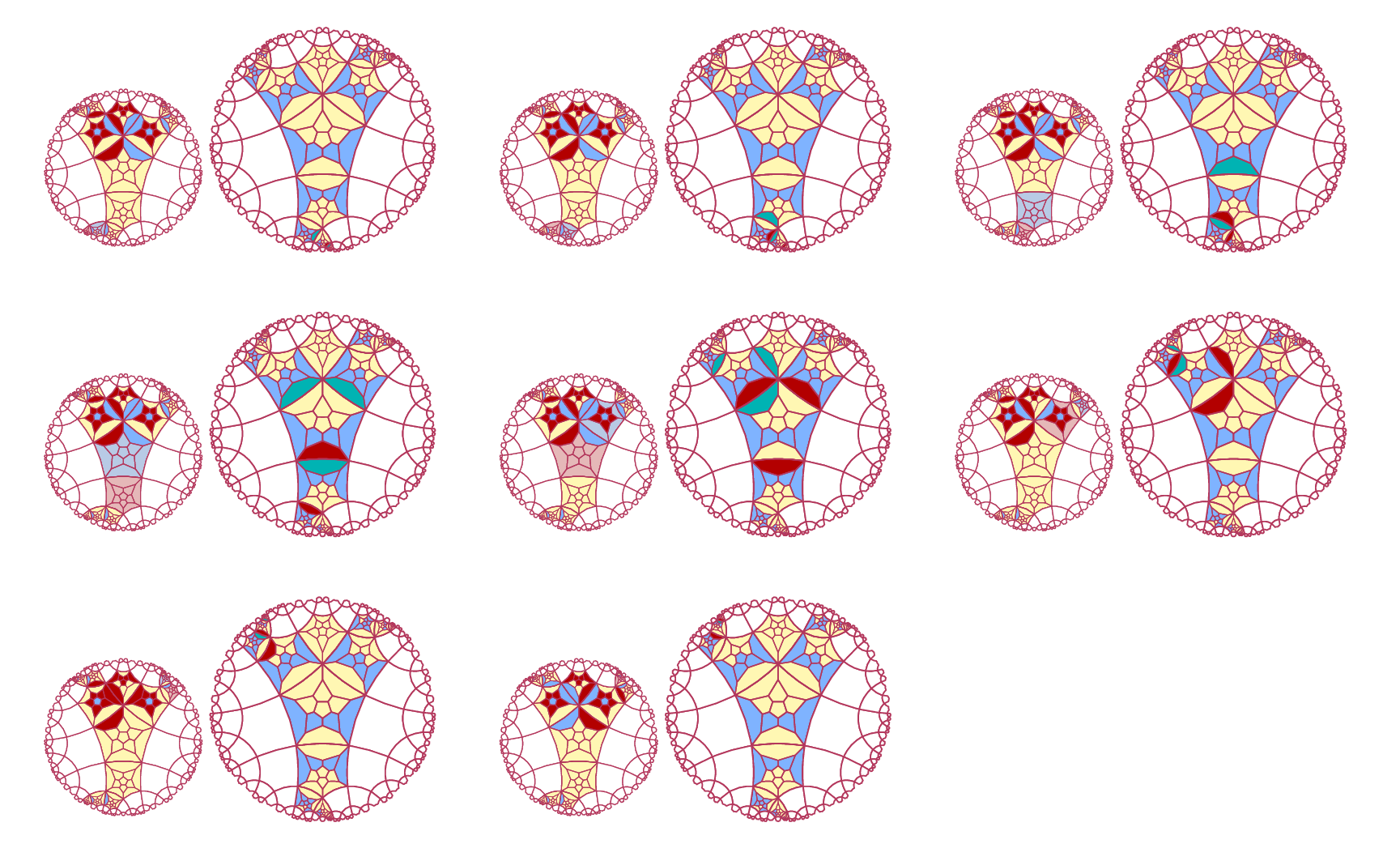,width=360pt}}
\vtop{
\vspace{10pt}
\ligne{\hfill
\PlacerEn {-350pt} {0pt} \box110
}
\vspace{-10pt}
\begin{fig}\label{loco_flip_flop_G}
\leurre
The active passage of the locomotive through a left-hand side flip-flop switch. Note that
the sensors are exchanged when the locomotive leaves the switch, now as a right-hand side
flip-flop switch.
\end{fig}
}
\vskip 7pt

\vtop{
\begin{tab}\label{exec_flip_flop_G}
\leurre
Run of the simulation programme corresponding to Figure~{\rm\ref{loco_flip_flop_G}}.
The active passage visits the cells of the selected track: cells~$1$ up to~$11$
in this order.
\end{tab}
\vspace{-12pt}
\grostrait
{\ttviii
\obeylines
\leftskip 0pt
\obeyspaces\global\let =\ \parskip=-2pt
active crossing of a left-hand side flip-flop switch :

          1  2  3  4  5  6  7  8  9 10 11 12 13 14 15 16 17 18 19 20 21 22

time 0 :  W  R  B  W  W  W  W  W  W  W  W  W  W  W  W  W  B  R  B  W  W  W
time 1 :  W  W  R  B  W  W  W  W  W  W  W  W  W  W  W  W  B  R  B  W  W  W
time 2 :  W  W  W  R  B  W  W  W  W  W  W  W  W  W  W  W  B  R  B  W  W  W
time 3 :  W  W  W  W  R  B  W  W  W  W  W  W  W  W  W  W  B  R  B  W  W  W
time 4 :  W  W  W  W  W  R  B  W  W  W  W  W  W  W  W  W  B  R  B  W  W  W
time 5 :  W  W  W  W  W  W  R  B  W  W  W  W  W  W  W  W  W  R  B  W  W  W
time 6 :  W  W  W  W  W  W  W  R  B  W  W  W  W  W  W  W  W  R  R  W  W  W
time 7 :  W  W  W  W  W  W  W  W  R  B  W  W  W  W  W  W  R  B  B  W  W  W
\par}
\demitrait
\vskip 7pt
}
\vskip 10pt
   This can be checked in Tables~\ref{exec_flip_flop_G} and~\ref{exec_flip_flop_D}.
The front of the locomotive is in the scanned cell at time~4, so that the blue sensor
is white at time~5. As in the case of the memory switch, as cell~17 and~18 do 
not see each other, the blue sensor cannot turn to red immediately. It becomes
white which triggers the flash of the controller at time~6 and the exchange of colours
between the sensors at time~7 only.

\setbox110=\hbox{\epsfig{file=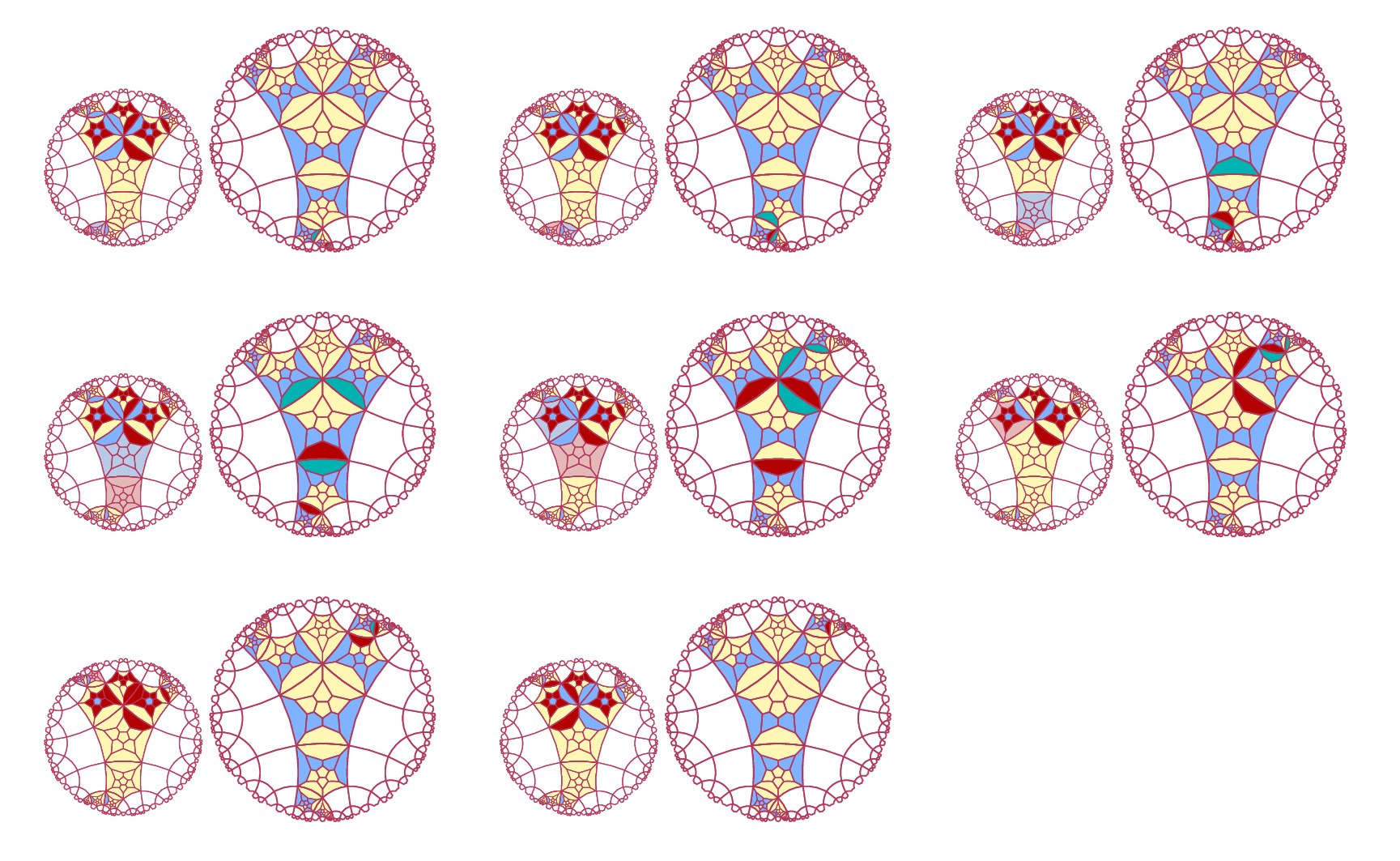,width=360pt}}
\vtop{
\vspace{10pt}
\ligne{\hfill
\PlacerEn {-350pt} {0pt} \box110
}
\vspace{-10pt}
\begin{fig}\label{loco_flip_flop_D}
\leurre
The active passage of the locomotive through a right-hand side flip-flop switch. Note that
the sensors are exchanged when the locomotive leaves the switch, now as a left-hand side
flip-flop switch.
\end{fig}
}
\vskip 7pt

\vtop{
\vspace{-10pt}
\begin{tab}\label{exec_flip_flop_D}
\leurre
Run of the simulation programme corresponding to Figure~{\rm\ref{loco_flip_flop_D}}
The active passage visits the cells of the selected track: , here cells~$1$ up to~$6$
and then cells~$12$ up to~$16$, both in this order.
\end{tab}
\vspace{-12pt}
\grostrait
{\ttviii
\obeylines
\leftskip 0pt
\obeyspaces\global\let =\ \parskip=-2pt
active crossing of a right-hand side flip-flop switch :

          1  2  3  4  5  6  7  8  9 10 11 12 13 14 15 16 17 18 19 20 21 22

time 0 :  W  R  B  W  W  W  W  W  W  W  W  W  W  W  W  W  R  B  B  W  W  W
time 1 :  W  W  R  B  W  W  W  W  W  W  W  W  W  W  W  W  R  B  B  W  W  W
time 2 :  W  W  W  R  B  W  W  W  W  W  W  W  W  W  W  W  R  B  B  W  W  W
time 3 :  W  W  W  W  R  B  W  W  W  W  W  W  W  W  W  W  R  B  B  W  W  W
time 4 :  W  W  W  W  W  R  W  W  W  W  W  B  W  W  W  W  R  B  B  W  W  W
time 5 :  W  W  W  W  W  W  W  W  W  W  W  R  B  W  W  W  R  W  B  W  W  W
time 6 :  W  W  W  W  W  W  W  W  W  W  W  W  R  B  W  W  R  W  R  W  W  W
time 7 :  W  W  W  W  W  W  W  W  W  W  W  W  W  R  B  W  B  R  B  W  W  W
\par}
\demitrait
\vskip 7pt
}

\subsection{The circuit}
\label{circuit}

   With the implementation of the tracks and the switches, it is possible to give
an idea of the implementation of the whole circuit in the hyperbolic $3D$~space.
Remember that the circuit is mainly a two-dimensional one and that the third dimension
is used to avoid crossings and to improve the working of the switch mechanisms.
This is why a projection on the plane~$\Pi_0$ of the face~5 of most the straight 
elements may give an idea of the general implementation. To this purpose, we 
give in Figure~\ref{univ_global} a schematic implementation of the toy program
of Figure~\ref{example} in the pentagrid, the same figure as in~\cite{mmbook2}.
This figure gives a correct idea of the projection of the implementation
onto~$\Pi_0$.
 
\setbox110=\hbox{\epsfig{file=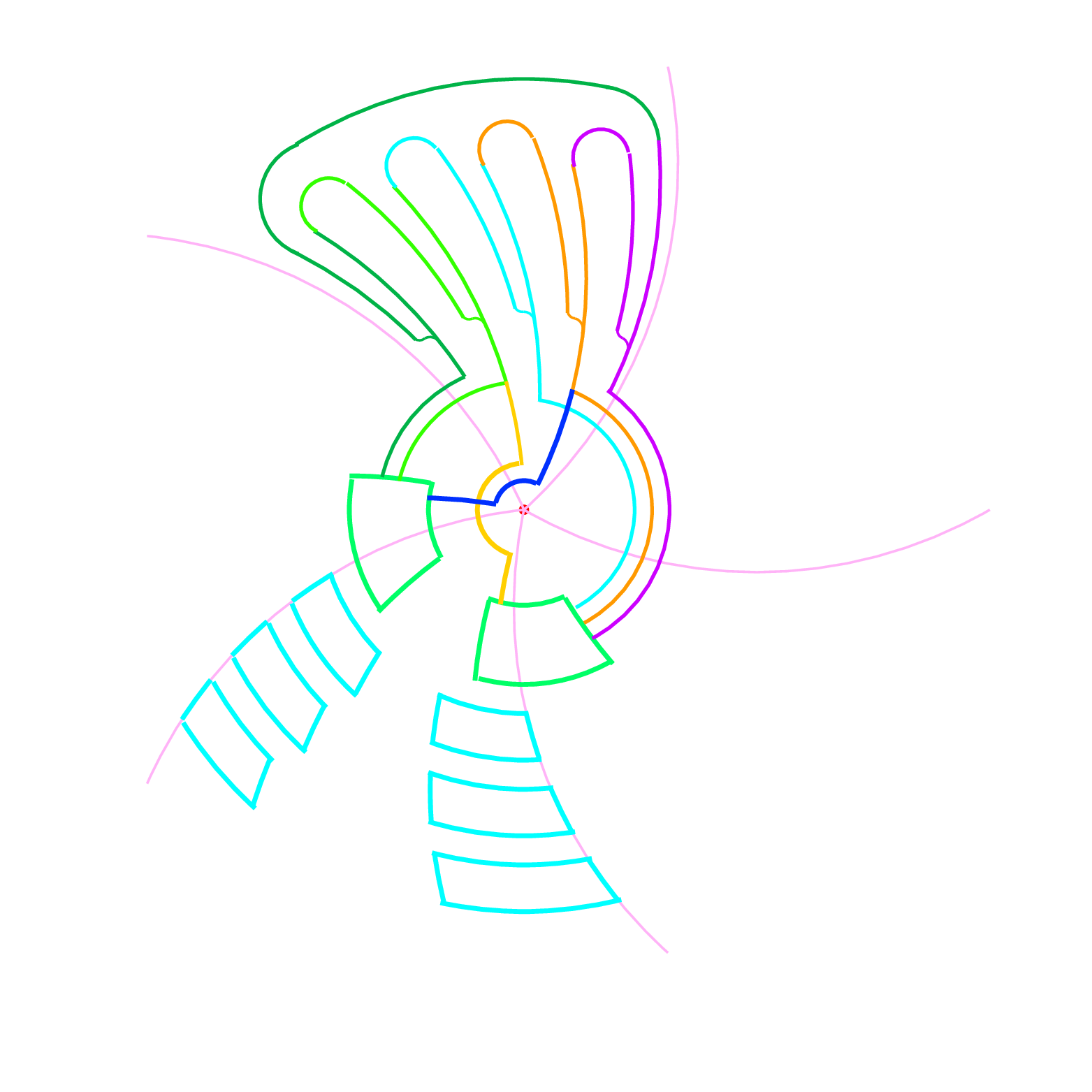,width=350pt}}
\vtop{
\vspace{-10pt}
\ligne{\hfill
\PlacerEn {-350pt} {0pt} \box110
}
\vspace{-40pt}
\begin{fig}\label{univ_global}
\leurre
A projection of an implementation of the example of a program of a register
machine in Figure~{\rm\ref{example}}. 
\end{fig}
}
\vskip 7pt

   From the figure, we can see that this general organization can be kept for
the circuit which we consider. Registers are installed in a quarter, the
sequencer of the instructions is installed in another quarter, almost opposite
to that of the registers.

\section{Devising the rules and checking their correctness}
\label{the_rules}

   Now we arrive to the important part of the paper which consists in listing the rules
and in proving their correctness.

   In order to write the rules of the cellular automaton, we shall use the numbering of
the faces of a dodecahedron which was introduced in Subsection~\ref{dodecagrid}
and which was used in Subsection~\ref{navigation} and in Sections~\ref{the_tracks}
and~\ref{the_switches}. However, there was no fixed rule to connect the numbering
of a cell to that of a neighbouring one except for the cells of the track, as
we did in Subsections~\ref{vert} and~\ref{horiz}. This is not a big problem as,
in fact, the rules which we shall devise have an important property: they are 
{\bf rotation invariant}, which means that they are not changed by a rotation
of the faces leaving the dodecahedron globally invariant. 

   In the plane, the characterization of rotation invariance in the rules is easy
to formulate: it is necessary and sufficient that the rules are not changed by a 
circular permutation on the neighbours. In the case of the pentagrid, this means
that once we fixed a rule, we automatically append to the table of rules the
other four permuted images of the rule. Here, the characterization is far less
trivial. In~\cite{mm3DJCA}, we could avoid this problem by imposing a stronger condition
on the rules, namely to be {\bf strongly lexicographically} different from each other.
This means that to each rule, we associate a word of the for $A_1^{k_1}..A_n^{_n}$
where $A_1$, ..., $A_n$ are the states and $k_1$ are non-negative numbers satisfying
$k_1+...+k_n = v$+1, where $v$ is the number of neighbours of the cell, the cell
being not counted. This was possible with 5~states and I could not keep this condition
for 3~states. This is why we first study how to check rotation invariance for our
cellular automaton in the hyperbolic $3D$~space.

\subsection{Rotation invariance}
\label{rotation_invariance}

   The question is the following: how does a motion which leaves the dodecahedron
globally invariant affect the numbering of its faces, an initial numbering being fixed
as in Subsection~\ref{dodecagrid}?

   In fact, it is enough to consider products of rotations as we do not consider
reflections in planes. The simplest way to deal with this problem is the following.
Consider a motion which preserves the orientation, we shall say a {\bf positive} motion. 
As it leaves the dodecahedron globally invariant, it transforms the face into another 
one. Accordingly, fix face~0. Then its image can be any face of the dodecahedron, 
face~0 included. Let $f_0$ be the image of face~0. Next, fix a second face which shares 
an edge with face~0, for instance face~1. Then its image $f_1$ is a face which shares 
an edge with~$f_0$. It can be any face sharing a face with~$f_0$. 
Indeed, let $f_2$ be another face sharing an edge with~$f_1$. Then, composing 
the considered positive motion with a rotation around $f_0$ transforming $f_1$ into~$f_2$, 
we get a positive motion which transforms $(0,1)$ into $(f_0,f_2)$. This proves that we 
get all the considered positive motion leaving the dodecahedron globally invariant,
by first fixing the image of face~0, say $f_0$ and then by taking any face~$f_1$
sharing an edge with~$f_1$. Note that once $f_0$ and~$f_1$ are fixed, the images of
the other faces are fixed, thanks to the preservation of the orientation.
Accordingly, there are 60 of these positive motions and the argument of the proof
shows that they are all products of rotations leaving the dodecahedron globally
invariant. 

   Figure~\ref{rotations} gives an illustrative classification of all these rotations.
The upper left picture represents the image of a Schlegel diagram of a dodecahedron
with the notation introduced in Subsection~\ref{dodecagrid}. Each image represents 
a positive motion. Its characterization is given by the couple of numbers under the
image: it has the form~$f_0\ f_1$, where $f_0$ is the image of face~0 and $f_1$ is 
the image of face~1. The figure represents two sub-tables, each one containing
30~images. Each row represents the possible images of~$f_1$, $f_0$ being fixed.
The image of face~0 is the back of the dodecahedron. The image of face~1
is the place of face~1{} in Figure~\ref{dodec}. As an example, $f_0=0$ 
for the first row of the left-hand side sub-table, and in the first row, the first
image gives $f_1=1$, so that it represents the identity. The other images
of the row represent the rotations around face~0.

\setbox110=\hbox{\epsfig{file=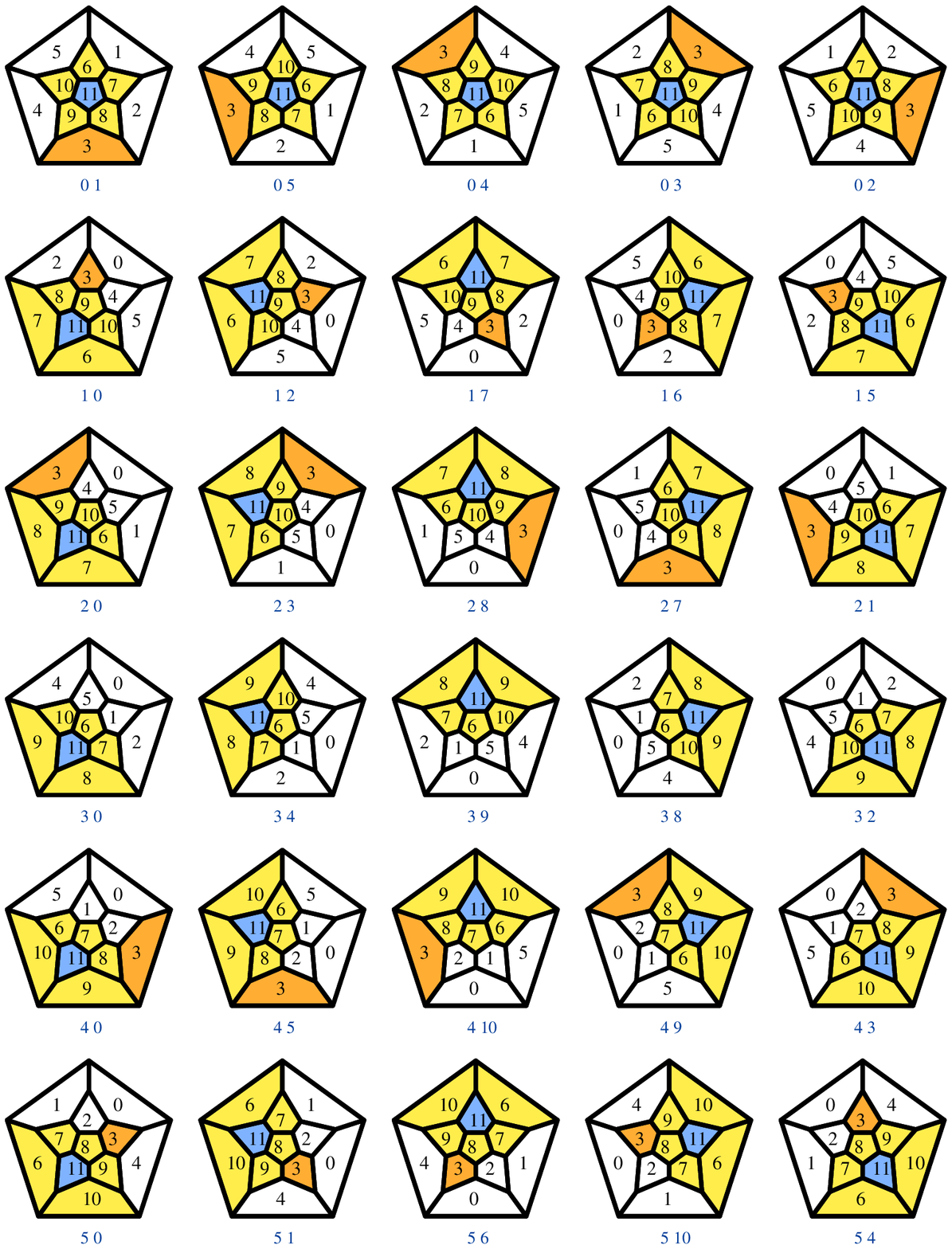,width=180pt}}
\setbox112=\hbox{\epsfig{file=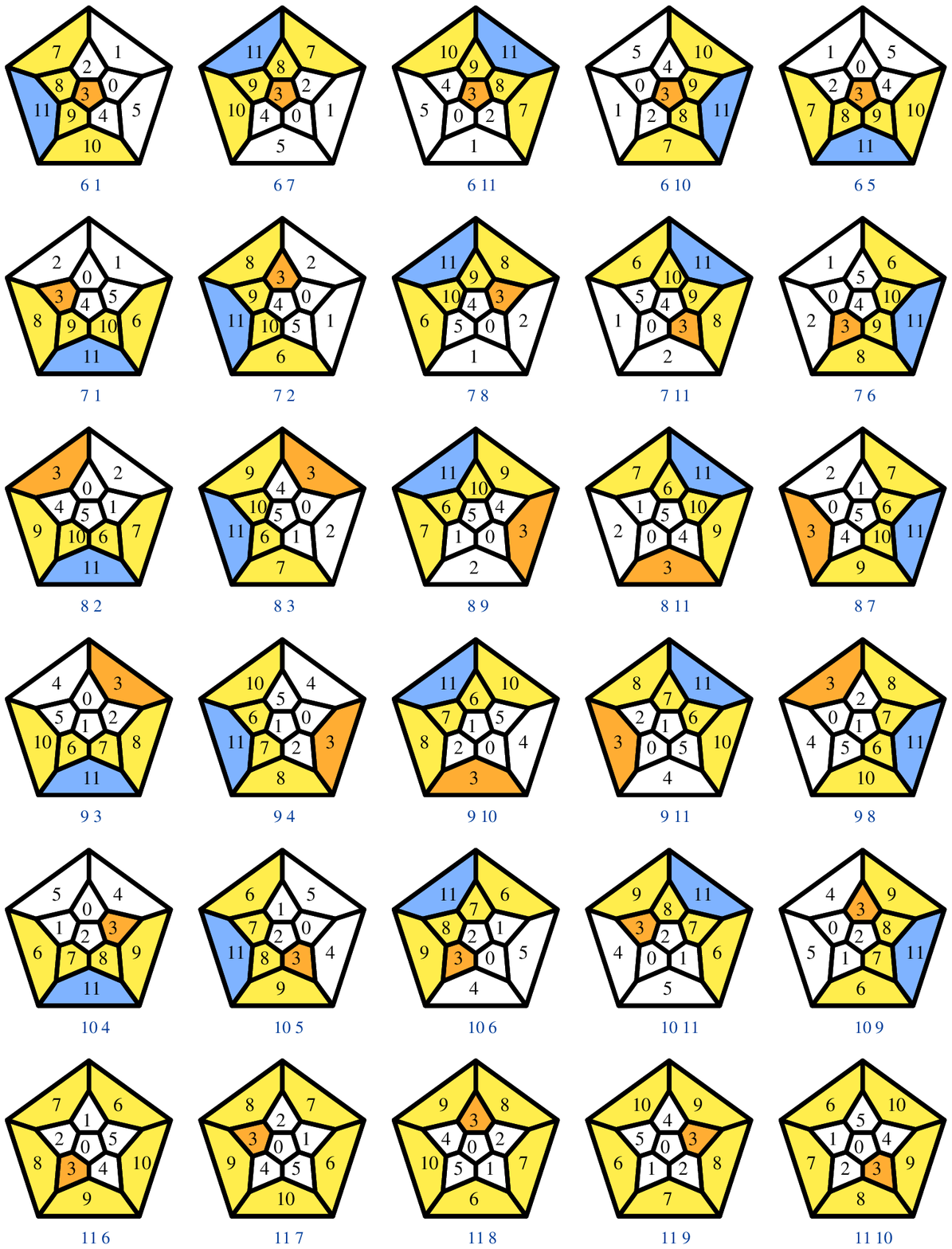,width=180pt}}
\vtop{
\vspace{-15pt}
\ligne{\hfill
\PlacerEn {-345pt} {0pt} \box110
\PlacerEn {-165pt} {0pt} \box112
}
\vspace{-20pt}
\begin{fig}\label{rotations}
\leurre
The map of the positive motions leaving the dodecahedron globally invariant.
\end{fig}
}
\vskip 7pt

\def\rangee #1 #2 #3 #4 #5 #6 {%
\ligne{\hbox to 30pt{\hfill#1\hfill}
       \hbox to 25pt{\hfill#2\hskip 7pt}
       \hbox to 25pt{\hfill#3\hskip 7pt}
       \hbox to 25pt{\hfill#4\hskip 7pt}
       \hbox to 25pt{\hfill#5\hskip 7pt}
       \hbox to 25pt{\hfill#6\hskip 7pt}
      }
\vskip 4pt
}
\vtop{
\begin{tab}\label{couronnes}
\leurre
The faces around a given face.
\end{tab}
\vspace{-12pt}
\ligne{\hfill
\vtop{\leftskip 0pt\parindent 0pt\hsize=190pt
\grostrait
\ligne{\hfill
       \vtop{\offinterlineskip\leftskip 0pt\parindent 0pt\hsize=168.3333pt
\rangee   { } 1  2   3   4   5
}\hfill}
\demitrait
\ligne{\hfill
       \vtop{\offinterlineskip\leftskip 0pt\parindent 0pt\hsize=168.3333pt
\rangee   0   1  5   4   3   2
\rangee   1   0  2   7   6   5
\rangee   2   0  3   8   7   1
\rangee   3   0  4   9   8   2
\rangee   4   0  5  10   9   3
\rangee   5   0  1   6  10   4
\rangee   6   1  7  11  10   5
\rangee   7   1  2   8  11   6
\rangee   8   2  3   9  11   7
\rangee   9   3  4  10  11   8
\rangee  10   4  5   6  11   9
\rangee  11   6  7   8   9  10
}
\hfill}
\demitrait
}
\hfill}
}

The construction of Figure~\ref{rotations} was performed by an algorithm
using Table~\ref{couronnes}. For each face of the dodecahedron, the table gives
the faces which surround it in the Schlegel diagram, taking the clockwise order
when looking at the face from outside the dodecahedron, this order coinciding with
increasing indices in each row. This coincides with the 
usual clockwise order for all faces as in Figure~\ref{dodec}, except for face~0
for which the order is counter-clockwise when looking above the plane of the projected
image. The principle of the drawings consists in placing~$f_0$ onto face~0 
and~$f_1$ onto face~1. The new numbers of the faces are computed by the algorithm
as follows. Being given the new numbers~$f_0$ and~$f_1$ of two contiguous 
faces~$\varphi_0$ and~$\varphi_1$ in the Schlegel diagram, the algorithm computes 
the position of $\varphi_1$ as a neighbour of~$\varphi_0$ in the table. This
allows to place~$f_1$ on the right face. Then, the algorithm computes the new
numbers of the faces which are around~$\varphi_1$ in the table: it is enough to take
the position of~$\varphi_0$ as a neighbour of~$\varphi_1$ and then to turn around
the neighbours of~$f_1$, looking at the new numbers in the row~$f_1$ of the table,
starting from the position of~$f_0$. This gives the new numbers of the faces
which surround face~1. It is easy to see that we have all faces
of the dodecahedron by turning around face~1, then around face~5, then around face~7 
and at last around face~8. As in these steps, each round of faces starts from a face 
whose new number is already computed, the algorithm is able to compute the new numbers 
for the current round of faces, using Table~\ref{couronnes} to find the
new numbers. Let us call this algorithm the {\bf rotation 
algorithm}.

   Thanks to the rotation algorithm, it is easy to compute the {\bf rotated forms}
of a rule of the cellular automaton.

   Let $\underline{\eta}\eta_0...\eta_{11}\underline{\eta'}$ be a rule of the automaton.
In this format, $\eta$ is the current state of the cell and $\eta(i)$ is the state 
of face~$i$ and $\eta'$ is the new state of the automaton. Remember that 
the current state of a cell is its sate at time~$t$ and that its new state is
its state at time~$t$+1. The number of a neighbour of the cell in the rule is the
number of the face which shares this face with the neighbour, the number of the 
face being defined by the numbering of the faces in the dodecahedron which supports 
the cell. By assumption, this numbering is a rotated image of the numbering defined by 
Figure~\ref{dodec} in Subsection~\ref{dodecagrid}. Later we shall call
$\underline{\eta}\eta_0...\eta_{11}$ the {\bf context} of the rule.
Let $\mu$ be a positive motion leaving the dodecahedron globally invariant. 
The {\bf rotated form} of the rule defined by~$\mu$ is 
$\underline{\eta}\eta_{\mu(0)}...\eta_{\mu(11)}\underline{\eta'}$ and, similarly,
$\underline{\eta}\eta_{\mu(0)}...\eta_{\mu(11)}$ is the {\bf rotated form} by~$\mu$
of the context of the initial rule.
We say that the cellular automaton is {\bf rotation invariant} if and only
two rules having contexts which are rotated forms of each other always produce the same
new state.

   Now, thanks to our study, we have a syntactic criterion to check this property.
We fix an order of the states. Then, for each rule, we compute its {\bf minimal form}. 
This form is obtained as follows. We compute all rotated forms of the rule and,
looking at the obtained contexts as words, we take their minimum in the lexicographic 
order. The minimal form of a rule is obtained by appending its new state to this minimum. 
Now it is easy to see that:

\begin{lem}
A cellular automaton on the dodecagrid is rotation invariant if and only if
for any pair of rules, if their minimal forms have the same context, they 
have the same new state too.
\end{lem}

   Now, checking this property can easily be performed thanks to the rotation algorithm.

   In the following, we shall successively deal with the rules regarding the tracks and
the motion of the locomotive on them, then the rules for the switches and their crossings
by the locomotive. In all cases, we shall distinguish between {\bf conservative} rules
in which the new state is the same as the current one and {\bf motion} rules in which
the new state is different, due to the contribution of the considered cell to the motion
of the locomotive. Also, due to the rotation invariance of our set of rules, we shall
give only one form for each rule, not necessarily the minimal one.

\subsection{The rules for the tracks and the motion of the locomotive}
\label{trackmotion_rules}

   The conservative rules for the tracks are given in Table~\ref{track_cons}.
The first rule is the rule of the quiescent state, symbolized by~$W$ in the tables
giving the rules. The next two rules are the rule for the cell of the track in a
straight element and then for the same cell in a corner. The other rules are
for the neighbours of a cell of the track: the rules for a blank neighbour and
the rules for a blue milestone. Also, as can be seen on Figures~\ref{idle_vert}
and~\ref{idle_horiz}, there are milestones, as those on face~5 and~6 which have
a common neighbour which is blank: this explains the rule in the last row of the table. 
Now, in the last two rules, a cell has at least ten blank neighbours. We shall
decide that in such a case, the new state of the rule is the same as its current
one and we shall no more display such rules. 

\vtop{
\vspace{-10pt}
\begin{tab}\label{track_cons}
\leurre
Conservative rules for the tracks: above, the straight element, below, the corner.
\end{tab}
\vspace{-12pt}
\grostrait
\vskip 2pt
{\ttviii
\obeylines
\leftskip 0pt
\obeyspaces\global\let =\ \parskip=-2pt
          -1   0   1   2   3   4   5   6   7   8   9  10  11  12
\par}
\vskip 4pt
{\ttviii
\obeylines
\leftskip 0pt
\obeyspaces\global\let =\ \parskip=-2pt

(0)        W   W   W   W   W   W   W   W   W   W   W   W   W   W
(0)        W   W   W   B   W   W   B   B   B   W   W   W   W   W
(0)        W   W   W   W   B   W   B   B   B   B   W   B   B   W
(0)        B   W   W   W   W   W   W   W   W   W   W   W   W   B
(0)        W   B   W   W   W   W   W   W   W   W   W   W   W   W
(0)        W   B   B   W   W   W   W   W   W   W   W   W   W   W
\par}
\demitrait
\vskip 7pt
\noindent\small\baselineskip 9pt
{\sc Note}: Above the table we have the number of the faces which number the neighbours
of the cell. Number~$-$$1$ denotes the entry for the current state of the cell itself
while number~$12$ denotes the new state of the cell.}
\vskip 7pt
   The motion rules for the track are given in Tables~\ref{straight_move} 
and~\ref{corner_move}. 

   In Table~\ref{straight_move}, first, we have four rules for the motion of the 
locomotive when it enters the cell through entry~1 and 
leaves it through exit~3. The next two lines deals with the same motion through
exit~4: the difference occurs only when the front of the locomotive is again outside
the cell, so that in this case, the first tow lines are the same as when the exit is~3.
The other six rules concern the reverse motion: the locomotive enters the cell
through exit~3 and then exits 
\ligne{\hfill}
\vskip-8pt
\vtop{
\vspace{-10pt}
\begin{tab}\label{straight_move}
\leurre
Motion rules for the straight element.
\end{tab}
\vspace{-12pt}
\grostrait
\vskip 2pt
{\ttviii
\obeylines
\leftskip 0pt
\obeyspaces\global\let =\ \parskip=-2pt
          -1   0   1   2   3   4   5   6   7   8   9  10  11  12
\par}
\vskip 5pt
{\ttviii
\obeylines
\leftskip 0pt
\obeyspaces\global\let =\ \parskip=-2pt
(1)        W   W   B   B   W   W   B   B   B   W   W   W   W   B
(2)        B   W   R   B   W   W   B   B   B   W   W   W   W   R
(3a)       R   W   W   B   B   W   B   B   B   W   W   W   W   W
(4a)       W   W   W   B   R   W   B   B   B   W   W   W   W   W
\vskip3pt
(3b)       R   W   W   B   W   B   B   B   B   W   W   W   W   W
(4b)       W   W   W   B   W   R   B   B   B   W   W   W   W   W
(3c)       R   W   W   B   W   W   B   B   B   B   W   W   W   W
(4c)       W   W   W   B   W   W   B   B   B   R   W   W   W   W
(3d)       R   W   W   B   W   W   B   B   B   W   W   B   W   W
(4d)       W   W   W   B   W   W   B   B   B   W   W   R   W   W
\vskip5pt
(1ra)      W   W   W   B   B   W   B   B   B   W   W   W   W   B
(2ra)      B   W   W   B   R   W   B   B   B   W   W   W   W   R
(3r)       R   W   B   B   W   W   B   B   B   W   W   W   W   W
(4r)       W   W   R   B   W   W   B   B   B   W   W   W   W   W
\vskip3pt
(1rb)      W   W   W   B   W   B   B   B   B   W   W   W   W   B
(2rb)      B   W   W   B   W   R   B   B   B   W   W   W   W   R
(1rc)      W   W   W   B   W   W   B   B   B   B   W   W   W   B
(2rc)      B   W   W   B   W   W   B   B   B   R   W   W   W   R
(1rd)      W   W   W   B   W   W   B   B   B   W   W   B   W   B
(2rd)      B   W   W   B   W   W   B   B   B   W   W   R   W   R
\par}
\demitrait
\vskip 3pt
\noindent\small\baselineskip 9pt
{\sc Note}: $1$, $2$, $3$ and~$4$ indicates the four times of the passage of the locomotive
through the element. At time~$1$, the front of the locomotive is seen by the element. At 
time~$2$, the front is in the element. At time~$3$ the rear is in the element.
At time~$4$ the rear is about to be out of scope of the element. The letter~$r$ indicates
the reverse motion. Letters~$a$, $b$, $c$ and~$d$ refer to exits~$3$, $4$, $8$ and~$10$
respectively.
}

\vtop{
\begin{tab}\label{corner_move}
\leurre
Motion rules for the corner.
\end{tab}
\vspace{-12pt}
\grostrait
\vskip 2pt
{\ttviii
\obeylines
\leftskip 0pt
\obeyspaces\global\let =\ \parskip=-2pt
          -1   0   1   2   3   4   5   6   7   8   9  10  11  12
\par}
\vskip 5pt
{\ttviii
\obeylines
\leftskip 0pt
\obeyspaces\global\let =\ \parskip=-2pt
(1a)       W   W   B   W   B   W   B   B   B   B   W   B   B   B
(2a)       B   W   R   W   B   W   B   B   B   B   W   B   B   R
(3a)       R   W   W   B   B   W   B   B   B   B   W   B   B   W
(4a)       W   W   W   R   B   W   B   B   B   B   W   B   B   W
--
(1r)       W   W   W   B   B   W   B   B   B   B   W   B   B   B
(2r)       B   W   W   R   B   W   B   B   B   B   W   B   B   R
(3r)       R   W   B   W   B   W   B   B   B   B   W   B   B   W
(4r)       W   W   R   W   B   W   B   B   B   B   W   B   B   W
\par}
\demitrait
\vskip 3pt
\noindent\small\baselineskip 9pt
{\sc Note}: $1$, $2$, $3$ and~$4$ have the same meaning as in 
Table~{\rm\ref{straight_move}}. The letter~$a$ indicates the motion from~$1$ to~$2$
and the letter~$r$ indicates the reverse motion.
}
\vskip 11pt
\noindent
through exit~1{} in the four first rows of this group
and it enters through exit~4{} in the last two lines, the required instructions
when the rear of the locomotive is in the cell and then leaves it being the same 
as when the locomotive enters the cell through exit~3. The rules for cells
using other exits, as~8 or~10 are exactly similar: the other exit is always~1 as 
already mentioned in Subsection~\ref{elements}.

   As visible from Table~\ref{corner_move}, there are less rules in this case.
There are only two exits: 1 and~2, which gives rise to one way and the reverse motion only.

\subsection{The rules for the memory switches}
\label{memory_rules}

   Again, the rules are split into conservative and motion ones.
Table~\ref{memo_cons} shows these rules for the memory switch, both for a left-hand side 
one or a right-hand side one. The first two rules concern cells~7 and~12 as can be seen
by the occurrence of a non blank neighbour through face~0. We can see the presence of
an additional milestone as this was announced in the figures regarding memory switches.
The next four rules concern the sensors under track~3 and under track~4. The next
two rules concern the upper controller, then the following two rules deal with the
lower controller. The last two rules concern the markers of the upper controller. 
Note that the central cell itself do not require additional rules as it is a straight 
element.

\vtop{
\vspace{-5pt}
\begin{tab}\label{memo_cons}
\leurre
Conservative rules for the switches.
\end{tab}
\vspace{-12pt}
\grostrait
\vskip 2pt
{\ttviii
\obeylines
\leftskip 0pt
\obeyspaces\global\let =\ \parskip=-2pt
          -1   0   1   2   3   4   5   6   7   8   9  10  11  12
\par}
\vskip 5pt
{\ttviii
\obeylines
\leftskip 0pt
\obeyspaces\global\let =\ \parskip=-2pt
(0-7)      W   B   W   B   W   W   B   B   B   W   W   W   B   W
(0-12)     W   R   W   B   W   W   B   B   B   W   W   W   B   W
\vskip5pt 
(0-17)     B   W   W   B   W   W   W   W   W   W   R   R   R   B
(0-17)     R   W   W   B   W   W   W   W   W   W   R   R   R   R
\vskip5pt 
(0-18)     B   W   W   W   W   W   B   W   W   R   R   W   R   B
(0-18)     R   W   W   W   W   W   B   W   W   R   R   W   R   R
\vskip5pt 
(0-20)     B   B   W   R   W   W   R   R   R   R   W   B   R   B
(0-20)     B   B   W   R   W   W   R   R   R   B   W   R   R   B
\vskip5pt 
(0-19)     B   B   W   R   B   R   R   R   R   W   W   W   R   B
(0-19)     B   B   W   R   R   B   R   R   R   W   W   W   R   B 
\vskip5pt 
(0-21-22)  B   B   W   W   W   W   W   W   W   W   R   R   R   B
(0-21-22)  R   B   W   W   W   W   W   W   W   W   R   R   R   R
\par}
\demitrait
\vskip 3pt
\noindent\small\baselineskip 9pt
{\sc Note}: The number indicated after~$0$ is the number of the cell to
which the rule applies. Remember that~$0$ itself is the mark of the conservative
rules
}
\vskip 10pt
   Note that the rules needed by the red neighbours of the controllers, the sensors
and the markers of the controllers have at least ten blank neighbours and,
consequently, they are conservative.
\vskip 5pt
   Now, let us turn to the rules needed by the motions of the locomotive.
They are displayed in Tables~\ref{memo_move_track}, \ref{memo_move_sensors}
and~\ref{memo_move_control}.   

   Table~\ref{memo_move_track} displays the rules of the motion of the
locomotive in cells~7 and~12 which are the cells contiguous to the exits of the
central cell. We know that the configuration of these cells is particular. With respect
to a straight element, we have an additional blue milestone on face~11 and we have
the addition of a sensor on face~0. Cells~7 and~12 have never a sensor of the same colour.
In an idle situation, one sensor is blue, the other is red. This situation remains
unchanged during an active crossing of the switch by the locomotive or when the latter
performs a passive crossing through the selected track. This can be seen in the table
where the rules are those of the straight element, see Table~\ref{straight_move}, 
except the occurrence of the new colours on face~11 and face~0.

\vtop{
\vspace{-5pt}
\begin{tab}\label{memo_move_track}
\leurre
Rules for the motions of the locomotive across a memory switch
for cells~$7$ and~$12$.
\end{tab}
\vspace{-12pt}
\grostrait
\vskip 2pt
{\ttviii
\obeylines
\leftskip 0pt
\obeyspaces\global\let =\ \parskip=-2pt
          -1   0   1   2   3   4   5   6   7   8   9  10  11  12
\par}
\vskip 5pt
{\ttviii
\obeylines
\leftskip 0pt
\obeyspaces\global\let =\ \parskip=-2pt
active crossing:
\vskip2pt
(1)        W   B   B   B   W   W   B   B   B   W   W   W   B   B
(2)        B   B   R   B   W   W   B   B   B   W   W   W   B   R
(3)        R   B   W   B   W   B   B   B   B   W   W   W   B   W
(4)        W   B   W   B   W   R   B   B   B   W   W   W   B   W
\vskip5pt
passive crossing through the selected track:
\vskip2pt
(1)        W   B   W   B   W   B   B   B   B   W   W   W   B   B
(2)        B   B   W   B   W   R   B   B   B   W   W   W   B   R
(3)        R   B   B   B   W   W   B   B   B   W   W   W   B   W
(4)        W   B   R   B   W   W   B   B   B   W   W   W   B   W
(5)        W   W   W   B   W   W   B   B   B   W   W   W   B   W
\vskip5pt
passive crossing through the non-selected track:
\vskip2pt
(1)        W   R   W   B   W   B   B   B   B   W   W   W   B   B
(2)        B   R   W   B   W   R   B   B   B   W   W   W   B   R
(4)        W   B   R   R   W   W   B   B   B   W   W   W   B   W
(5)        W   W   R   R   W   W   B   B   B   W   W   W   B   W
\vskip2pt
(6)        W   W   R   B   W   W   R   B   B   W   W   W   B   W
(7)        W   B   R   B   W   W   R   B   B   W   W   W   B   W
\vskip5pt
blocking effect of the red sensor:
\vskip2pt
(1c)       W   R   B   B   W   W   B   B   B   W   W   W   B   W
(2c)       W   R   R   B   W   W   B   B   B   W   W   W   B   W
\vskip5pt
blocking effect of a blank upper controller:
\vskip2pt
(1d)       W   B   B   B   W   W   W   B   B   W   W   W   B   W
(2d)       W   B   R   B   W   W   W   B   B   W   W   W   B   W
(1e)       W   B   B   W   W   W   B   B   B   W   W   W   B   W
(2e)       W   B   R   W   W   W   B   B   B   W   W   W   B   W
\par}
\demitrait
\vskip 3pt
\noindent\small\baselineskip 9pt
{\sc Note}: Note that the rule 
{\ttviii R   W   B   W   W   W   B   B   B   W   W   W   B   W} does not
appear in the rules for the passive crossing through the non-selected track as it is
a rotated form of a motion rule of a straight element.
}
\vskip 10pt
   The last instructions of the table, starting from those devoted to the passive
crossing of the locomotive through the non-selected track are more intricate.

  The instructions of this group control the blocking effect of the red sensor
in the active crossing and in the passive crossing through the selected track.
They also control the passive effect of the red sensor during the crossing when the
locomotive arrives from the non-selected track. It is interesting to remark that
the rule which allows the rear of the locomotive to leave cell~12 to the central cell 
is not present in the table because it is a rotated form of a rule already used in
the straight element. Indeed, the rule is of the form
\hbox{\ttviii 
$\underline{\hbox{\ttviii R}}$   X   B   W   W   Y   B   B   B   W   W   W   B   
$\underline{\hbox{\ttviii W}}$},
where {\ttviii X} and {\ttviii Y} are the respective state of cell~18 and cell~20.
Now, when {\ttviii X} and~{\ttviii Y} are both~{\ttviii W}, we have
the rule \hbox{\ttviii
$\underline{\hbox{\ttviii R}}$   W   B   W   W   W   B   B   B   W   W   W   B   
$\underline{\hbox{\ttviii W}}$} which is a rotated form of the rule
\hbox{\ttviii 
$\underline{\hbox{\ttviii R}}$  W   B   B   W   W   B   B   B   W   W   W   W   
$\underline{\hbox{\ttviii W}}$}, which occurs at the line $(3r)$ of 
Table~\ref{straight_move}. We can notice that the latter form is obtained from
the former one by the rotation labelled $(10\ 9)$ in Figure~\ref{rotations}.

At last, note the structure of the rules labelled $(1d)$, $(2d)$, $(1e)$ and~$(2,e)$.
These rules show the blocking effect of cell~20 when it is white: this prevents the
locomotive to enter on the wrong track.

\vtop{
\vspace{-5pt}
\begin{tab}\label{memo_move_sensors}
\leurre
Rules for the motions of the locomotive across a memory switch
for the sensors: cells~$17$ and~$18$.
\end{tab}
\vspace{-12pt}
\grostrait
\vskip 2pt
{\ttviii
\obeylines
\leftskip 0pt
\obeyspaces\global\let =\ \parskip=-2pt
          -1   0   1   2   3   4   5   6   7   8   9  10  11  12
\par}
\vskip 5pt
{\ttviii
\obeylines
\leftskip 0pt
\obeyspaces\global\let =\ \parskip=-2pt
cell 17:
\vskip2pt
(1a)       B   B   W   B   W   W   W   W   W   W   R   R   R   B
(2a)       B   R   W   B   W   W   W   W   W   W   R   R   R   B
(3a)       B   W   W   R   W   W   W   W   W   W   R   R   R   R
\vskip2pt
(1b)       R   B   W   B   W   W   W   W   W   W   R   R   R   W
(2b)       W   R   W   B   W   W   W   W   W   W   R   R   R   W
(3b)       W   W   W   R   W   W   W   W   W   W   R   R   R   B
\vskip5pt
cell 18:
\vskip2pt
(2a)       B   R   W   W   W   W   B   W   W   R   R   W   R   B
\vskip2pt
(1b)       R   B   W   W   W   W   B   W   W   R   R   W   R   W
(2b)       W   R   W   W   W   W   B   W   W   R   R   W   R   W
(4b)       W   W   W   W   W   W   B   W   W   R   R   W   R   W
\par}
\demitrait
\vskip 3pt
}
\vskip 10pt
   Table~\ref{memo_move_sensors} gives the rules which control the behaviour of
the sensors in the memory switch. There are two kinds of rules: one for cell~17, 
the other for cell~18, although the working of these cells is the same. However, 
their configuration are not rotated from one another: they are symmetric. Indeed, 
cell~19 is seen from face~5 by cell~18 and from face~2 by cell~17. This is why we 
have two sets of rather similar rules. However, note that a few rules for cell~17 
also work for cell~18 as we have less rules for the latter. On the rules of 
Table~\ref{memo_move_sensors}, we can see the implementation of the scenario 
depicted in Subsection~\ref{memory} explaining the reason why a blue sensor cannot 
directly turn to red.

   Table~\ref{memo_move_control} gives the complement of this implementation
by giving the rules which govern both controllers. First, the lower controller,
and then the upper one. We also have four rules for the markers of the upper controller:
the rules which do not change the colours when cell~20 is blank and the rules 
needed to exchange the colours when cell~20 is red.

   It should not be surprising to see that there are many rules for the upper
controller. In fact, the change of position in the colours of the markers, which
are symmetric with respect to the plane of self-reflection of cell~20 considered 
without its markers, forces us to produce two sets of very similar rules: one for
each position of the markers. Also, the upper controller sees both cells~7 and~12
which requires additional rules. In comparison, the lower controller sees only
cells~17 and~18 which makes its job simpler and requires a smaller number of rules.

\vtop{
\vspace{-5pt}
\begin{tab}\label{memo_move_control}
\leurre
Rules for the motions of the locomotive across a memory switch
for the controllers and their markers: cells~$19$, $20$, $21$ and~$22$.
\end{tab}
\vspace{-12pt}
\grostrait
\vskip 2pt
{\ttviii
\obeylines
\leftskip 0pt
\obeyspaces\global\let =\ \parskip=-2pt
          -1   0   1   2   3   4   5   6   7   8   9  10  11  12
\par}
\vskip 5pt
{\ttviii
\obeylines
\leftskip 0pt
\obeyspaces\global\let =\ \parskip=-2pt
cell 19:
\vskip2pt
(1a)       B   B   W   R   B   W   R   R   R   W   W   W   R   R
(2a)       R   B   W   R   B   W   R   R   R   W   W   W   R   B
\vskip2pt
(1b)       B   B   W   R   W   B   R   R   R   W   W   W   R   R
(2b)       R   B   W   R   W   B   R   R   R   W   W   W   R   B
\vskip2pt
(1c)       B   W   W   R   B   W   R   R   R   W   W   W   R   R
(2c)       R   R   W   R   B   W   R   R   R   W   W   W   R   B
\vskip2pt
(1d)       B   W   W   R   W   B   R   R   R   W   W   W   R   R
(1d)       R   R   W   R   W   B   R   R   R   W   W   W   R   B
\vskip5pt
cell 20:
\vskip2pt
(1a)       B   B   W   R   W   B   R   R   R   R   W   B   R   B
(2a)       B   B   W   R   W   R   R   R   R   R   W   B   R   B
\vskip2pt
(1b)       B   B   W   R   B   W   R   R   R   R   W   B   R   W
(2b)       W   B   W   R   R   W   R   R   R   R   W   B   R   R
(3b)       R   B   W   R   W   W   R   R   R   R   W   B   R   B
\vskip2pt
(1c)       B   B   W   R   B   W   R   R   R   B   W   R   R   B
(2c)       B   B   W   R   R   W   R   R   R   B   W   R   R   B
\vskip2pt
(1b)       B   B   W   R   W   B   R   R   R   B   W   R   R   W
(2b)       W   B   W   R   W   R   R   R   R   B   W   R   R   R
(3b)       R   B   W   R   W   W   R   R   R   B   W   R   R   B
(4b)       R   R   W   R   W   B   R   R   R   R   W   B   R   B
(5b)       R   R   W   R   W   W   R   R   R   R   W   B   R   B
(6b)       B   R   W   R   W   W   R   R   R   B   W   R   R   B
(7b)       R   R   W   R   W   W   R   R   R   B   W   R   R   B
\vskip5pt
cells 21 and 22:
\vskip2pt
(1a)       B   W   W   W   W   W   W   W   W   W   R   R   R   B
(2a)       R   W   W   W   W   W   W   W   W   W   R   R   R   R
(1b)       B   R   W   W   W   W   W   W   W   W   R   R   R   R
(2b)       R   R   W   W   W   W   W   W   W   W   R   R   R   B
\par}
\demitrait
\vskip 3pt
}

\subsection{The rules for the fixed switch}
\label{fixed_rules}

   As mentioned in Subsection~\ref{fixed}, the working of the fixed switches makes
us of the mechanism of an upper controller whose markers are fixed: it will do the
job to prevent the locomotive arriving from the non-selected track to  be duplicated
in the wrong way. As the markers are fixed, the controller cannot change it, even if it
flashes. As now markers are milestones, they are indifferent to the flashes of the
controller. This is why we have a very few rules to append to what has already been
settled for the memory switches, see Table~\ref{fixed_misc}. Now, the sensors are 
also changed to milestones and the lower controller disappeared. This entails a 
few changes in the rules as a few rules of cells~7 and~12 have to be changed as well 
as a few rules of cell~20 whose neighbour sharing its face~0 is now always blank.

\vtop{
\vspace{-5pt}
\begin{tab}\label{fixed_misc}
\leurre
Rules for the motions of the locomotive across a memory switch
for the controllers and their markers: cells~$19$, $20$, $21$ and~$22$.
\end{tab}
\vspace{-12pt}
\grostrait
\vskip 2pt
{\ttviii
\obeylines
\leftskip 0pt
\obeyspaces\global\let =\ \parskip=-2pt
          -1   0   1   2   3   4   5   6   7   8   9  10  11  12
\par}
\vskip 5pt
{\ttviii
\obeylines
\leftskip 0pt
\obeyspaces\global\let =\ \parskip=-2pt
cells 7 and 12:
\vskip2pt
(3)        R   R   B   W   W   W   B   B   B   W   W   W   B   W
(4)        W   R   R   R   W   W   B   B   B   W   W   W   B   W
\vskip5pt
cell 20:
\vskip2pt
(0)        B   W   W   R   W   W   R   R   R   R   W   B   R   B
(1)        B   W   W   R   W   B   R   R   R   R   W   B   R   B
(2)        B   W   W   R   W   R   R   R   R   R   W   B   R   B
(3)        B   W   W   R   B   W   R   R   R   R   W   B   R   W
(4)        W   W   W   R   R   W   R   R   R   R   W   B   R   R
(5)        R   W   W   R   W   W   R   R   R   R   W   B   R   B
\par}
\demitrait
\vskip 3pt
}

\subsection{The rules for the flip-flop switch}
\label{flip_flop_rules}

   In Table~\ref{flip_flop_cons}, we have the conservative rules which are specific
to the flip-flop switches. This comes from the fact that although the flip-flop switch
takes from the memory switch its lower mechanism, it also changes a bit the working
of this mechanism.

\vtop{
\vspace{-5pt}
\begin{tab}\label{flip_flop_cons}
\leurre
Rules for the motions of the locomotive across a flip-flop switch. 
\end{tab}
\vspace{-12pt}
\grostrait
\vskip 2pt
{\ttviii
\obeylines
\leftskip 0pt
\obeyspaces\global\let =\ \parskip=-2pt
          -1   0   1   2   3   4   5   6   7   8   9  10  11  12
\par}
\vskip 5pt
{\ttviii
\obeylines
\leftskip 0pt
\obeyspaces\global\let =\ \parskip=-2pt
cell 17:
\vskip2pt
(0)        B   W   W   B   W   W   W   R   R   R   R   R   B   B
(0)        R   W   W   B   W   W   W   R   R   R   R   R   B   R
\vskip5pt
cell 18:
\vskip2pt
(0)        B   W   W   W   W   W   B   R   R   R   R   R   B   B
(0)        R   W   W   W   W   W   B   R   R   R   R   R   B   R
\vskip5pt
cell 19:
\vskip2pt
(0)        B   W   W   R   R   B   R   R   R   W   W   W   R   B
(0)        B   W   W   R   B   R   R   R   R   W   W   W   R   B
\par}
\demitrait
\vskip 3pt
}
\vskip 10pt

We have
seen in Subsection~\ref{flip_flop} that the sensors, again red and blue, work in a 
different way: the role of the colours is somehow reversed. This means that the 
rules for cells~17 and~18 must be new: the change of the blue sensor to white is
now triggered by seeing the front of the locomotive through face~0. This is also 
why the patterns of the neighbours of these cells is different from their patterns 
in the memory switch. Also, as the 
upper controller disappeared, the neighbour of cell~19 through face~0 is always blank.
Even if its rules are basically the same as in memory switches, they have to be
changed for what is the neighbour through face~0. All these features
appear in Table~\ref{flip_flop_move}. 
 
\vtop{
\vspace{-5pt}
\begin{tab}\label{flip_flop_move}
\leurre
Rules for the motions of the locomotive across a flip-flop switch. 
\end{tab}
\vspace{-12pt}
\grostrait
\vskip 2pt
{\ttviii
\obeylines
\leftskip 0pt
\obeyspaces\global\let =\ \parskip=-2pt
          -1   0   1   2   3   4   5   6   7   8   9  10  11  12
\par}
\vskip 5pt
{\ttviii
\obeylines
\leftskip 0pt
\obeyspaces\global\let =\ \parskip=-2pt
cells 7 and 12:
\vskip2pt
(4)        R   W   W   B   W   B   B   B   B   W   W   W   B   W
(5)        W   W   W   B   W   R   B   B   B   W   W   W   B   W
\vskip5pt
cell 17:
\vskip2pt
(1)        B   B   W   B   W   W   W   R   R   R   R   R   B   W
(2)        W   R   W   B   W   W   W   R   R   R   R   R   B   W
(3)        W   W   W   R   W   W   W   R   R   R   R   R   B   R
(4)        R   W   W   R   W   W   W   R   R   R   R   R   B   B
\vskip5pt
cell 18:
\vskip2pt
(1)        B   B   W   W   W   W   B   R   R   R   R   R   B   W
(2)        W   R   W   W   W   W   B   R   R   R   R   R   B   W
(3)        W   W   W   W   W   W   R   R   R   R   R   R   B   R
\vskip5pt
cell 19:
\vskip2pt
(1)        B   W   W   R   W   R   R   R   R   W   W   W   R   R
(1a)       B   W   W   R   R   W   R   R   R   W   W   W   R   R
(1a)       B   W   W   R   B   W   R   R   R   W   W   W   R   R
(2b)       R   W   W   R   W   B   R   R   R   W   W   W   R   B
(2c)       R   W   W   R   B   W   R   R   R   W   W   W   R   B
(1d)       B   W   W   R   W   B   R   R   R   W   W   W   R   R
(6a)       R   W   W   R   W   R   R   R   R   W   W   W   R   B
(6b)       R   W   W   R   R   W   R   R   R   W   W   W   R   B

\par}
\demitrait
\vskip 3pt
\noindent\small\baselineskip 9pt
{\sc Note}: There is one rule less for cell~$18$ than for rule~$17$. This comes
from the fact that the rule~$(4)$ for cell~$17$ also works for cell~$18$ as it
is a rotated image of the rule needed for cell~$18$ in a similar situation. 
}

\subsection{About the computer program}
\label{computer}

   As indicated in the introduction, I wrote a computer program in order to
check the correctness of the rules. The program was written in $ADA95$ and
it implements the algorithms mentioned in the paper.

   Before giving a short account on the program itself, I would like to stress
that using a simulation program for this purpose is mandatory. The computations
are so complex for a man, at least for me, that the help of the computer allows
me to check that the rules are correct. What is meant by this latter expression?
We mean two things: a syntactical one and a semantic one. The syntactical
correctness is that there is no pair of rules with the same context giving rise
to different new states. This is the minimal condition when working with
deterministic cellular automata, which is of course the case here. In this work,
we reinforced the condition by checking that two rules with the same minimal
context always give rise to the same new state: this guarantees that the automaton
is not only correct, but that it is also rotation invariant.

   Now, the semantic correctness means that the rules do what we expect from them to
do. This is far more complex to check and this cannot be completely ascertained
by proof. Again, the computer program is useful in this regard. We can implement the
simulation in the program and then run it. If everything goes smoothly through,
we can believe that the implementation is correct. There is no guarantee of that.
There is no automatic checking that the implementation is a correct hyperbolic
implementation. There is also no proof that the program itself is correct. However, the
setting is rather involved and while adjusting the program, many errors in my
first table of rules were found by the program. As an example, the program also
indicated me the need to cover face~11 of cells~7 and~12 with a blue milestone: 
otherwise, the set of rules would not be rotationally invariant. 

\setbox110=\vtop{\hsize=167pt
\parindent 0pt
{\ttvi
\obeylines
\leftskip 0pt
\obeyspaces\global\let =\ \parskip=-2pt
  6   -1  0  1  2  3  4  5  6  7  8  9 10 11
       W  W  W  B  W  W  B  B  B  W  W  W  W
       v  f  v  f  v  v  f  f  f  f  f  f  f
             5     7 12
  7    W  B  W  B  W  W  B  B  B  W  W  W  B
       v  v  v  f  f  v  v  f  f  f  f  f  f
         17  6        8 20
  8    W  W  W  B  W  W  B  B  B  W  W  W  W
       v  f  v  f  f  v  f  f  f  f  f  f  f
             7        9
  9    W  W  W  B  W  W  B  B  B  W  W  W  W
       v  f  v  f  f  v  f  f  f  f  f  f  f
             8       10
 10    W  W  W  B  W  W  B  B  B  W  W  W  W
       v  f  v  f  f  v  f  f  f  f  f  f  f
             9       11
 11   -1  0  1  2  3  4  5  6  7  8  9 10 11
       W  W  W  B  W  W  B  B  B  W  W  W  W
       v  f  v  f  f  f  f  f  f  f  f  f  f
            10
 12    W  R  W  B  W  W  B  B  B  W  W  W  B
       v  v  v  v  f  v  f  f  f  f  f  f  f
         18  6 20    13
 13    W  W  W  B  W  W  B  B  B  W  W  W  W
       v  f  v  f  f  v  f  f  f  f  f  f  f
            12       14
\par}
}

\setbox112=\vtop{\hsize=167pt
\parindent 0pt
{\ttvi
\obeylines
\leftskip 0pt
\obeyspaces\global\let =\ \parskip=-2pt
17    B  W  W  B  W  W  W  W  W  W  R  R  R
       f  v  f  v  f  f  f  f  f  f  f  f  f
          7    19
 18    R  W  W  W  W  W  B  W  W  R  R  W  R
       f  v  f  f  f  f  v  f  f  f  f  f  f
         12             19
 19    B  B  W  R  B  R  R  R  R  W  W  W  R
       f  v  f  f  v  v  f  f  f  f  f  f  f
         20       17 18
 20    B  B  W  R  W  W  R  R  R  R  W  B  R
       f  v  f  f  v  v  f  f  f  v  f  v  f
         19       12  7          21    22
 21   -1  0  1  2  3  4  5  6  7  8  9 10 11
       R  B  W  W  W  W  W  W  W  W  R  R  R
       v  v  f  f  f  f  f  f  f  f  f  f  f
         20
 22    B  B  W  W  W  W  W  W  W  W  R  R  R
       v  v  f  f  f  f  f  f  f  f  f  f  f
         20
\par}
}
\vtop{
\vspace{-5pt}
\begin{tab}\label{exec}
\leurre
Two pieces of the big trace of the program, corresponding to the simulation of
an active passage of the locomotive through a left-hand side memory switch,
at initial time.
\end{tab}
\vspace{-12pt}
\grostrait
\vskip 2pt
\ligne{\hfill
\box110
\hskip 5pt
\box112
\hfill}
\vskip 4pt
\demitrait
}
\vskip 10pt

   About the program implementation itself.

   First, I implemented the algorithms to compute the minimal form of a rule
and, taking advantage of this implementation, the programme computed the
PostScript program for Figure~\ref{rotations}. All traces given in the tables of
Section~\ref{the_switches} were computed during the execution of the program by the 
program itself.

   Each test of a crossing of the switch by the locomotive was performed
within the same implementation frame: the cells of the tracks were gathered
in a table of tables. The big table has 22 entries corresponding to the numbering
of the cells explained in Section~\ref{the_switches}. For each index of the big table,
a table of 14 entries gives various information on the cell in its current state,
and its neighbours. The neighbours correspond to faces and are numbered from~0
up to~11 as in the paper. For each face, it is indicated whether the neighbour 
through the face has a permanent state or a variable one. As an example, a milestone
seen from a face of a cell of the track is permanently blue. When the neighbour
has a variable state, the table indicates the index of this neighbour in the big table.
In fact, the big table is first a list of the variable cells and, at this occasion,
it collects a useful information about each cell. The program also computes a bigger
trace where at each time, the big table is dispatched in full detail. Table~\ref{exec}
gives two short pieces of this trace, taken during an active passage of the locomotive,
at time~1. We can see the information which was just indicated, $v$ meaning 'variable'
and $f$ meaning 'fixed'. The other indications are self-explaining.

  \vskip 5pt
  As the program performed a successful execution of all possible crossings and
also along various vertical and horizontal segments with some mix of them,
we can conclude that the proof of theorem~\ref{mainthm} is complete.
\cqfd

\section{Conclusion}

   With this result, we are very close to the limit which amounts to two states.
As noticed in the introduction, the result proved in this paper can be obtained
very 'easily' by using a rather trivial implementation of rule~110 which is now
known as weakly universal. The present proof is much simpler, although its checking
involved a computer program. By the way, it is also the case of rule~110 and if we 
compare the programming effort needed for rule~110 together with cost in time and
space of the executions of the programme with the corresponding issues needed for 
this paper, the programming effort and the execution costs of the program in 
the case of this paper are much less than what was needed for rule~110. This is 
also why the present proof is much simpler, which of course, does not diminish at
all the merit of the discoverers of rule~110. 

   Moreover, the automaton constructed in this proof is truly a $3D$-one. As 
mentioned in the introduction, the third dimension is used to avoid the crossings, 
which spares a lot of instructions. Also, as noticed in Section~\ref{the_switches}, 
the third dimension allows us to perform the fixed and the memory switches without 
the side effect of inconveniences which were noticed in the plane.

   Accordingly, there is some work ahead, probably the hardest as we are now close 
to the goal.

\end{document}